\documentclass[a4paper,fleqn,usenatbib]{mnras}

\usepackage{natbib}
\usepackage{multicol}

\usepackage{graphicx}	
\usepackage{amsmath}	
\usepackage{amssymb}	
\usepackage{subcaption}
\usepackage{float}
\usepackage{multirow}

\DeclareGraphicsExtensions{.pdf,.png,.jpg,.eps}
\graphicspath{{figures/}}

\newcommand{\mean}[1]{\ensuremath{\left\langle #1 \right\rangle}}

\newcommand{\lcdm}{\ensuremath{\Lambda\text{CDM}}}

\newcommand{\mpch}{\ensuremath{h^{-1}\text{Mpc}}}

\def\reff@jnl#1{{\rm#1\/}}

\def\aj{\reff@jnl{AJ}}                  
\def\araa{\reff@jnl{ARA\&A}}            
\def\apj{\reff@jnl{ApJ}}                        
\def\apjl{\reff@jnl{ApJ}}               
\def\apjs{\reff@jnl{ApJS}}              
\def\apss{\reff@jnl{Ap\&SS}}            
\def\aap{\reff@jnl{A\&A}}               
\def\aapr{\reff@jnl{A\&A~Rev.}}         
\def\aaps{\reff@jnl{A\&AS}}             
\def\baas{\reff@jnl{BAAS}}              
\def\jrasc{\reff@jnl{JRASC}}            
\def\memras{\reff@jnl{MmRAS}}           
\def\mnras{\reff@jnl{MNRAS}}            
\def\physrep{\reff@jnl{Phys.Rep.}}
\def\pra{\reff@jnl{Phys.Rev.A}}         
\def\prb{\reff@jnl{Phys.Rev.B}}         
\def\prc{\reff@jnl{Phys.Rev.C}}         
\def\prd{\reff@jnl{Phys.Rev.D}}         
\def\prl{\reff@jnl{Phys.Rev.Lett}}      
\def\pasp{\reff@jnl{PASP}}              
\def\pasj{\reff@jnl{PASJ}}              
\def\skytel{\reff@jnl{S\&T}}            
\def\solphys{\reff@jnl{Solar~Phys.}}    
\def\sovast{\reff@jnl{Soviet~Ast.}}     
\def\ssr{\reff@jnl{Space~Sci.Rev.}}     
\def\nat{\reff@jnl{Nature}}             

\newcommand{\beq}{\begin{equation}}
\newcommand{\eeq}{\end{equation}}
\newcommand{\beqa}{\begin{eqnarray}}
\newcommand{\eeqa}{\end{eqnarray}}

\usepackage{color}

\newcommand{\dev}{{de Vaucouleurs}}
\newcommand{\vx}{\ensuremath{\textbf{x}}}

\definecolor{sdeep}{rgb}{0., 0.6, 0.}

\title[BOSS FP]{Fundamental Plane of BOSS galaxies: Correlations with galaxy properties, density field and impact on RSD measurements.
}

\author[S.~Singh et al.]{
   Sukhdeep Singh$^{1}$\thanks{E-mail: sukhdeep1@berkeley.edu},
   Byeonghee Yu$^{1}$,
   Uro\v{s} Seljak$^{1}$
\\
   $^{1}$ Berkeley Center for Cosmological Physics, University of California, Berkeley, CA 94720, USA
}

\date{Accepted XXX. Received YYY; in original form ZZZ}

\pubyear{2019}

\begin{document}
\label{firstpage}
\pagerange{\pageref{firstpage}--\pageref{lastpage}}
\maketitle

\begin{abstract}
	Fundamental plane of elliptical galaxies can be used to predict the intrinsic size of galaxies and has a number of plausible application to study cosmology and galaxy physics.
	We present a detailed analysis of the fundamental plane of the SDSS-III BOSS LOWZ and CMASS galaxies. For the standard fundamental plane, we find a strong redshift evolution for
	the mean residual and show that it is primarily driven by the redshift evolution of the surface brightness of the galaxies. After correcting for the redshift evolution, 
	the FP residuals are strongly correlated with the galaxy properties and some observational systematics. We show that the variations in the FP between the central and satellite galaxies,
	that have been observed  in the literature, can primarily be explained by the correlation of the FP with the galaxy luminosity. We also measure the cross correlations of the FP residuals with the 
	galaxy density field. The amplitude of the cross correlations depends on the galaxy properties and environment with brighter and redder galaxies showing stronger correlation. In general, 
	galaxies in denser environments (higher galaxy bias ) show stronger correlations. We also compare FP amplitude with the amplitudes of intrinsic alignments of galaxy shapes (IA), finding 
	the two to be correlated. Finally, using the FP residuals we also study the impact of intrinsic 
	alignments on the constraint of growth rate using redshift space distortions. We do not observe any significant trends in measurements of the growth rate $f$ as function of the amplitude
	of FP-density correlations, resulting in null detection of the effects of IA on the RSD measurements. 
\end{abstract}

\begin{keywords}
cosmology: observations
  --- large-scale structure of Universe\ --- gravitational
  lensing: weak
\end{keywords}

\section{Introduction}
	Fundamental Plane (FP) of galaxies, an empirical relation between the size, surface brightness and the velocity dispersion of early type galaxies, 
	has been proposed as a cosmological probe to estimate distances to galaxies, galaxy velocities \citep{Strauss1995}, weak gravitational lensing 
	magnification \citep{Bertin2006}, doppler magnification of galaxies \citep{Bonvin2017}, impact of intrinsic alignments on galaxy selection functions 
	\citep{Hirata2009}; in addition to its value as probe for galaxy physics.
	
	For elliptical galaxies, a relation between the size, surface brightness and velocity dispersion can be derived from the virial theorem assuming constant
	mass to light ratio for the galaxies. Such relations have been observed for a long time \citep[eg. ][]{Dressler1987,Djorgovski1987, 
	Bernardi2003,Saulder2013,Saulder2019}, though the observed FP deviates significantly from the virial theorem predictions as the 
	galaxies do not follow the simplified 
	underlying assumptions. Furthermore FP has been observed to be a function of galaxy properties \citep[eg. ][]{Scodeggio1998,Netro2009,Saulder2019} and 
	their environment \citep{Carvalho1992,Joachimi2015,Saulder2019}.
	 
	The cosmological prowess of the FP arises from its ability to provide an estimate of the true intrinsic size of the galaxies (with some scatter). 
	The observed galaxy size is affected by several 
	processes, including estimates of cosmological distances, peculiar motion of galaxies (since we estimate distance through redshift), 
	relativistic effects including the 
	doppler shift, gravitational lensing and the effects of projecting three dimensional shapes onto the plane of the sky. 
	Once the intrinsic size of a galaxy is known, the difference between the
	observed and the true size of the galaxies (hereby size residual or FP residual) can be used to study several of these effects. The size residuals can provide a 
	(noisy) estimate of the peculiar velocity of individual galaxies which can be used to map the cosmological flows \citep[][]{Strauss1995}. Cross 
	correlations of the size residuals with the foreground galaxies (or clusters) can be also be used to measure the galaxy-lensing cross correlations 
	\citep{Bertin2006,Huff2011}. \cite{Bonvin2017} also suggested measuring the dipole of the galaxy-size residual cross correlations to estimate the 
	doppler magnification of the galaxies. \cite{Hirata2009} pointed out that due to the radial intrinsic alignment of galaxies and the projection effects,
	the size residuals are correlated with the local galaxy environment and if the selection function for a galaxy survey is sensitive to such 
	residuals, it can introduce biases into measurements of redshift anisotropy of the galaxy auto correlations. Such an effect was tentatively 
	detected by \cite{Martens2018} using the fundamental plane of SDSS-III BOSS galaxies.
	
	However, as pointed out earlier, the FP depends on the galaxy properties and their environment 
	\citep[eg. ][]{Carvalho1992,Scodeggio1998,Netro2009,Joachimi2015,Saulder2019}. 
	\cite{Joachimi2015} detected the cross correlations between the FP residuals and the galaxy density field implying that the FP residuals are influenced by
	the galaxy environment. They further detected the dependence of the FP residuals on the galaxy type, with the brightest galaxies in groups having 
	larger sizes than 
	predicted by the FP while the satellite galaxies have smaller sizes. Such dependence of the FP residuals on the galaxy environment complicates the cosmological
	applications of the FP and detailed studies are required to understand such dependencies and avoid possible contamination to the cosmological inferences.
	
	In this work,
	we extend the FP analysis from \cite{Joachimi2015} to the BOSS LOWZ and CMASS sample galaxies. We estimate the fundamental plane for 
	these samples as well as for sub-samples using splits based on color, luminosity and environment of the galaxies. We also study the cross
	correlations between FP residuals and galaxy density field and compare these cross correlations to those expected from the effects of intrinsic alignments as pointed out
	by \cite{Hirata2009}.
	Finally we also 
	perform a detailed study of contamination to the galaxy clustering measurements from the radial alignments, similar to the analysis by \cite{Martens2018}, in an
	attempt to confirm their results.
	
	Throughout this work we will use flat \lcdm\ model with the Planck 2015 \citep{Planck2015cosmo} cosmology with $h=0.677$, $\Omega_m=0.307$. 
	Unless mentioned otherwise, 
	distances are measured in unit of comoving \mpch. To compute the matter power spectrum, we use \texttt{CLASS} code \citep{Lesgourgues2011} with halofit 
	\citep{Takahashi2012} prescription for the non-linear matter power spectrum.
	For redshift-space galaxy power spectrum calculations, we employ the FFT-based algorithm implemented in \texttt{nbodykit} \citep{Hand:2017pqn}, and use \texttt{pyRSD} \citep{Hand:2017ilm} to compute the theoretical predictions of the redshift-space power spectrum of galaxies and run a likelihood analysis to find the best-fit theory model parameters.


\section{Formalism}\label{sec:formalism}
	In this section we describe our formalism used in the estimation of fundamental plane and the residuals over it (size or FP residuals); the 
	estimators and models used to study the cross-correlation between FP residuals and the density field; the power spectra and modeling of the redshift space 
	galaxy clustering.
    \subsection{Fundamental Plane}\label{ssec:formalism_FP}
    	To estimate the fundamental plane of galaxies (FP), 
		we closely follow the methodology from \cite{Saulder2013} and \cite{Joachimi2015}.
    	We define the FP as 
        \begin{equation}
            \log R_0=a\log\sigma_0+b\log I_0+c+\sum_{i=1}^{N_z} d_i z_{cor}^i,
            \label{eq:FP_definition}
        \end{equation}
        where $R_0$ is the physical radius of the galaxy, $I_0$ is the surface brightness and $\sigma_0$ is the velocity dispersion. Following \cite{Joachimi2015},
        we also introduce polynomial terms dependent on the redshift of the galaxies. 
        $z_{cor}$ is the redshift of the galaxies 
        in the CMB rest frame (correcting for the motion of the earth with rest to CMB rest frame) and is estimated as detailed in \cite{Saulder2013}.
        
        The physical size of the galaxy, $R_0$, is measured as (in units of $kpc/h$) 
        \begin{align}
    	    r_{cor}&=r_0\sqrt{q_{b/a}}\\
        	R_0&=D_A(z_{cor})\tan{(r_{cor})}\times 1000
        \end{align}
        where $r_0$ is the angular galaxy size and $q_{b/a}$ is the axis ratio which is used to measure the circularized galaxy size, $r_{cor}$ \citep{Bernardi2003,Saulder2013}, with both $r_0$ and $q_{b/a}$ measured using the \dev\ 
        profile. 
        $D_A(z)$ is the angular diameter distance in units of Mpc/h.
        
        The surface brightness, $I_0$, is computed as
        \begin{equation}
            \log I_0=-\frac{1}{2.5}\left[M_{ke}+5\log\left(\frac{D_L}{D_{L0}}\right)\right]-\log(2\pi R_0)+4\log(1+z_{cor})
            \label{eq:logI}
        \end{equation}
        $M_{ke}$ is the $k+e$ corrected absolute magnitude as defined in \cite{Singh2015}, $\frac{D_L}{D_{L0}}$ is the correction to the  
        luminosity distance due to redshift correction ($D_L$ is estimated using $z_{cor}$ while $D_{L0}$ is estimated using measured redshift in observer frame).
        $4\log(1+z_{cor})$ factor corrects for the cosmological dimming of the surface brightness \citep{Tolman1930}.
    
        We also correct the velocity dispersion, $\sigma_0$, for the
        effects of the fiber size (different correction for BOSS and SDSS spectra) as \citep{Saulder2013}
        \begin{equation}
        	\sigma_0=\sigma \left(\frac{r_\text{fiber}}{r_{cor}/8}\right)^{0.04}
        \end{equation}
        where $r_\text{fiber}=1\arcsec$ for BOSS and $r_\text{fiber}=1.5\arcsec$ for SDSS spectrographs. We identify the spectrograph from the date the
        spectra for the given galaxy was obtained and then apply the relevant correction.
        
        FP residual for a galaxy is defined in terms of the fractional difference between the measured size and the size predicted using FP,
        \begin{equation}
            \lambda_{N_z}=\ln \frac{R_0}{R_{\text{FP}, N_z}}=\log R_0-a\log\sigma_0-b\log I_0-c-\sum_{i=1}^{N_z} d_i z_{cor}^i,
            \label{eq:FP_residual}
        \end{equation}
        where $N_z$ refers to the order of polynomial used for fitting the FP as defined in eq.~\eqref{eq:FP_definition}.
        We find that the upto third order polynomials in $z$ in eq.~\eqref{eq:FP_definition} 
        are not necesasrily sufficient to fully null the redshift evolution of the $\lambda$. To further reduce the effects of redshift evolution,
        we also fit the FP in redshift bins and will denote $\lambda$ from such fits as $\lambda^{zb}$. We typically use bins with width 
        $\delta z=0.02$ for such fits to obtain $\lambda^{zb}_{N_z}$ for all galaxies and will subsequently carry out the following analysis in
        the same manner as $\lambda$ (i.e. ignoring z binning).
        
        \cite{Martens2018} ignored the velocity dispersion measurements in their FP analysis. To study the influence of velocity dispersion, we
        also define 
        \begin{align}
            \log R_0&=b^I\log I_0+c^I+\sum_{i=1}^{N_z} d_i^I z^i\\
            \lambda^I_{N_z}&=\ln \frac{R_0}{R^I_{\text{FP},N_z}}
        \end{align}
        where we used superscript $I$ to denote that the FP is only dependent on surface brightness and not the velocity dispersion. We note that this FP definition is
        not strictly equivalent to the FP definition used by \cite{Martens2018} who defined the FP using magnitudes instead of surface brightness. We will present detailed comparisons with the 
        \cite{Martens2018} results in appendix~\ref{appendix:Martens_comparison}.
    
    \subsection{Galaxy-Galaxy correlations}\label{sssec:corr_galaxy}
		\subsubsection{Estimator}
			To compute the galaxy-galaxy cross-correlation function between two different samples, we use the Landy-Szalay estimator \citep{Landy1993,Singh2017cov}
			\begin{equation}
               \xi_{gg}(r_p,\Pi)=\frac{SD-DR_S-SR_D+R_SR_D}{R_SR_D},
            \end{equation}
            where $r_p$ is the projected separation for a pair of galaxies, $\Pi$ is the line of sight separation, $D$ and $S$ refers to the dataset 
            (galaxies) being cross correlated ($D=S$ in case of auto correlations) and $R_S$ and $R_D$ refer to the set of random points corresponding to
            $S$ and $D$ samples. Product $XY$ (eg. $SD$) refers to the binned weighted count of pairs across two samples with distances that are within the
            $(r_p,\Pi)$ range of the given bin. The weight of a pair is the product of the galaxy weights that are described in section~\ref{sec:data}.
            
            The projected correlation function is then obtained by integrating $\xi_{gg}$ over the bins in 
            $\Pi$
            \begin{equation}
               w_{gg}(r_p)=\sum_{-\Pi_\text{max}}^{\Pi_\text{max}}\Delta\Pi\, \xi_{gg}(r_p,\Pi).
               \label{eq:wgg_estimator}
            \end{equation}
            Large values of $\Pi_\text{max}$ are required to reduce the impact of redshift space distortions (RSD) on measured correlation function,
            even though measurement noise increases with larger $\Pi_\text{max}$ \citep{Singh2017cov}.
            To reduce the impact of redshift space distortions on the projected correlations, 
            we use $\Pi_\text{max}=100\mpch$, with 20  bins of size $\Delta\Pi=10\mpch$. 
            
            Separately, to analyze the line of sight anisotropy, we also compute the multipoles of the correlation function as 
            \begin{equation}
               \xi_{gg,2\ell}(s)=\frac{2\ell+1}{2}\int d\mu \xi_{gg}(s,\mu)L_{2\ell}(\mu)d\mu
               \label{eq:xi_multipole_estimator}
            \end{equation}
            where $s=\sqrt{r_p^2+\Pi^2}$ is the separation between pair of galaxies in the redshift space and $\mu=\Pi/s$.

			\subsubsection{Modelling}
			The galaxy cross correlation function between samples $S$ and $D$ in redshift space is given by
			\begin{align}
				\xi_{gg}(r_p,\Pi)=&\int \mathrm{d}z W(z)b_{g,S}(s,z)b_{g,D}(s,z) r_{gg}(s,z)\int \frac{\mathrm{d}^2k_\perp\mathrm{d}k_z}{(2\pi)^3}\nonumber\\
				&\times P_{\delta\delta}(\vec{k},z)(1+\beta_S\mu_k^2)(1+\beta_D\mu_k^2) e^{\mathrm{i}(\vec{r}_p.\vec{k}_\perp+\Pi k_z)}\label{eqn:xi}.
	    	\end{align}
			where $s=\sqrt{r_p^2+\Pi^2}$, $b_g$ is the galaxy bias and is in general a function of redshift and scale, 
			$P_{\delta\delta}$ is the matter power spectrum. The Kaiser factor $(1+\beta\mu_k^2)$ accounts for the effects of redshift space 
			distortions \citep{Kaiser1995} with $\beta=f(z)/b_g$, $f$ is the growth function. We also 
			introduced the cross correlation coefficient, $r_{gg}(s,z)$, between 
			the two samples of galaxies but we will assume that $r_{gg}(s,z)=1$ on all scales used for fitting the model ($r_p>5\mpch$).
			$W(z)$ is the redshift weight accounting for the effective contributions from different redshifts to the 
			measured correlation function and is given by \citep{Mandelbaum2011}
			\begin{equation}
				W(z)=\frac{p(z)^2}{\chi^2 (z)\mathrm{d}\chi/\mathrm{d}z} \left[\int \frac{p(z)^2}{\chi^2
				(z)\mathrm{d}\chi/\mathrm{d}z} \mathrm{d}z\right]^{-1}.
			\end{equation}
			$p(z)$  is the redshift probability distribution for the galaxy sample.
			
			To compute the projected correlation function, we will assume a scale independent bias, $b_g$
			and use the effective redshift, $z$, 
			for our sample computed by integrating over weights $W(z)$. We then  
			integrate over the correlation function multipoles to obtain the projected correlation function as \citep{Baldauf2010}
			\begin{equation}
				w_{gg}(r_p)=\sum_{\ell=0}^2 2
				\int_0^{\Pi_\text{max}}d\Pi\xi_{gg,2\ell}(r) L_{2\ell}({\Pi}/{r})
				\label{eq:w_model}
			\end{equation}
			where $L_{2\ell}$ are the Legendre polynomials, prefactor 2 arises because we assume symmetry around $\Pi=0$ and change the limits of 
			integration.
			$\xi_{gg,2\ell}(r)$ are the correlation function multipole given as  
			\begin{equation}
				\xi_{gg,2\ell}(r)=(-1)^\ell\alpha_{2\ell}(\beta(z))\frac{b_{g,S}b_{g,D}}{2\pi^2}\int dk k^2 P_{\delta\delta}(k)j_{2\ell}(kr)
				\label{eq:xi_multipole_model}
			\end{equation}
			where $j_{2\ell}$ are the spherical bessel functions. We use the package \texttt{mcfit} \citep{Li2019} to compute the correlation function 
			multipoles.
			The coefficients $\alpha_{2\ell}(\beta)$ are given by \citep{Baldauf2010}
			\begin{align}
				\alpha_0(\beta)&=1+1/3(\beta_S+\beta_D)+1/5\beta_S\beta_D\label{eq:rsd_alph0}\\
				\alpha_2(\beta)&=2/3(\beta_S+\beta_D)+4/7\beta_S\beta_D\label{eq:rsd_alph2}\\
				\alpha_4(\beta)&=8/35\beta_S\beta_D\label{eq:rsd_alph4}
			\end{align}
		
	\subsection{Galaxy-FP residual cross correlations}\label{sssec:corr_FP}
		\subsubsection{\it Estimator}
				To compute the cross correlations between the galaxy density and the FP residuals, we use
				\begin{equation}
        	       \xi_{g\lambda}(r_p,\Pi)=\frac{\lambda_S D-\lambda_S R_D}{R_SR_D},
	            \end{equation}
	            $\lambda_S$ is the FP residuals for sample $S$ , 
	            $D$ is the sample of galaxies used as galaxy density tracers and $R_S,R_D$ are the 
	            corresponding randoms sample.
	            $\lambda_S D$ effectively refers to the pair counts, weighted with FP residuals $\lambda$, 
	            \begin{equation}
	            	\lambda_S D (r_p,\Pi)=\sum_{S,D} \lambda_{S}w_{SD}.
	            \end{equation}
	            The $\sum_{S,D}$ is over all the galaxy pairs with separation within the ($r_p,\Pi$ ) bin limits, $w_{SD}$ is the weight assigned to the
	            pair of galaxies and $\lambda_S$ is the FP residual from sample $S$.
	            $S_\lambda R_D$ measures the same quantity with density tracer sample being replaced by the random points. 
	            Randoms subtraction can remove potential 
	            additive systematics that donot correlate with underlying galaxy density and also leads to optimal covariance \citep{Singh2017cov}.
	            
	            We caution that this estimator can be biased if the $\mean{\lambda}$ is not zero, even after including the randoms subtraction. 
	            This is because the $\lambda$ is estimated at the 
	            position of the galaxies and is hence weighted by the galaxy density field, which results in contribution from the galaxy clustering
	            in case the $\mean{\lambda}$ is not zero, i.e.
	            \begin{align}
        	       \xi_{g\lambda}(r)=&\mean{\left[(\lambda_0+\mean{\lambda})(1+\delta_g^S)\right]\delta_g^D}(r)\nonumber\\
	       							=&\mean{\lambda_0(1+\delta_g^S)\delta_g^D}(r)+\mean{\lambda}\mean{\delta_g^S
								\delta_g^D}(r)\label{eq:xigl_density_weight}
	            \end{align}
	            where we used $\lambda_0$ to explicitly define the mean zero quantity. Thus before 
	            computing the correlation function, we subtract out
	            $\mean{\lambda}$ even though FP definition and our fitting procedure ensures that it is very small.
	            
	            The projected correlation function $w_{g\lambda}$ is then obtained by integration over line of sight as in eq.~\eqref{eq:wgg_estimator} 
	            and the multipoles are obtained as in eq.~\eqref{eq:xi_multipole_estimator}.
	            
 		\subsubsection{\it Modelling}
			Following \cite{Hirata2009}, we assume that the deviations from fundamental plane are 
			correlated with the 
			tidal field due to the effects of intrinsic alignments of galaxy shapes, i.e. galaxy shapes are aligned with tidal field in the three dimensions and the 
			projection effects then lead to correlations between tidal field and the projected shape and size of galaxies. Galaxy sizes are affected by the 
			intrinsic alignments along the line of sight.
			$\lambda$ can then be described in terms of matter field as 
			\begin{align}
				\lambda&=-A_\lambda {\zeta}\left[\nabla_z\nabla_z\nabla^{-2}-\frac{1}{3}\right]\delta_m\\
				\lambda&=A_\lambda {\zeta}\left[\frac{1}{3}-\frac{k_z^2}{k^2}\right]\delta_m\\
				\lambda&=A_\lambda \frac{\zeta}{3}\left[1-3\mu^2\right]\delta_m \label{eq:lambda}
			\end{align}
			where we used $\zeta=\frac{C_1\rho_{crit}\Omega_m}{D(z)}$ and $\mu_{\vec k}=k_z/\vec k$. Our sign convention implies that for $A_\lambda>0$ galaxies
			in higher overdensities (larger $\delta_m$) have larger size. Following convention of intrinsic alignments studies \citep[eg. ][]{Joachimi2011}, 
			we will use $C_1\rho_{crit}=0.0134$.

			We note here that in general it is  plausible that additional galaxy 
			environment effects also affect the projected galaxy sizes, in which case the deviations from the fundamental plane can be written in terms of the 
			trace of the tidal field, $\lambda\propto\nabla^2\phi\propto\delta_m$. This formalism also results in the similar form for $\lambda$ as in 
			eq.~\eqref{eq:lambda}, but with different constants and different line of sight anisotropy term as compared to $(1-3\mu^2)$. 
			Such a model was assumed by \cite{Joachimi2015} when modeling $\lambda$.
			We will use the form in eq.~\eqref{eq:lambda}
			to fit the measurements of projected correlation functions (where line of sight anisotropy has negligible effect due to large line of sight integration)
			and study the deviations from the model by comparing $A_\lambda$ to the amplitude of intrinsic alignments of galaxies $A_{IA}$, where the 
			expectation under the model is $A_\lambda=A_{IA}/2$ \citep{Hirata2009}. 
			The primary difference between our model and that
			of \cite{Joachimi2015} is that fitted values of $A_\lambda$ are rescaled by a constant $\zeta/3 = C_1\rho_{crit}\Omega_m/3D(z)$.
			
			To check for the impact of the $(1-3\mu^2)$ term, we will also compute the multipole moments of the galaxy-$\lambda$ cross correlations
			and we will replace this factor with $(1+\beta_\lambda\mu^2)$, i.e.
			\begin{equation}
				\lambda=A_\lambda \frac{\zeta}{3}\left[1+\beta_\lambda\mu^2\right]\delta_m
				\label{eq:lambda_beta}
			\end{equation}
			where $\beta_\lambda$ is a free parameter to be fit, with fiducial value set to $\beta_\lambda=-3$.


			The cross correlation function of $\lambda$ with galaxies in redshift space is given by
			\begin{align}
				\xi_{g\lambda}(r_p,\Pi)=&A_\lambda \frac{\zeta}{3}\int \mathrm{d}z W(z)b_g(r,z)r_{cc}(r,z)\nonumber\\
				\int \frac{\mathrm{d}^2k_\perp\mathrm{d}k_z}{(2\pi)^3}
				& P_{\delta\delta}(\vec{k},z)(1+\beta_g\mu_k^2)(1+\beta_\lambda\mu_k^2) e^{\mathrm{i}(\vec{r}_p.\vec{k}_\perp+\Pi k_z)}\label{eqn:xi},
	    	\end{align}
			where $r_{cc}(r,z)$ is the cross correlation coefficient between galaxies and matter. In this work, we will assume $r_{cc}(r,z)=1$ on the scales used to
			fit the model ($r_p>5\mpch$). 
			Note that
			Kaiser factor for RSD $(1+\beta\mu_k^2)$ is different from the galaxy clustering as we assume that only galaxy positions are affected 
			by RSD 
			and $\lambda$ carries the ($1+\beta_\lambda\mu^2$) term 
			(we are ignoring the fact that the FP residuals 
			are affected by RSD. RSD effects on $\lambda$ scale as $\Delta\lambda\propto \frac{v}{H(z)D(z)}$, $v$ is galaxy velocity and $D(z)$ is the line of 
			sight distance to the galaxy).
			
			The projected correlation function and multipoles are then computed using similar transforms as in galaxy clustering, 
			eq.~\eqref{eq:w_model} and eq.~\eqref{eq:xi_multipole_model}. When computing the projected correlation function, $w_{g\lambda}$, we
			fix $\beta_\lambda=-3$, while when fitting the multipoles $\beta_\lambda$ is a free parameter.
			
			As shown in eq.~\eqref{eq:xigl_density_weight}, since the FP residuals are sampled at the position of the galaxies, the measured 
			$\xi_{g\lambda}=\lambda_S(1+\delta_g^S)\delta_g^D$ is weighted by the galaxy density and thus includes higher order terms \citep[see ][ for 
			detailed study of this effect in measurements of intrinsic alignments of galaxies]{Blazek2015}. 
			Detailed modeling of this effect is outside the scope of current work 
			and we will ignore it in our models.

	\subsection{Galaxy-Intrinsic shear}\label{sssec:corr_IA}
		We follow \cite{Singh2015} and \cite{Singh2016ia} for the measurements and modeling of intrinsic alignments. We only briefly describe the methodology 
		here and refer the reader to \cite{Singh2015} for more details.
		\subsubsection{Estimator}
			The cross correlations between galaxy shapes and the galaxy density field are measured as 
			\begin{equation}
        	       \xi_{g+}(r_p,\Pi)=\frac{S_+D-S_+R_D}{R_SR_D},
	         \end{equation}
	         where $S_+D$ refers to the summation over the radial shear, $\gamma_{+,S}$, measured in the coordinate frame defined by the pair of galaxies 
	         \begin{equation}
	         	S_+D=\sum_{S,D} \gamma_{+,S}w_{SD}.
	         \end{equation}
	         where $\gamma_{+,S}$ is positive for radial alignment and negative for tangential alignments.

	         The projected correlation function $w_{g+}$ is obtained by integration over line of sight as in 
	            eq.~\eqref{eq:wgg_estimator}
		\subsubsection{Modeling}
			We assume the nonlinear-alignment model \citep{Hirata2003} for modeling the alignment signal
				\begin{align}
				\xi_{g+}(r_p,\Pi)=&A_I \zeta\int \mathrm{d}z W(z)b_g(r,z)r_{cc}(r,z)\nonumber\\
				\int \frac{\mathrm{d}^2k_\perp\mathrm{d}k_z}{(2\pi)^3}
				&\times P_{\delta\delta}(\vec{k},z)(1+\beta\mu_k^2)(1-\mu_k^2) e^{\mathrm{i}(\vec{r}_p.\vec{k}_\perp+\Pi k_z)}\label{eqn:xi},
	    	\end{align}
			with the line of sight anisotropy $1-\mu_k^2$ accounting for the projection effects \citep[see ][ for a detailed analysis]{Singh2016ia}. We
			will only use the projected correlation functions for intrinsic alignments where these terms have negligible effect.
		
		\subsection{Redshift-space galaxy power spectrum}\label{sssec:rsd}
    	
    	The model for the galaxy power spectrum in redshift space is based on \cite{Hand:2017ilm}. We only briefly summarize the formalism here, referring the reader to \cite{Hand:2017ilm} for more details. 
	
	In this model, we follow the halo model formalism in \cite{Okumura:2015fga} and separately model the 1-halo (correlations of galaxies in the same halo) and 2-halo (correlations of galaxies in different halos) contributions to the clustering of galaxies. To achieve such modeling, it is convenient to decompose the galaxy density field in redshift space into contributions from centrals and satellites:
    	\begin{equation}
    		\delta_g(\bold{k}) = (1-f_s)\delta_c(\bold{k}) + f_s \delta_s(\bold{k}),
    	\end{equation}
where $f_s$ is the satellite fraaction, and $\delta_c$ and $\delta_s$ are the density field of centrals and satellites, respectively.
The total galaxy power spectrum in redshift space, in turn, can be modelled as:
    	\begin{equation}
    		P_{gg}(\bold{k}) = (1-f_s)^2 P_{cc}(\bold{k}) + 2f_s(1-f_s)P_{cs}(\bold{k}) + f_s^2P_{ss}(\bold{k}),
    	\end{equation}
    	where $P^{cc}, P^{cs}$, and $P^{ss}$ are the centrals auto power, central-satellite cross power, and satellite auto power, respectively. We then separate 1-halo and 2-halo terms by further decomposing the galaxy sample into the following four subsamples: centrals without satellites (denoted as ``type A" centrals), centrals with satellites (``type B" centrals), satellites with no other satellites (``type A" satellites), and satellites with other neighboring satellites (``type B" satellites). We also account for the Fingers-of-God (FoG) effect when modeling 1-halo and 2-halo terms in redshift space, by separately modeling the FoG effect from each subsample.
	
	The model for the dark matter halo power spectrum in redshift space is based on the distribution function expansion \citep{Seljak:2011tx, Okumura12JCAP, 2012JCAP...02..010O, 2012JCAP...11..009V, Vlah:2013lia, Blazek2011}, and Eulerian perturbation theory and halo biasing model is applied to model the halo velocity correlator terms \citep{Vlah:2013lia}. The results of N-body simulations are also used to calibrate key terms in the model.
	
	The resulting galaxy power spectrum model depends on 13 physically-motivated parameters, which include: the Alcock-Paczynski (AP) effect parameters $\alpha_{||}, \alpha_{\perp}$, the growth rate $f$ and the amplitude of matter fluctuations $\sigma_8$ evaluated at the effective redshift of the sample $z_{\mathrm{eff}}$, the linear bias of the type A centrals and type A and B satellites [$b_{1\, c_A}$, $b_{1\, s_A}$, $b_{1\, s_B}$], the satellite fraction $f_s$, the fraction of type B satellites $f_{s_B}$, the mean number of satellite galaxies in halos with more than one satellite $\langle N_{>1,s}\rangle$, the centrals FoG velocity dispersion $\sigma_c$, The type A satellites FoG velocity dispersion $\sigma_{s_A}$, and normalization nuisance parameter for the 1-halo amplitude $f^{1h}_{s_B s_B}$. We follow the notations introduced in \cite{Hand:2017ilm}. In this work, we fix the AP parameters to their fiducial values, 1.
	
	We measure the clustering of galaxies using the multipole moments of the power spectrum $P_l(k)$. In this work, we take the FFT-based algorithm presented in \cite{Hand:2017irw}, built upon the methods proposed in \cite{Bianchi:2015oia} 
	and \cite{Scoccimarro:2015bla}, and this allows fast evaluation of the estimator in \cite{Yamamoto:2005dz}. Using the spherical harmonic addition theorem to expand the Legendre polynomials into spherical harmonics, we write the multipole estimator as:
    	\begin{equation}
    		P_l(k) = \frac{2l+1}{A}\int \frac{d\Omega_k}{4\pi}F_0(\bold{k})F_l(\bold{-k}),
    	\end{equation}
    	where $\Omega_k$ is the solid angle in Fourier space, $\mathcal{L}_l$ is the Legendre polynomial, $w$ is the weight, $A$ is the normalization defined as $A \equiv \int d\bold{r} [n(\bold{r})w(\bold{r})]^2$, and
    	\begin{align}
    	\begin{split}
    	F_l(\bold{k}) &= \int d\bold{r}F(\bold{r})e^{i\bold{k}\cdot\bold{r}}\mathcal{L}_l(\bold{\hat{k}}\cdot\bold{\hat{r}}) \\
    	&= \frac{4\pi}{2l+1}\sum_{m=-l}^{l}Y_{lm}(\bold{\hat{k}})\int d\bold{r}F(\bold{r})Y_{lm}^*(\bold{\hat{r}})e^{i\bold{k}\cdot\bold{r}}.
    	\end{split}
    	\end{align}
	The weighted galaxy density field $F(\bold{r})$ is given by
    	\begin{equation}
    		F(\bold{r}) = \frac{w(\bold{r}) }{A^{1/2}}[n(\bold{r}) - \alpha n_s(\bold{r})],
    	\end{equation}
	where $n$ and $n_s$ are the number density field for the galaxy and randoms catalogs respectively, and the factor $\alpha$ normalizes $n_s$ to $n$.

        
        \subsubsection{IA effects on RSD}\label{sssec:IARSD}
        
	 To account for the effects of intrinsic alignments of galaxies on the selection functions of galaxies and hence the RSD measurements, we follow the formalism in 
	 \cite{Hirata2009,Martens2018}, defining the bias in observed galaxy density as:
	 
	 \begin{equation}
	 	\widehat{\delta}_g(\vx,\lambda)=\delta_g(\vx)+\epsilon(\lambda(\vx))
	 \end{equation}
	 To derive the error $\epsilon$, we assume that the probability of a galaxy in the observed sample, $O$, is given as 
	 \begin{equation}
	 	P(O|T,\lambda)=P(T)P(\lambda) (1+S(\lambda))
	 \end{equation} 
	Where $T$ is the target sample, $\lambda$ is the FP residual for the given galaxy and $S(\lambda)$ is the size dependent selection function.
	We can also write the number of galaxies with the observed value of $\lambda$ as
	\begin{equation}
		N_\lambda=\frac{dN}{d\lambda}=N P(\lambda)(1+S(\lambda)),
	\end{equation}
	$N$ is the total number of galaxies.
	Following the ansatz in \cite{Hirata2009,Martens2018}, we assume that the some galaxies are missed when they are aligned with the plane of the sky, i.e. 
	have positive $\lambda$. Under this assumption and assuming that the intrinsic distribution $\lambda$ is symmetric within $T$, we can write 
	\begin{equation}
		S(\lambda)=\frac{N_{\lambda}-N_{-\lambda}}{N_{\lambda}+N_{-\lambda}}
	\end{equation}
	where $N_{-\lambda}$ is the number of galaxies with a negative value of $\lambda$.
	Note that the mean of $S(\lambda)$ is zero by construction. $S(\lambda)$ is also zero if there are no size dependent selection effects, since we assume the 
	$\lambda$ distribution to be symmetric ($N_\lambda=N_{-\lambda}$ when no size dependent selection).
	
	Galaxies have some random $\lambda$ values, $\lambda_R$, due to intrinsic variations in galaxy properties as well as random projections. The additional $\lambda_I$ 
	sourced by intrinsic alignments acts as a small perturbation on top of the $\lambda_R$ (we assume $\lambda_I<<\lambda_R$). 
	We can then write the error $\epsilon$ that is relevant for cosmological inferences as 
	\begin{equation}
		\epsilon(\lambda_R(\vx)+\lambda_I(\vx))=S(\lambda_R)+\lambda_I\frac{\partial S(\lambda)}{\partial \lambda}|_{\lambda=\lambda_R}
	\end{equation}
	
	 %
Since $\lambda_R$ and hence $S(\lambda_R)$ does not correlate with the density field,
the relevant part of $\epsilon$ that correlates with the density field can be written as 
\begin{equation}
	\epsilon(\textbf{x}) \approx \gamma A_\lambda {\zeta}\left[\frac{1}{3}-\mu^2\right]\delta_m(\vx)
\end{equation}
where we used the relation of $\lambda$ to tidal field as defined in eq.~\eqref{eq:lambda} and defined the ensemble response $\gamma$ as
\begin{equation}
	\gamma=\mean{\frac{\partial S(\lambda)}{\partial \lambda}}=\frac{1}{\int d\lambda N_\lambda}\int d\lambda N_\lambda\frac{\partial S(\lambda)}{\partial \lambda}
\end{equation}
The error in the growth rate and galaxy bias measurements is given by
\begin{align}
	\Delta f&=\widehat{f}-f_0=-\gamma A_\lambda\zeta \label{eq:IARSD_eq} \\ 
	\widehat{b}_g&=b_{g,0}+\frac{1}{3} \gamma A_\lambda\zeta=b_{g,0}-\frac{1}{3}\Delta f\label{eq:IARSD_eq_b}
\end{align}
We will determine $\gamma$ from the distribution of FP residuals. 
We note that since the datasets we will be using are already affected by the selection effects, $S(\lambda)$, this can introduce a bias the 
estimations of the $\gamma$ (the mean of $\lambda$ distribution is shifted which biases $S$). 
We will work under the assumption that the selection effects are small and hence the shift in the mean of the distribution and the bias in $\gamma$ is also small.
$A_\lambda$ will be determined using the cross correlations between the FP residuals and the galaxy density field and $\zeta$ is a cosmology dependent constant. We will also compute the variations in $f$ and $b$ from our RSD analysis and these measurements will allow us to test the eq.~\eqref{eq:IARSD_eq} and eq.~\eqref{eq:IARSD_eq_b}.

In section~\ref{subsubsection:FPcut}, we will also split the galaxy sample into two subsamples according to the sign of FP residuals, following \cite{Martens2018}. These subsamples are expected to have almost identical $\gamma$ values: $\gamma^+ \approx \gamma^-$, where $+$ and $-$ indicates samples with positive and negative FP residuals, respectively. That is because $\partial S(\lambda)/\partial \lambda$ is mirrored across the y-axis, as shown in figure~\ref{fig:Slambda}. Therefore, we expect that the growth rate measurements between the two FP residual subsamples is consistent: $\Delta f^+ - \Delta f^- = -(\gamma^+ - \gamma^-)A_\lambda \zeta \approx 0$, unlike the model assumed by \cite{Martens2018}.


%

		\subsection{Covariance matrices}
		 For the correlation function measurements, we compute the covariance matrices using Jackknife resampling method by splitting the sample into 100 
		 approximately equal area patches (68 patches in North Galactic Cap (NGC) and 32 in South Galactic Cap (SGC) of BOSS data described in section~\ref{sec:data}). 
		 Since the jackknife covariances are noisy, leading to biased inverse matrix, we also
		 apply the Hartlap correction \citep{Hartlap2007} when computing the inverse covariance matrix used in the likelihood functions.
		
		For the multipoles in Fourier space, we assume the theoretical Gaussian covariance, following \cite{Grieb:2015bia}:
		\begin{equation}\label{cov_multipole}
		C_{\ell_1 \ell_2}(k_i, k_j) = \frac{2 (2\pi)^4}{V^2_{k_i}} \delta_{ij}\int_{k_i - \Delta k/2}^{k_i + \Delta k /2} \sigma^2_{\ell_1 \ell_2}(k) k^2 \mathrm{d}k,
		\end{equation}
		where $V_{k_i} = 4\pi[(k_i+\Delta k/2)^3 -(k_i-\Delta k/2)^3)^3]/3$ is the volume of the shell in $k$-space. When the expected mean number density $\bar{n}(z)$ is varying, the per-mode covariance $\sigma_{\ell_1 \ell_2}^2(k)$ is given by \cite{Yamamoto:2005dz}:
		\begin{align} 
		\begin{split}
		\sigma^2_{\ell_1 \ell_2}(k) = &\frac{(2\ell_1+1)(2\ell_2+1)}{A^2}
          \int_{-1}^1 \mathrm{d} \mu \int_{V_s} \mathrm{d} \boldsymbol{r} \;  \bar{n}^4(\boldsymbol{r}) 
                     w^4(\boldsymbol{r})\\ &\left[P(k,\mu,z) + \bar{n}^{-1}(\boldsymbol{r}) \right]^2 \mathcal{L}_{\ell_1}(\mu) \mathcal{L}_{\ell_2}(\mu),
        \end{split}
        \end{align}
        where the normalization terms $A$ is defined as $A \equiv \int d\bold{r} [\bar{n}(\bold{r})w(\bold{r})]^2$, following the notations in section~\ref{sssec:rsd}.
        
        For the galaxy subsamples we define in section~\ref{sec:data}, we compute their mean number densities in order to obtain correct covariance matrices using eq.~\ref{cov_multipole}.

	\section{Data}\label{sec:data}
	In this work we use the LOWZ ($0.16<z<0.43$) and CMASS ($0.43<z<0.7$) spectroscopic samples from SDSS-III BOSS
        \citep{Blanton:2003,Bolton:2012,Ahn:2012,Dawson:2013,Smee:2013} data
        	release 12
            \citep[DR12;][]{Alam2015},
			 which are selected using the imaging data from SDSS-I and SDSS-II surveys.
			 The SDSS survey \citep{1998AJ....116.3040G,2000AJ....120.1579Y,2001AJ....122.2129H, 2004AN....325..583I,1996AJ....111.1748F,
	         2002AJ....123.2121S,2001AJ....122.2267E,Gunn2006,
			  2002AJ....123.2945R,2002AJ....124.1810S} 
			  \citep{Lupton2001,
		  	2003AJ....125.1559P,2006AN....327..821T,2009ApJS..182..543A,2011ApJS..193...29A,2008ApJ...674.1217P}.
			To fit the FP, we also obtain the photometric measurements for our galaxies, specifically the Radii, axis ratios and magnitudes from \dev\ fits
			from the SDSS photometric catalog. The magnitudes are corrected for the extinction and we also apply k-corrections using the formalism of 
			\cite{Wake2006}.
					  
		  	When computing galaxy clustering and galaxy-FP cross correlations, we apply weights to the galaxies, where the weights are
            given by \citep{Ross2012}
            \begin{equation}
            	w=w_\text{sys}(w_{no-z}+w_{cp}-1),
            \end{equation}
            where $w_\text{sys}$ weights correct for the variations in the selection
            function on the
            sky (important for CMASS) and $w_{no-z}$, $w_{cp}$ correct for missing redshifts due to
            failure to obtain redshift (no-z) or fiber collisions for close pairs, $cp$.
            
            For intrinsic alignments of galaxies, the shape sample used to estimate the shear
			is described in more detail in \cite{Reyes2012} and \cite{Singh2015}.
            
            Furthermore, to study the dependence of FP and residuals on galaxy properties, we also split the BOSS samples based on color and luminosity. We follow 
            the procedure from \cite{Singh2015}, whereby we split the samples in narrow redshift bins based on the percentiles of the color and luminosity. We make 5 
            color samples, $C_1-C_5$, with each sub-sample containing 20\% of the galaxies such that $C_1$ starts with reddest galaxies and the subsequent samples
            contain progressively bluer galaxies. Split in the redshift bins ensures that each sample has the same redshift distribution. We follow the similar 
            procedure for luminosity and make 4 luminosity subsamples, $L_1-L_4$ with $L_1$ being brightest and $L_4$ being faintest. $L_1-L_3$ contain 20\% of the 
            sample each while $L_4$ contains 40\% of galaxies.
            
            For the LOWZ sample, we also identify the galaxies in groups using the counts-in-cylinders technique \citep{Reid2009} using the same procedure as was 
            followed in \cite{Singh2015}. Galaxies that are not in groups (or are in group of 1) are designated as `Field' galaxies, the brightest galaxy in a group
            of more than 1 galaxy is designated as BGG (brightest group galaxy) while all non-BGG galaxies are designated as satellite galaxies. 
            
            Since we are splitting the samples by color, luminosity and also FP residuals, the variations in the photometry across the sky can lead to some 
            variations in selection function of the galaxies for these sub-samples. The variations can introduce biases when computing the galaxy auto correlation
            functions for these galaxies. To avoid this issue, we will use cross correlations when computing the correlation functions, 
            where we cross correlate the sub-samples with the full sample
            from which they were selected. As a result the biases in the signal are reduced though the covariance will still be affected \citep{Singh2017cov}. 
            In the power spectrum measurements, we will use the auto correlations.
            To reduce the impact of selection functions on sub samples, we reweight the randoms provided by BOSS to correspond to these samples. We compute the weight 
            using the ratio of number 
            galaxies within the sub-samples to number of galaxies in the full sample, within each $\sim80\deg^2$ patch we generated for the jackknife calculations. 
            We also tried the weights computed in much smaller patches, but those weights biased the measurements on small scales. 
             More details of randoms re-weighting are presented in the appendix~\ref{appendix:randoms_reweight}. 
           
\section{Results}\label{sec:results}
	In this section we present our results, beginning with the fits of FP to LOWZ and CMASS as well as various subsamples and 
	analysis of FP residuals based on some galaxy properties. 
	Then we present the measurements of cross correlations 
	between FP residuals and galaxy density and comparison of these cross correlations with the Intrinsic alignments of galaxies. Following this
	we present the measurements of redshift space distortions (RSD) and the correlations between RSD constraints and FP residuals.
	Some additional details, analysis, and tests based on systematics are presented in the appendices of the paper.
	\subsection{FP Fits}\label{ssec:results_FP_fits}
		In this section we present results of fitting FP to full LOWZ and CMASS samples and an analysis of the FP residuals based on the redshift, luminosity and environment
		of the galaxies. Note that unless mentioned otherwise, in this subsection, FP residuals are obtained from fitting FP to the full samples and not
		the sub-samples.
		
		\begin{figure}
    	    		\centering
        	 		\includegraphics[width=\columnwidth]{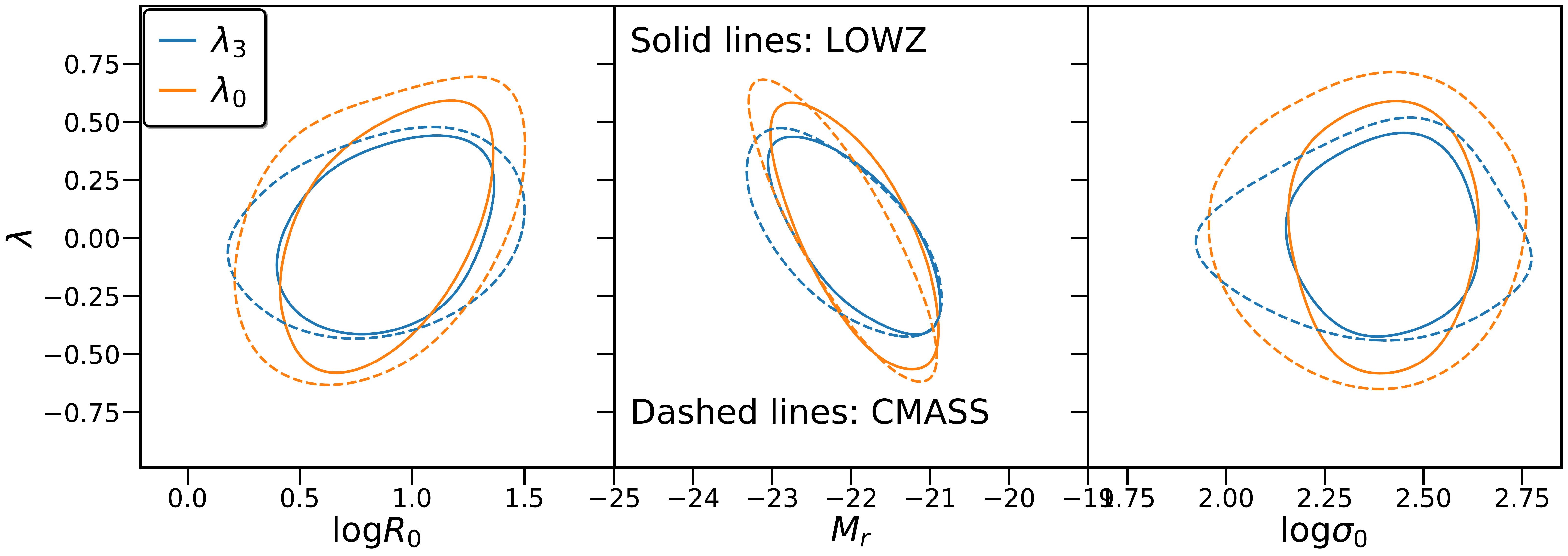}		
					\caption{The residuals over the FP for standard FP, $\lambda_0$ and the redshift dependent FP $\lambda_3$ as function of 
					galaxy size, $R_0$, $r$ band magnitude $M_r$ and the velocity dispersion $\sigma_0$. We show the contours covering 95\% of the sample for
					both LOWZ (solid lines) and CMASS (dashed lines) samples.}
					\label{fig:lambda_gp_hist}
		\end{figure}
		In figure~\ref{fig:lambda_gp_hist}, we show the contour plots of the FP residuals as function of galaxy size, magnitude and velocity dispersion.
		For the standard FP, $\lambda$, we obtain the RMS values of $\lambda_{rms}=0.22$ for LOWZ and $\lambda_{rms}=0.26$ for CMASS sample and for the redshift
		dependent FP, $\lambda_3$, we obtain $\lambda_{rms}=0.15$ for LOWZ and $\lambda_{rms}=0.16$ for CMASS samples. 
		Note that $\lambda$ is defined in $\log_e$ 
		($\ln$) 
		base and hence $\lambda_{rms}$ is larger than the scatter in $\log_{10} R$ space by a factor of $\ln 10\sim2.3$. After accounting for this effect, our 
		results are consistent with the FP scatter of $\sim0.1$ obtained by \cite{Saulder2019,Joachimi2015}, albeit for somewhat different samples.
		
		Also note in figure~\ref{fig:lambda_gp_hist} that the FP residuals are correlated with the galaxy properties. We now investigate some of these correlations in 
		more detail.
		
		\subsubsection{Redshift dependence}
			\begin{figure*}
				\begin{subfigure}[t]{\columnwidth}
    	    		\centering
        	 		\includegraphics[width=\columnwidth]{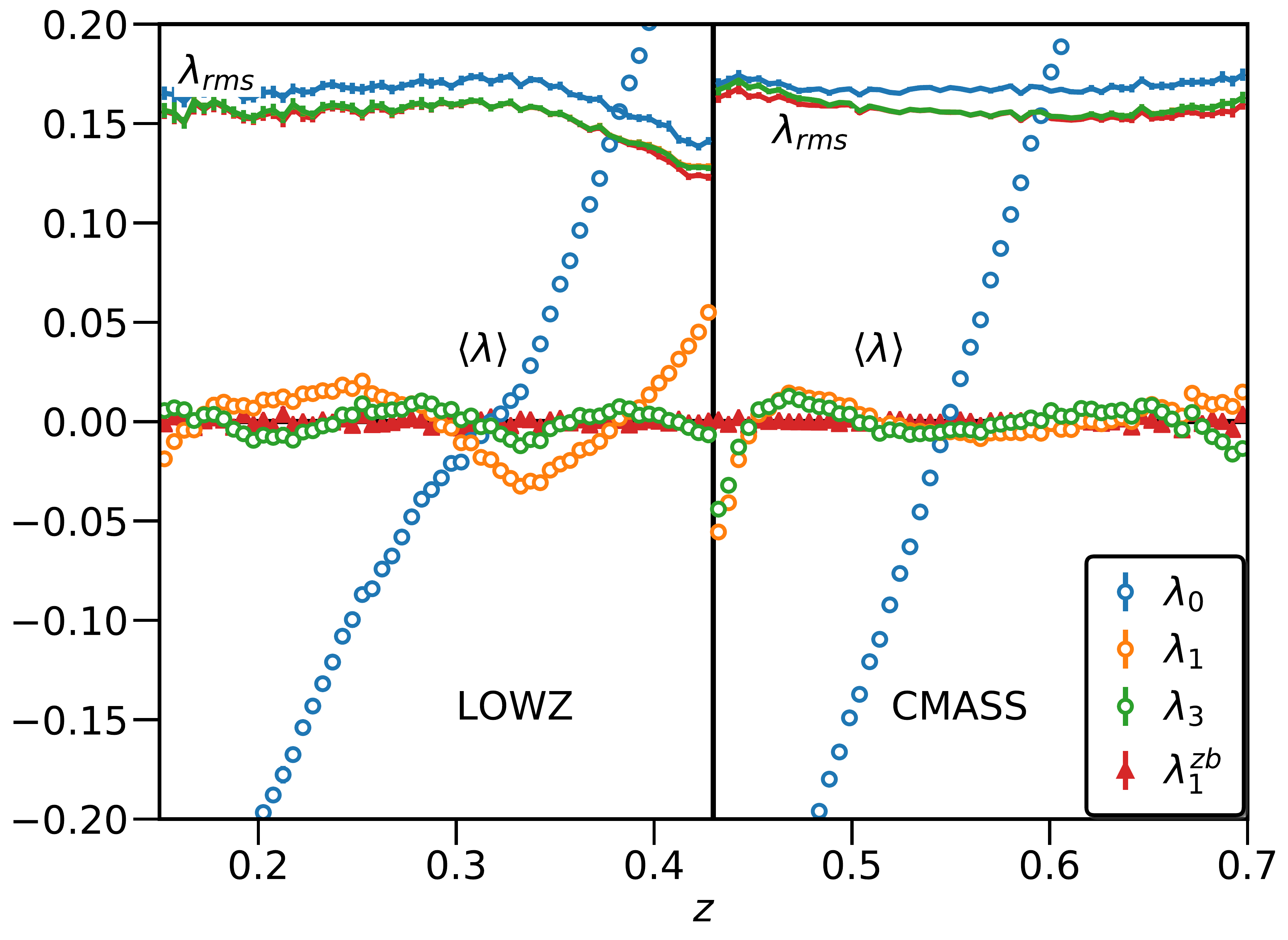}
				
					\caption{}
					\label{fig:lambda_mean_z}
				\end{subfigure}
				\begin{subfigure}[t]{\columnwidth}
	         		\includegraphics[width=\columnwidth]{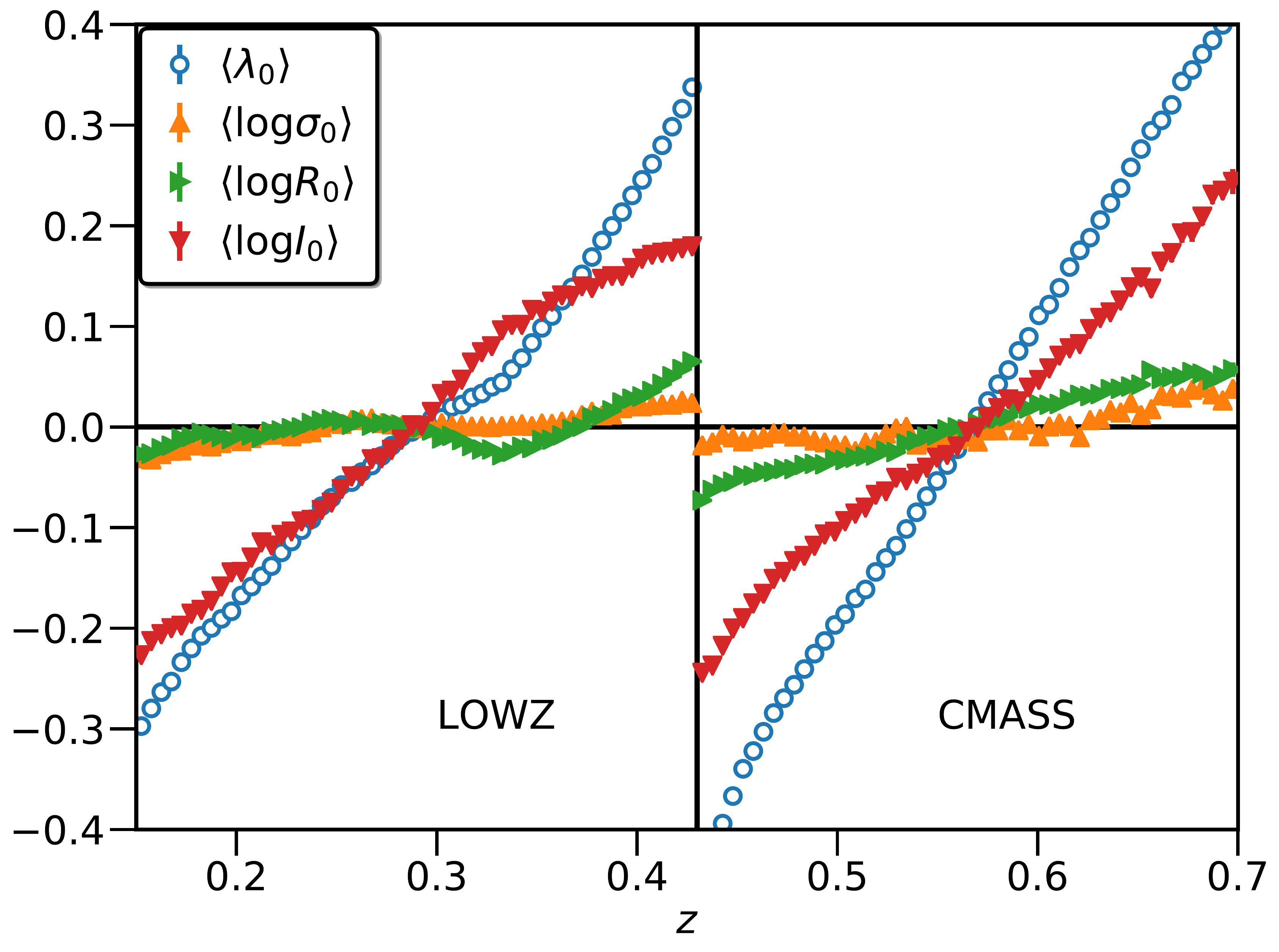}
					\caption{
					}
					\label{fig:gp_mean_z}
				\end{subfigure}
				\caption{ a) Redshift dependence of the mean and standard deviation (mean subtracted RMS) of the FP residuals for both LOWZ and CMASS samples.
					The subscript, $\lambda_i$, $i \in[0,1,3]$ denotes the order of the polynomial in $z$ used for fitting FP. The superscript `$zb$' 
					denotes the sample fitted in narrow redshift bins $\Delta z=0.02$. The residuals over for the standard fundamental plane, 
					$\lambda_0$, have strong dependence 
					on the redshift and including redshift polynomials in the FP reduce this dependence as well as the scatter. Fitting FP within
					small redshifts further reduces the mean of FP (though there can be evolution within the bins). 
					b)Redshift evolution of galaxy properties that are included in the FP. The redshift dependence of the residuals for standard FP 
					can be explained by the redshift dependence of these properties, especially the surface brightness of galaxies, $\log I$, which
					has the strong and monotonic dependence on redshift, driven by the $\log(1+z)$ correction for the Tolman dimming.
					}
	    	     \label{fig:gp_lambda_mean_z}
		     \end{figure*}

		In figure~\ref{fig:gp_lambda_mean_z}\subref{fig:lambda_mean_z} 
		we present the redshift dependence of the mean and the RMS of the FP residuals for both LOWZ and CMASS samples.
		For the standard FP we find a strong correlation of the mean residuals, $\lambda_0$, with the redshift. Including redshift dependence within the
		FP using polynomials reduces the dependence of the mean by a large magnitude and also reduces the RMS (as function of $z$) by $\sim10\%$. 
		The change (reduction) in the mean is 
		largest when using the first order polynomial with some further improvement when going to the third order polynomial. There are still small
		residuals correlations between the mean and redshift and such correlations can potentially be important for the correlation functions we present in 
		section~\ref{ssec:results_FP_correlations} and
		for the  cosmological applications of the FP in general. To further reduce correlation between mean and the redshift, we instead fit the FP in
		narrow redshift bins, $\Delta z=0.02$ (this choice is motivated to keep the bin size small but have large enough line of sight size 
		so as to not bias the cross correlation measurements presented in section~\ref{ssec:results_FP_correlations}). Fits in the bins further reduce the impact of 
		the redshift dependent residuals. We will use $\lambda_3$ as 
		our fiducial FP (unbinned with third order polynomial in redshift), but we will test the cross correlation results with the binned FP for comparison. 
		
		To study the source of the redshift dependence of FP, 
		we show the redshift dependence of the galaxy properties in figure~\ref{fig:gp_lambda_mean_z}\subref{fig:gp_mean_z}. 
		The velocity dispersion and physical radius only have mild dependence on redshift for both LOWZ and CMASS sample. The surface
		brightness of the galaxies on the other hand evolves strongly with redshift and is the primary driver for the redshift evolution of the standard 
		FP residuals, $\lambda_0$. The redshift evolution of surface brightness is driven by redshift dimming \citep{Tolman1930} of the surface 
		brightness, as a result of which we only observe galaxies with larger surface brightness at higher redshifts. The evolution of surface brightness 
		is very similar to 
		the $\log(1+z)$ correction that was included in the eq.~\ref{eq:logI}. Hence, when including the 
		redshift polynomials into the FP, we are essentially undoing the $\log(1+z)$ correction.
		 Including $\log(1+z)$ dependence in the FP is also a possibility
		instead of the polynomials in $z$, but we opt for the polynomials (or fit in narrow $z$ bins) 
		as they provide extra degrees of freedom which can account for the small redshift dependence of the velocity dispersion, physical radius and luminosity.
		
		Our results on the redshift 
		dependence of the FP are qualitatively consistent with those from \cite{Joachimi2015}, where such trends were observed for the SDSS main galaxy
		sample. Due to the differences in the samples used in this study and \cite{Joachimi2015}, a more quantitative comparison is difficult but we have 
		tested our pipelines on the samples used in \cite{Joachimi2015} and we are able to reproduce their results.
			
		\subsubsection{Luminosity and environment dependence}\label{ssec:results_FP_lum}
			\begin{figure*}
				\begin{subfigure}[t]{\columnwidth}
         			\includegraphics[width=\columnwidth]{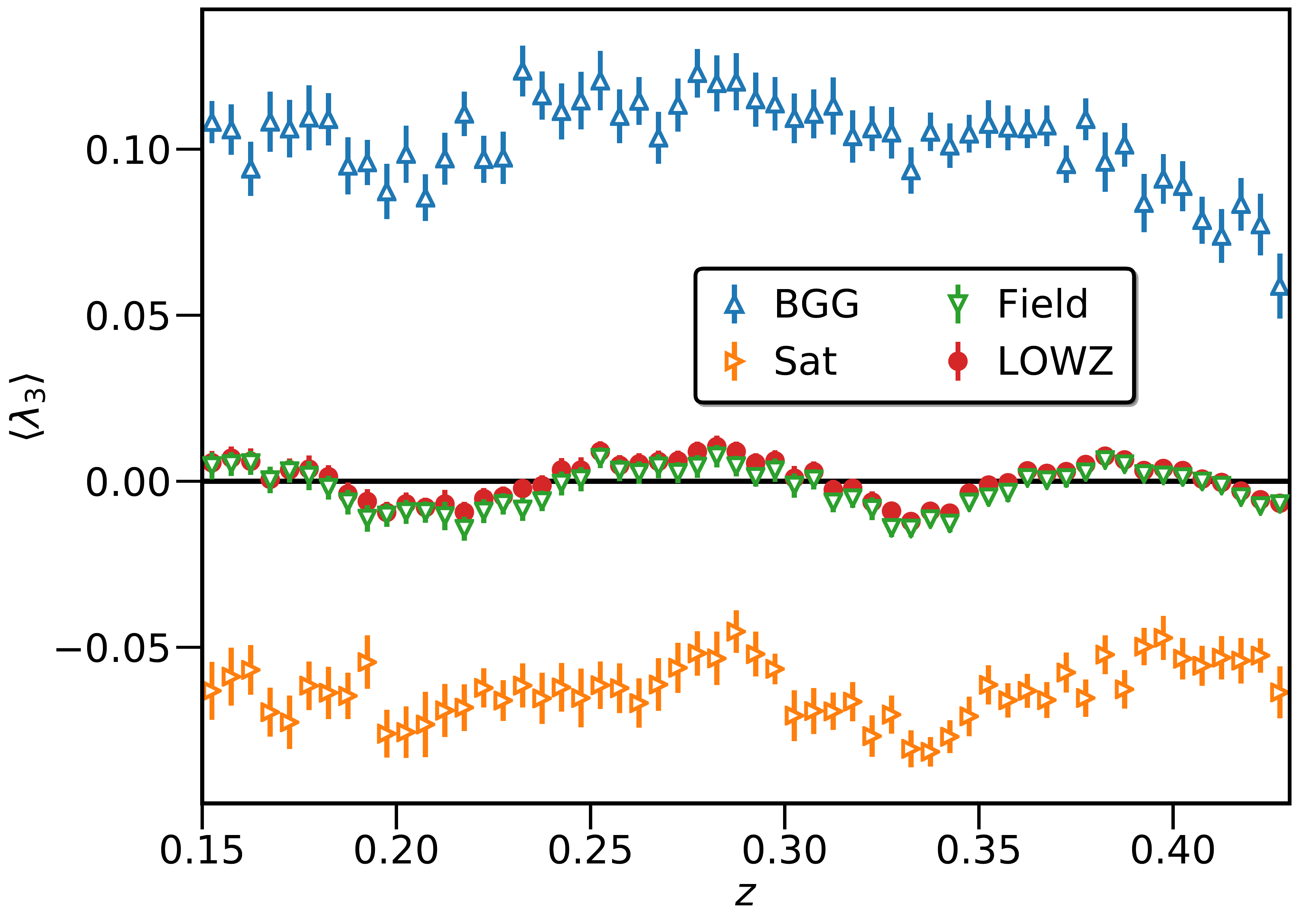}
					\caption{}
		    	     \label{fig:lambda_mean_env}
			     	
		    	 \end{subfigure}
			     \begin{subfigure}[t]{\columnwidth}
         			\includegraphics[width=\columnwidth]{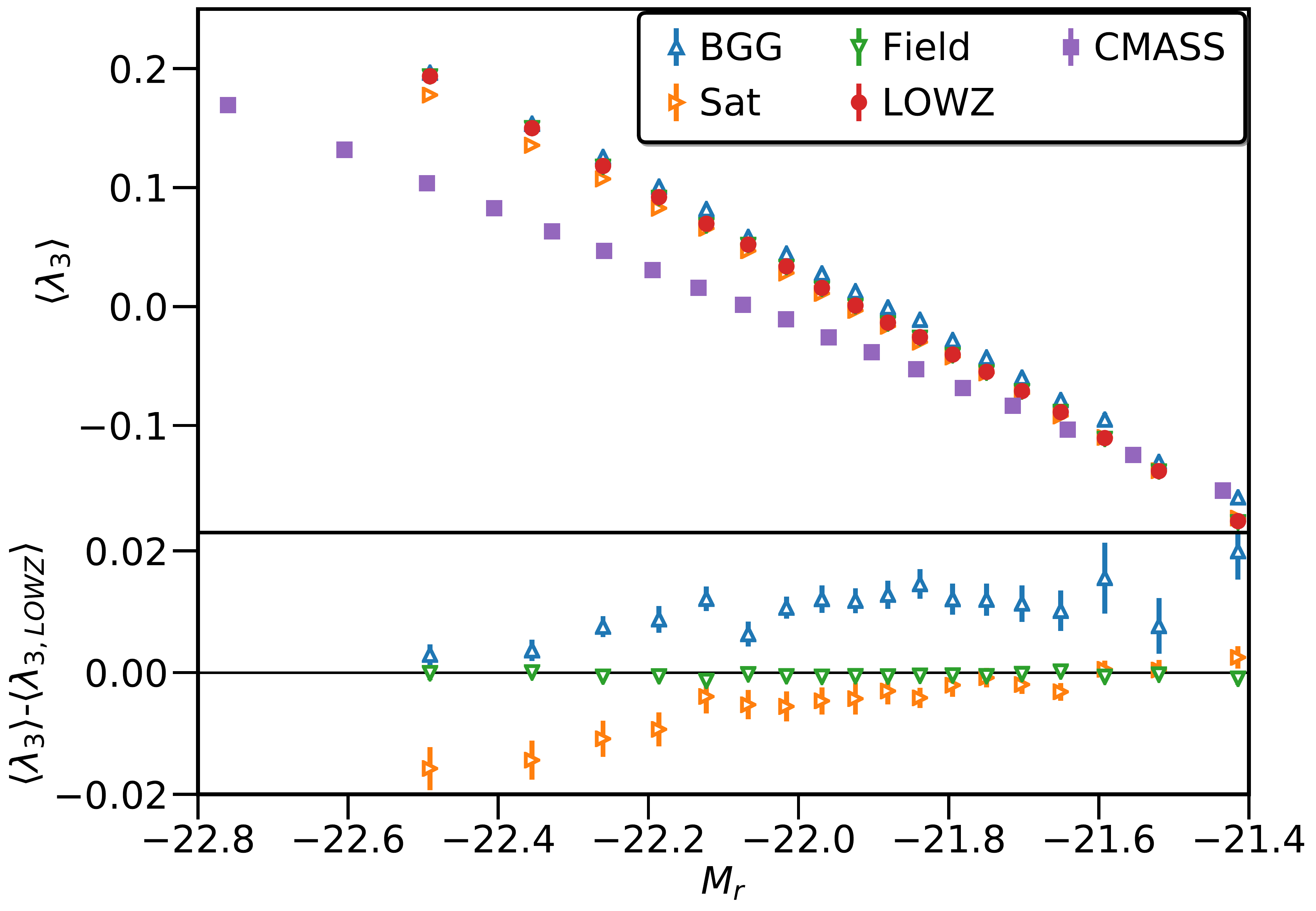}
					\caption{}
	    	    	 \label{fig:lambda_mean_Mr}		     
				\end{subfigure}
				\caption{a)  Dependence of the FP residuals on galaxy types. Similar to \citet{Joachimi2015}, we observe that Brightest group galaxies (BGG), 
				satellites and field galaxies have different FP residuals, with BGGs being larger than FP predictions while satellites being smaller.
				b) Mean FP residuals as function of the $r$ band magnitude. Different types of galaxies, BGGs, Satellites, Field, all give very similar 
				relation which suggests that the
				 dependence in a) can be explained largely by magnitude (or luminosity) dependence of the FP residuals. In the lower panel 
				we show the difference in the mean FP residuals of different samples relative to the full LOWZ sample. BGGs (satellites) are still higher 
				(lower) than the full sample, though the differences are much smaller than in a).
				We do not observe
				any significant dependence of RMS of $\lambda$ with luminosity.
				}
				\label{fig:lambda_mean_env_Mr}
		     \end{figure*}
			In figure~\ref{fig:lambda_mean_env_Mr}\subref{fig:lambda_mean_env}, 
			we show the dependence of the FP residuals on the environment as function of redshift. As in 
			\cite{Joachimi2015}, we observe that the brightest 
			group galaxies, $BGGs$ have positive residuals on average (they are larger in size than predicted by FP) while satellites 
			have negative residuals implying they are smaller than average. It is tempting to interpret these results based on the environment dependence
			of galaxies, whereby centrals or BGGs tend to be larger but less concentrated while satellites under going tidal stripping tend to be smaller 
			and more concentrated. However, in figure~\ref{fig:lambda_mean_env_Mr}\subref{fig:lambda_mean_Mr} 
			we observe that the FP residuals are strongly correlated with the galaxy 
			luminosity, where brighter galaxies have larger (or more positive) residuals. These trends can explain most of the variations between different galaxy types in
			figure~\ref{fig:lambda_mean_env_Mr}\subref{fig:lambda_mean_env}, with BGGs (satellites) being only marginally larger (smaller) after accounting 
			for the luminosity evolution, as shown in the lower panel of figure~\ref{fig:lambda_mean_env_Mr}\subref{fig:lambda_mean_Mr}.
			
			Given that $\lambda\propto -b\log{I}\propto b M_r$ ($b$ is negative), the luminosity dependence of FP residuals $\lambda$ is expected in case 
			the various galaxy
			properties, namely luminosity, radius and velocity dispersion are not perfectly correlated in a way to cancel such a dependence.
			While it is possible to include the higher order luminosity dependence in the FP as well, we opt not to do so as FP residuals are correlated with 
			multiple galaxy properties and systematics (see appendices) and including dependence on too many variables complicates the interpretation of FP and
			its residuals (inclusion of z-dependence is necessary to avoid biases in cross correlation measurements presented in section~\ref{ssec:results_FP_correlations}). 
			Instead, we split our sample based on luminosity, color and environment as described in section~\ref{sec:data} and fit the FP separately to those 
			subsamples when studying the dependence of cross correlations on these galaxy properties. 
		     
		
	\subsection{Cross correlations with galaxy density}\label{ssec:results_FP_correlations}
		In this section we present the measurements of galaxy clustering and cross correlations between the galaxy positions and FP residuals. Throughout 
		this section we will use FP with third order polynomial as the primary FP and when considering the subsamples, we will fit FP to each of the
		subsamples separately. 
		
			\begin{figure*}
				\begin{subfigure}[t]{\columnwidth}
         			\includegraphics[width=\columnwidth]{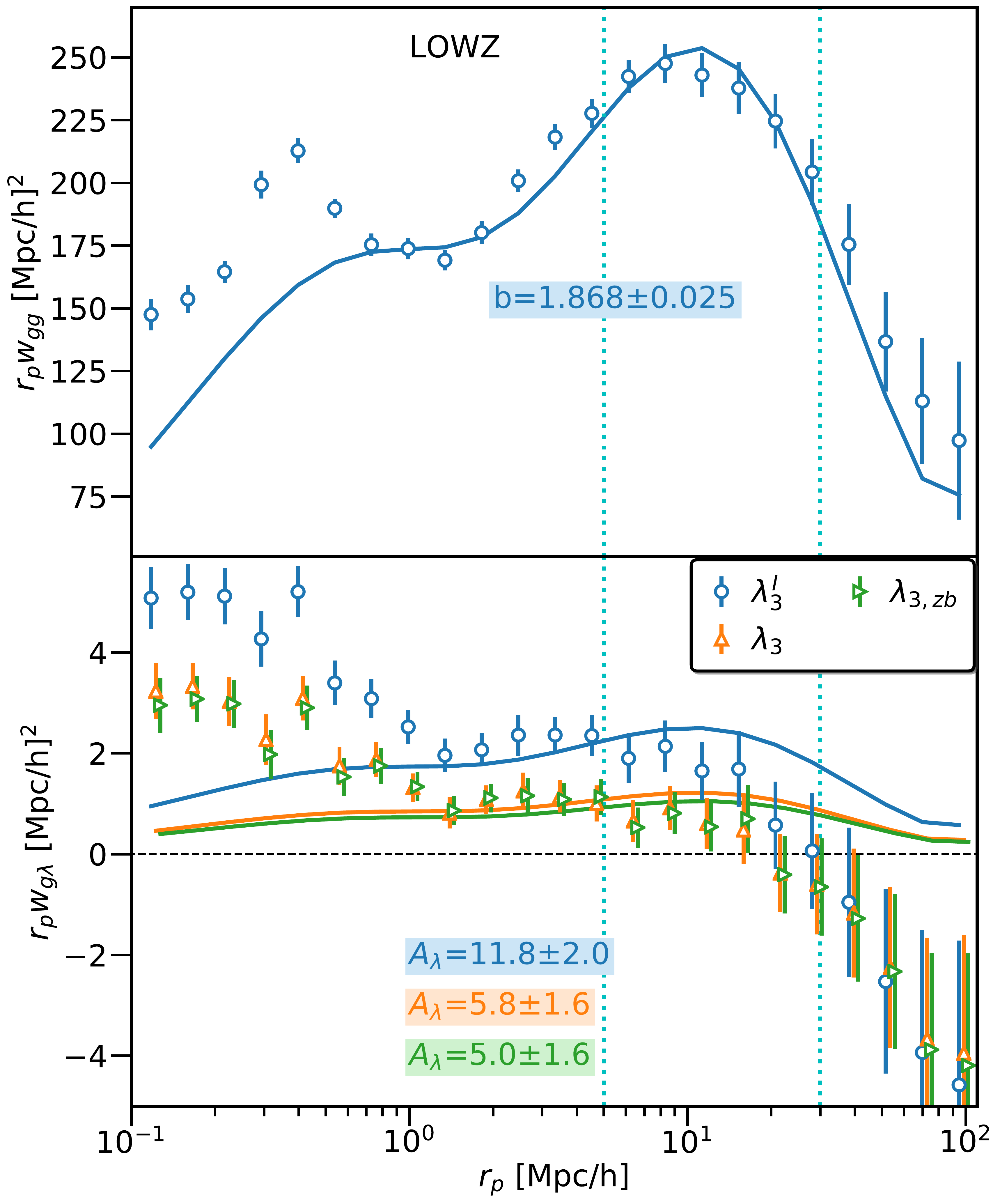}
		    	     \label{fig:lowz_w}
			     	\caption{}
		    	 \end{subfigure}
			     \begin{subfigure}[t]{\columnwidth}
         			\includegraphics[width=\columnwidth]{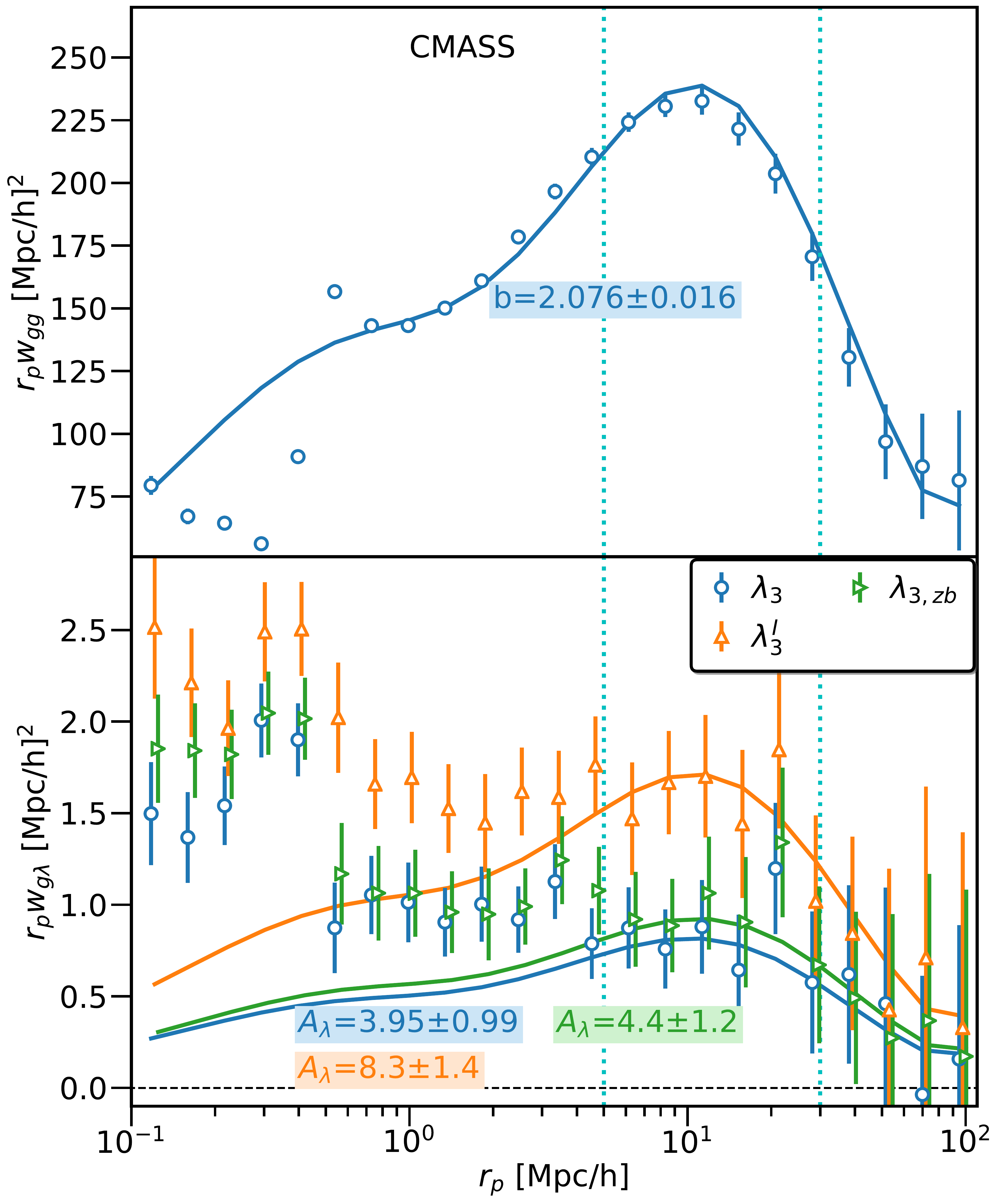}
	    	    	 \label{fig:cmass_w}
				     \caption{}
				\end{subfigure}
				\caption{Measurements of galaxy clustering (upper panels) and galaxy-$\lambda$ (lower panels) 
					cross correlation functions for (a) LOWZ  and (b) CMASS samples. Lower panel shows measurements with three different
					FP definitions, $\lambda_3^I$ (blue, FP without velocity dispersion), $\lambda_3$ (orange) and FP fitted in $z$-bins $\lambda_{3,zb}$.
					We measure strong correlations between
					galaxies and FP residuals for both CMASS and LOWZ samples, with $\lambda_3^I$ signal being factor of $\sim2$ larger than 
					the $\lambda_3$. Given that the mean of $\lambda_3^I$ and $\lambda_3$ are very similar and $\lambda_3^I$ rms is larger by $\sim10\%$,
					this difference is originating from the intrinsic differences between the two FP planes. Also, the consistency between $\lambda_3$ and
					$\lambda_{3,zb}$ suggests that the negative signal in LOWZ at large scales is unlikely due to any redshift dependent additive 
					systematics in the FP (as observed in figure 1). Numbers quoted in the plots are the best fit galaxy bias and $\lambda$ amplitude 
					$A_\lambda$ obtained by fitting the model in range $5<r_p<30\mpch$ (marked by vertical dashed cyan lines). For the LOWZ sample, the $\chi^2_{dof}\sim0.7$, 
					even though the fit looks inconsistent with the data. This is due to strong correlations between the bins on large scales, likely driven by systematics 
					(see also appendix~\ref{appendix:FP_correlations}).
				}
				\label{fig:lowz_cmass_w}
		   \end{figure*}
		   
		   We begin by presenting the measurements of projected correlation functions using full LOWZ and CMASS samples in figure~\ref{fig:lowz_cmass_w}, 
		   with  FP fit to the whole sample, $\lambda_3$, FP fit in narrow redshift bins, $\lambda_{3,zb}$ and the FP without velocity dispersion, 
		   $\lambda_3^I$. We fit the models described in section~\ref{sec:formalism} to both galaxy clustering and galaxy-$\lambda$ cross correlations in
		   the range $5<r_p<30\mpch$. We do not fit the scales below $r_p<5\mpch$ as our model with assumption of linear bias 
		   is not expected to work well on these non-linear scales and we also do not use $r_p>30\mpch$, as there is some evidence of systematics in the 
		   galaxy-$\lambda$ cross correlations (see also appendix~\ref{appendix:FP_correlations}). We have checked that including scales between $30<r_p<70\mpch$ 
		   does not significantly change the best fit parameters as measurements on those scales are correlated and also have  lower signal to noise compared to 
		   smaller scales. 
		 
		   For LOWZ sample, we obtain the linear galaxy bias, $b_g=1.868\pm0.025$ and for CMASS sample we obtain $b_g=2.096\pm0.019$.
		   Using galaxy-$\lambda_3$ cross correlations , we obtain $A_\lambda=5.8\pm1.6$ for LOWZ sample and $A_\lambda=5.3\pm1.3$ for CMASS sample, with 
		   CMASS sample have lower noise due to larger volume of the sample, even though $\lambda_\text{rms}$ is larger for CMASS sample. The values of 
		   $A_\lambda$ do not change significantly if we fit FP in narrow redshift bins ($\lambda_{3,zb}$ in figure~\ref{fig:lowz_cmass_w}) and/or if we 
		   fit the FP for North and South regions of BOSS separately (measurements not shown). However, fitting the FP without velocity dispersion 
		   , $\lambda_3^I$, leads to significantly larger amplitude, with $A_\lambda$ being larger by factor of $\sim2-5$ depending upon the sample and the FP 
		   definition.
		   
		   \begin{figure}
         			\includegraphics[width=\columnwidth]{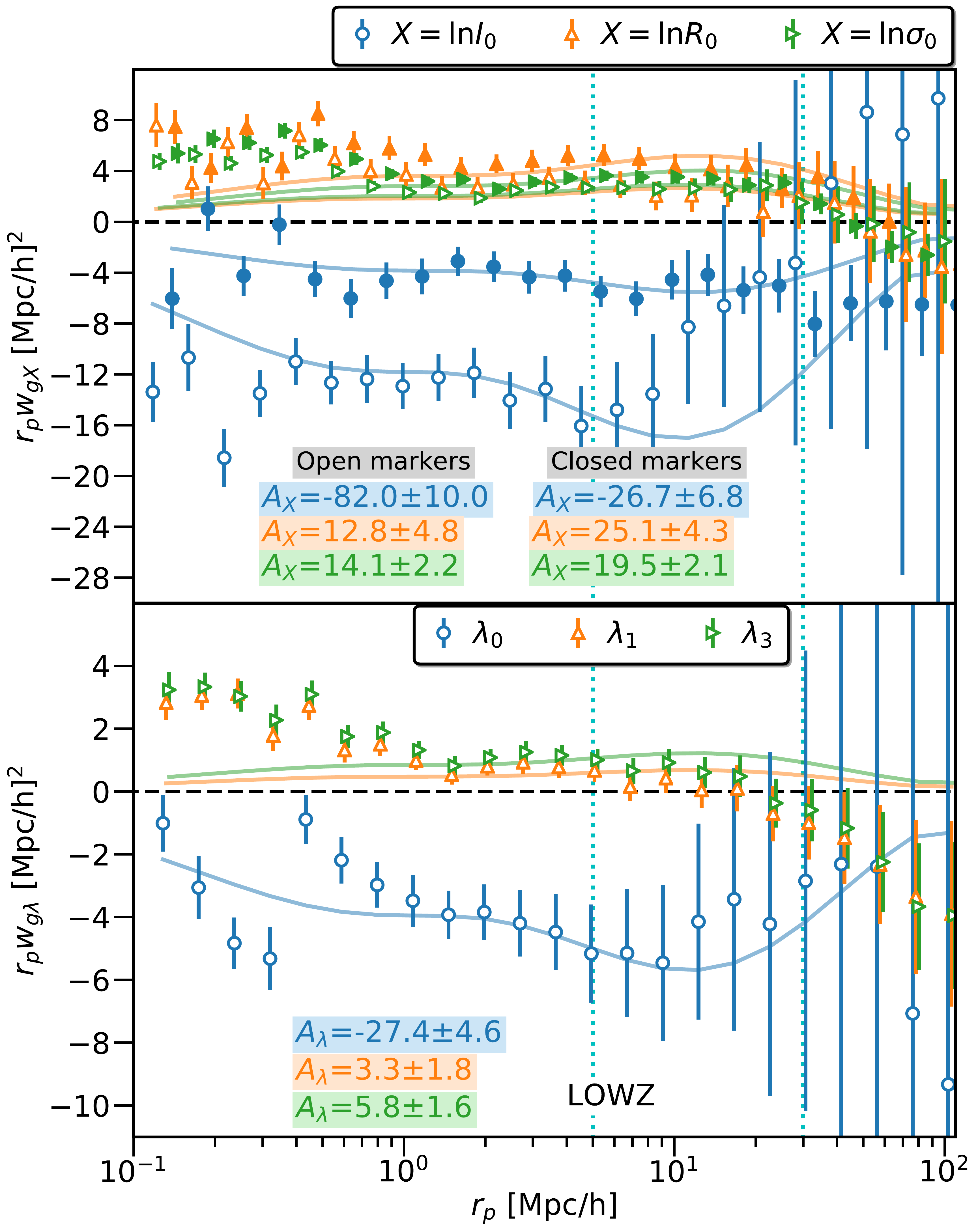}
				\caption{ Upper panel: Galaxy-$\lambda$ like cross correlation functions, where $\lambda$ is replaced with different galaxy properties, 
						namely 
						the surface brightness, $\log I$ (blue), physical radius, $\log R$ (orange) and velocity dispersion, $\log v$ (green). For the
					open points, we set the mean of these galaxy properties to be zero only at the level of full sample, while for the closed points, the 
					mean is set to zero within small redshift bins, $\Delta z=0.02$. Size and velocity 
					dispersion are positively correlated with the density field, though surface brightness shows negative correlations which are also stronger 
					when redshift evolution is not corrected for. Lower panel: Cross correlation measurements using residuals from 
					different definitions of fundamental plane. 
					Standard FP residuals , $\lambda_0$ is negatively correlated driven by the effects of surface brightness, while FP corrected for redshift 
					evolution show positive correlations with density.
					}
				\label{fig:lowz_fp_gp}
		\end{figure}
		   To understand the source of the correlations, in figure~\ref{fig:lowz_fp_gp} we show the cross correlations of galaxy properties, the surface brightness
		   $\log I$, physical 
		   radius, $\log R$ (orange) and velocity dispersion, $\log v$ in figure~\ref{fig:lowz_fp_gp}. The surface brightness shows strong correlations with the 
		   density field, with a large fraction of the signal driven by its strong evolution with redshift. 
		   These correlations provide interesting insights that the larger galaxies and ones with higher velocity dispersion tend to reside in the over dense 
		   regions but galaxies in over dense regions tend to have lower surface brightness.
		   
		   The impact of these parameters can be observed in the lower panel of figure~\ref{fig:lowz_fp_gp}. The correlations of the FP residuals, $\lambda$, are 
		   essentially the weighted sum of the correlations of the galaxy properties, where the weights are the parameters of the FP.
		   For $\lambda_0$, surface brightness dominates given that 
		   the redshift evolution has not been corrected for and leads to large negative correlations.
		   Once redshift evolution is corrected, the correlations of
		   surface brightness decrease and hence the FP cross correlations become positive for $\lambda_1$ and $\lambda_3$. This further justifies our choice to
		   include the redshift evolution in the FP, as the correlation functions otherwise are dominated by the redshift evolution of the FP which itself is dominated
		   by the redshift evolution of the surface brightness. 
		    
		   We also note that our measurements of galaxy-$\lambda_3$ cross correlations are not consistent with the results of \cite{Joachimi2015}.
		   This is because of the very different samples used in the two studies. As observed in section~\ref{ssec:results_FP_lum}, the FP residuals are
		   strongly correlated with the galaxy luminosity, with fainter galaxies having negative $\lambda$. As we will show in the appendix~\ref{appendix:lcdep}, 
		   fainter sub-samples also show negative correlations between FP residuals and density, in our measurement. 
		   Given that SDSS main sample used in \cite{Joachimi2015} was fainter than BOSS samples, we hypothesize that the
		   different measurements in the two studies are primarily due to the different galaxy samples. In order to rule out pipeline differences, we also 
		   reanalyzed the data of \cite{Joachimi2015} with our current pipeline and reproduced their results\footnote{Both pipelines were developed by S. Singh} (not shown).
		   				   
		\subsubsection{Environment dependence}
			\begin{figure}
         		\includegraphics[width=\columnwidth]{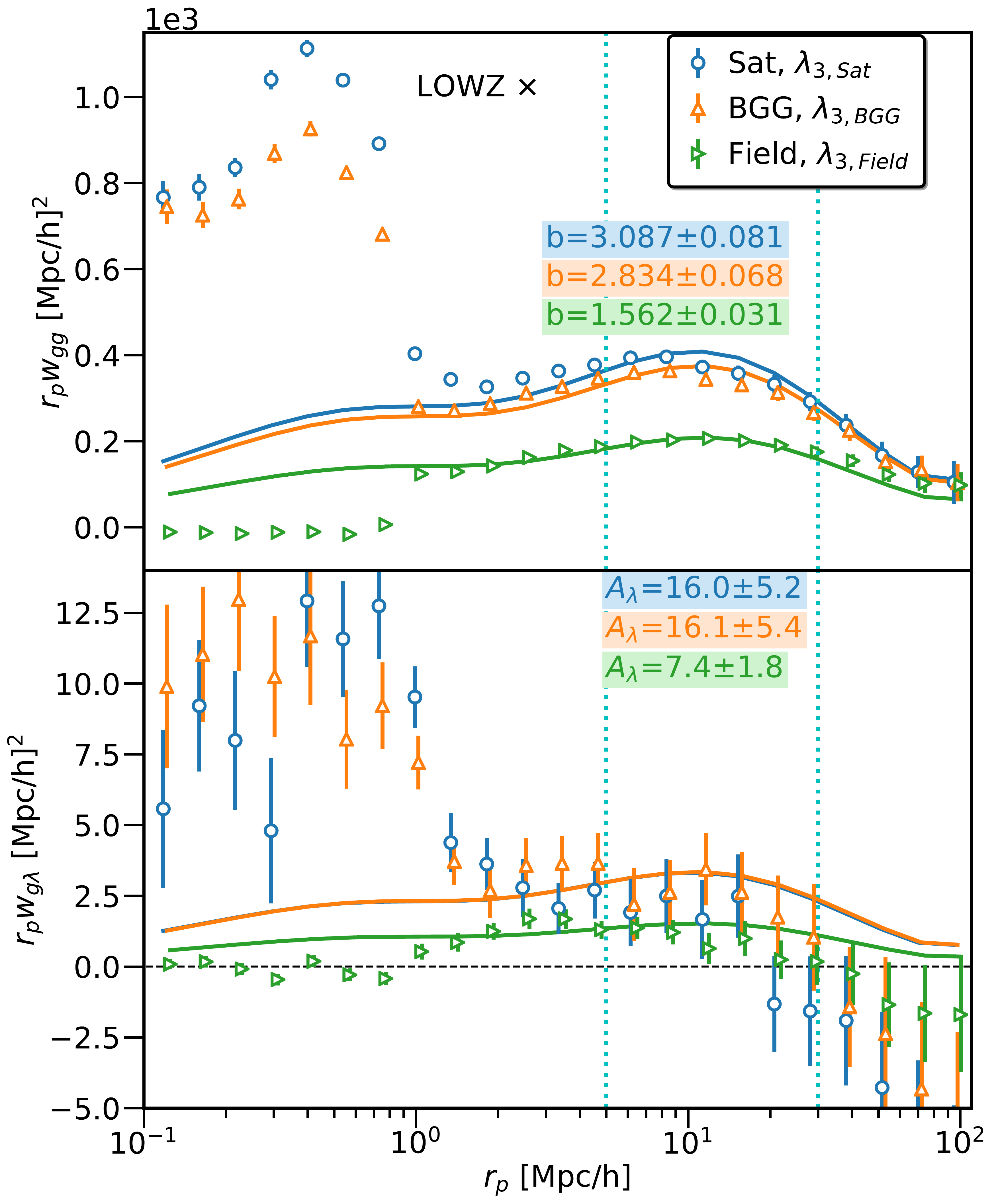}
				\caption{Galaxy clustering (cross correlation with LOWZ) 
					and galaxy-$\lambda$ cross correlation measurements for different environment samples, with full LOWZ sample
					as density tracers. Group galaxies, BGGs and Satellites, have similar clustering and galaxy-size cross correlations, 
					unlike intrinsic alignments, where satellites do not show large scale shape alignments.
					}
				\label{fig:lowz_env_w}
		   \end{figure}
		   
		   In figure~\ref{fig:lowz_env_w}, we show the correlations of FP residuals for group galaxies (satellites and BGGs) as well as the field galaxies.
		   Satellites and BGGs have higher galaxy bias as expected since they are on average in more massive halos and hence more dense environments. 
		   More interestingly, FP residuals for both satellites and BGGs show much stronger correlations with the density field as compared to the 
		   field galaxies. 
		   
		   $A_\lambda$ is rather large and also very similar for both Satellites and BGGs though the uncertainties for both samples are also large. 
		   Primary concern with such large signal is that since these 
		   galaxies reside in crowded regions, there can be some residual   
		   systematics in the photometry leading to such correlations. Since we subtract the randoms signal, additive systematics are unlikely
		   to be able to lead to such large signals unless they strongly correlate with the galaxy density field. We have some evidence for systematics affecting the measurements, especially for
		   LOWZ sample, but those systematics are predominately on large scales and do not lead to such large and significant $A_\lambda$ values (see 
		   appendix~\ref{appendix:FP_correlations}). We have also tested for the flags in SDSS photometry for blending and other photometry issues and BGGs, 
		   satellites and field galaxies all have very similar (low) rate of problematic photometry flags. Thus it is unlikely that satellite and BGGs 
		   results in particular are affected by the photometry problems. 
		   
\begin{figure}
         		\includegraphics[width=\columnwidth]{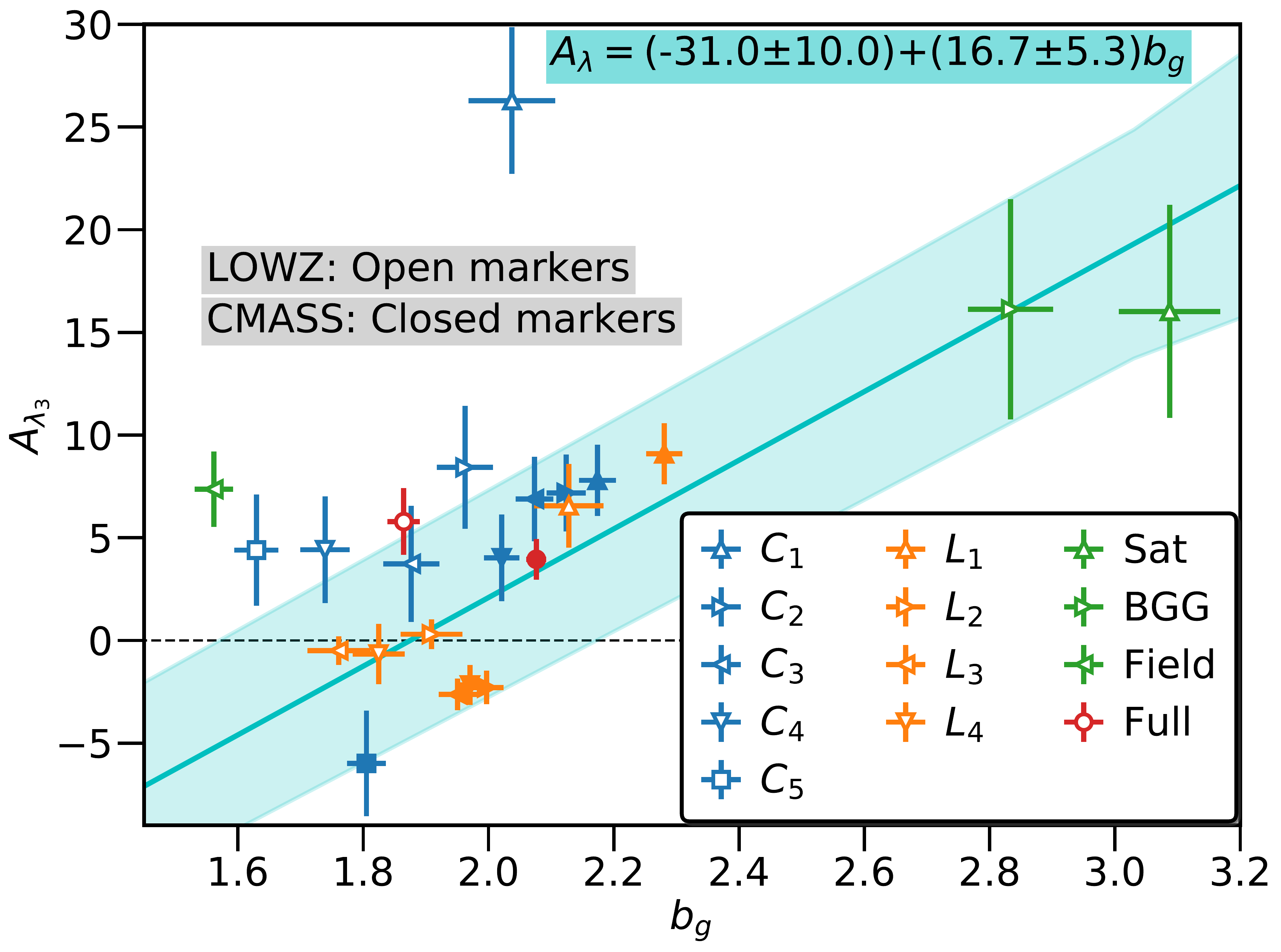}
				\caption{Environment dependence (as characterized by linear galaxy bias) of FP residuals.
				{In more biased (overdense) environments, scatter about FP has stronger correlations with the environment.}
					}
				\label{fig:lowz_cmass_A_bias}
		   \end{figure}

		   Another possible explanations for such similarities 
		   is that the galaxy environment plays a strong role in determining the galaxy size, in addition to galaxy properties such as luminosity and 
		   size. To further study the impact of the environment on FP residuals, in figure~\ref{fig:lowz_cmass_A_bias} we show the $A_\lambda$ as function of 
		   galaxy bias, $b_g$, where bias is a proxy for the galaxy environment (measurements for samples based on luminosity and color are presented in 
		   appendix~\ref{appendix:lcdep}). Though there is considerable scatter, we observe that the galaxies with 
		   larger bias, i.e., the galaxies in over dense regions, tend to have larger $A_\lambda$. The observed high $A_\lambda$ for satellite galaxies is 
		   consistent with this trend as these galaxies also have higher bias. 
		   These results are are not straight forward to interpret within the context of the tidal stripping of satellite galaxies as was used as an explanation in 
		   \cite{Joachimi2015}. Our results in figure~\ref{fig:lambda_mean_Mr} suggests that FP residuals of the satellite galaxies can primarily be explained 
		   by the luminosity dependence of the FP. However, when computing the correlation functions, we fit FP only to the satellite galaxies and within the satellite 
		   sample it is possible that stronger tidal stripping in denser environment can imprint some environment dependence on FP residuals leading to stronger 
		   correlations.
%
		   In either case, our results suggest that environment plays a dominant role in determining the FP 
			residuals of a galaxy.		   
		   More detailed interpretation of these results will require a study using the realistic galaxy simulations to understand
		   the relative importance of various processes involved in determining the galaxy sizes. We leave such a study for the future work.
		   
		\subsubsection{Correlations with IA}

			\begin{figure*}
				\begin{subfigure}[t]{\columnwidth}
         			\includegraphics[width=\columnwidth]{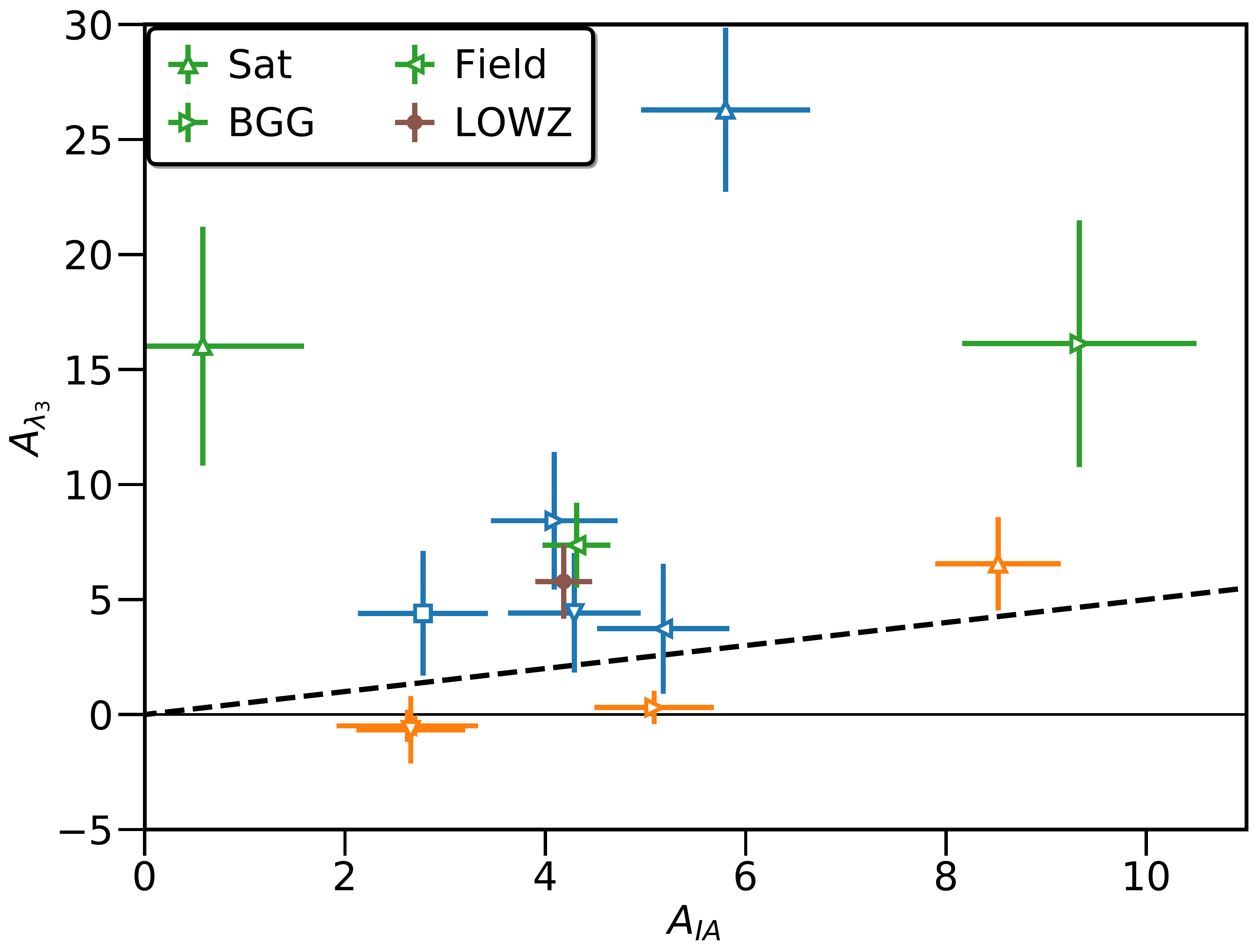}
		    	     \label{fig:}
			     	\caption{}
		    	 \end{subfigure}
			     \begin{subfigure}[t]{\columnwidth}
         			\includegraphics[width=\columnwidth]{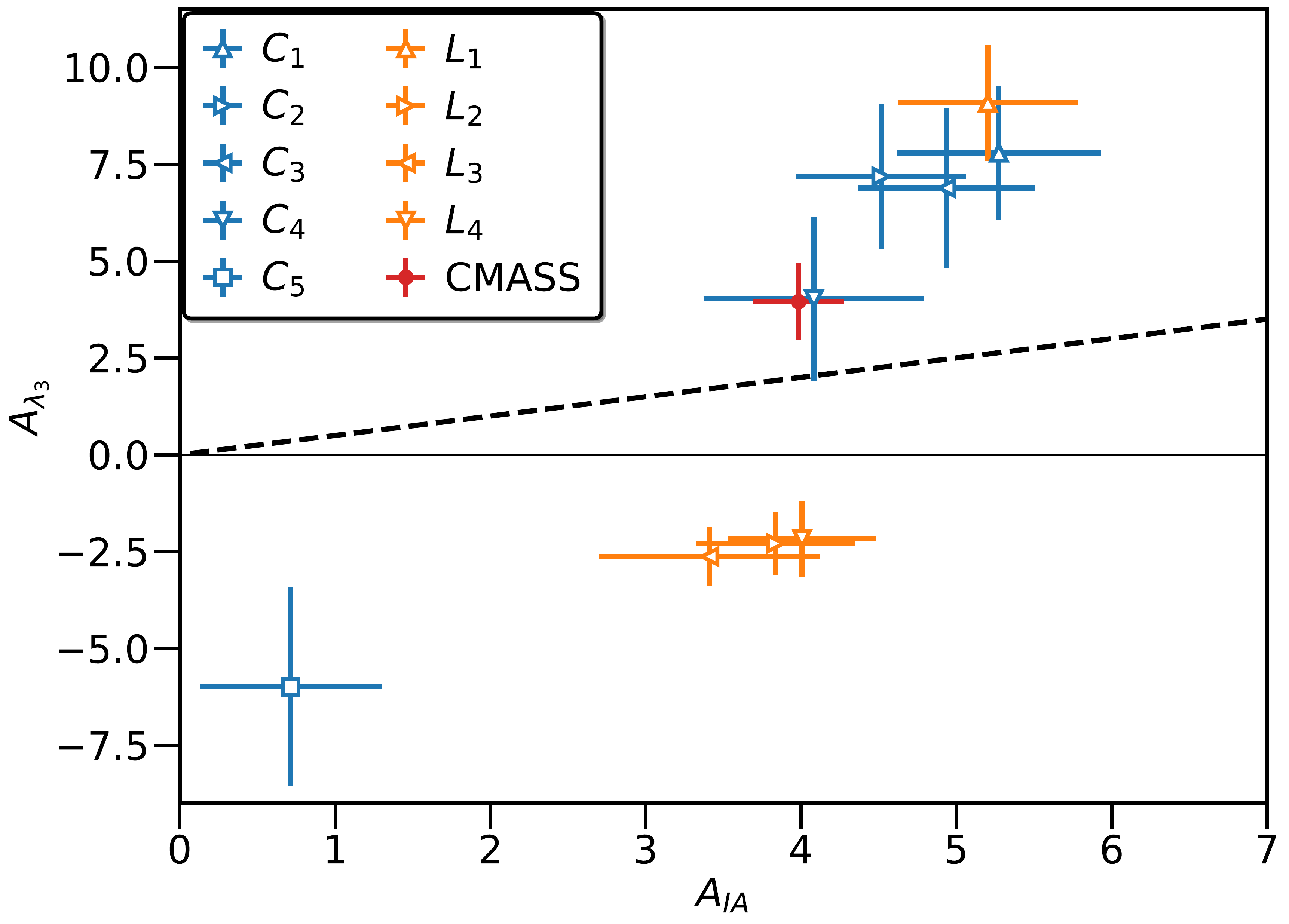}
	    	    	 \label{fig:}
				     \caption{}
				\end{subfigure}
				\caption{ Comparison of the intrinsic alignments amplitude ($A_{IA}$) measured using galaxy shear and the amplitude of galaxy size 
					correlations measured using the fundamental plane residuals ($A_{\lambda_3}$), for different subsamples of LOWZ (a) and CMASS (b).
					Different colors represent different splits and different markers represent different subsamples (labels for color and luminosity 
					subsamples are consistent across two panels).
					Under the model assumed in section~\ref{sssec:corr_FP}, 
					size correlations are caused by intrinsic alignments in conjunction with projection effects and 
					we expect $A_{\lambda_3}\propto A_{IA}/2$ (shown by dashed black line). Solid cyan line shows the best fit linear 
					model with parameter as shown in the figures. Data prefers $A_{\lambda_3}\propto 4 A_{IA}$, which suggests that in addition to 
					projection effects, galaxy sizes themselves are affected by the tidal field, such that $\lambda_3\propto\nabla^2\phi$, with similar 
					constants as the IA model.
				}
				\label{fig:A_ia_fp}
		   \end{figure*}		   
		   As discussed in section~\ref{sec:formalism}, intrinsic alignments (IA) of galaxies coupled with the projection effects can lead to the scatter over the 
		   FP and the correlations between the FP residuals and the density field. The model in used in fitting the cross correlations between the FP residuals 
		   and the galaxy density field accounts for this effects and indeed if IA is the only cause of correlations of FP residuals, we expect $A_\lambda\sim A_I/2$.
		   
		   In figure~\ref{fig:A_ia_fp} we present the comparison of the the intrinsic alignments amplitude (a detailed analysis of IA measurements was presented in
		   \cite{Singh2015} and in this work we repeat those measurements using BOSS DR12 data) 
		   and the FP residual amplitude $A_\lambda$ derived from 
		   the cross correlations with the density field. For both LOWZ and CMASS samples we observe positive correlations between $A_I$ and $A_\lambda$, samples
		   with stronger IA also showing stronger correlations for FP. This is also consistent with the observations that IA and FP correlations have similar 
		   environment dependence (see previous section and \cite{Singh2015}). However, our measurements are inconsistent with the model predictions of 
		   $A_\lambda=A_{IA}/2$. In addition to considerable scatter in the measurements, the best fit linear models we obtained (not shown) deviated significantly
		   from the model.
		   
		   
		   Our results suggest that the galaxy size correlation (as measured by FP) include contributions beyond the effects of intrinsic alignments and the 
		   projection effects. These contributions can come from physical processes such as stronger feedback in over dense regions or observational systematics 
		   affecting the estimation of size, magnitude and velocity dispersions of galaxies (eg. errors in PSF modeling, small fiber size used in spectroscopic 
		   measurements). The interpretation of these results is 
		   further complicated by the fact that IA also depends on the shape measurement methods. As shown in \citep{Singh2016ia}, \dev\ shape results in 
		   $\sim 15-20\%$ larger IA amplitude though \dev\ shapes were also shown to be affected by systematics in the same study because of which we use the 
		   re-gaussianzation shapes to measure the IA in this paper. Hence, it is difficult to fully explain the origin and the magnitude of the size correlation 
		   amplitudes and a detailed exploration of the physical origin of these effects will require further study with realistic simulations.
		   
		  \subsubsection{Multipoles}	
			\begin{figure}
         		\includegraphics[width=\columnwidth]{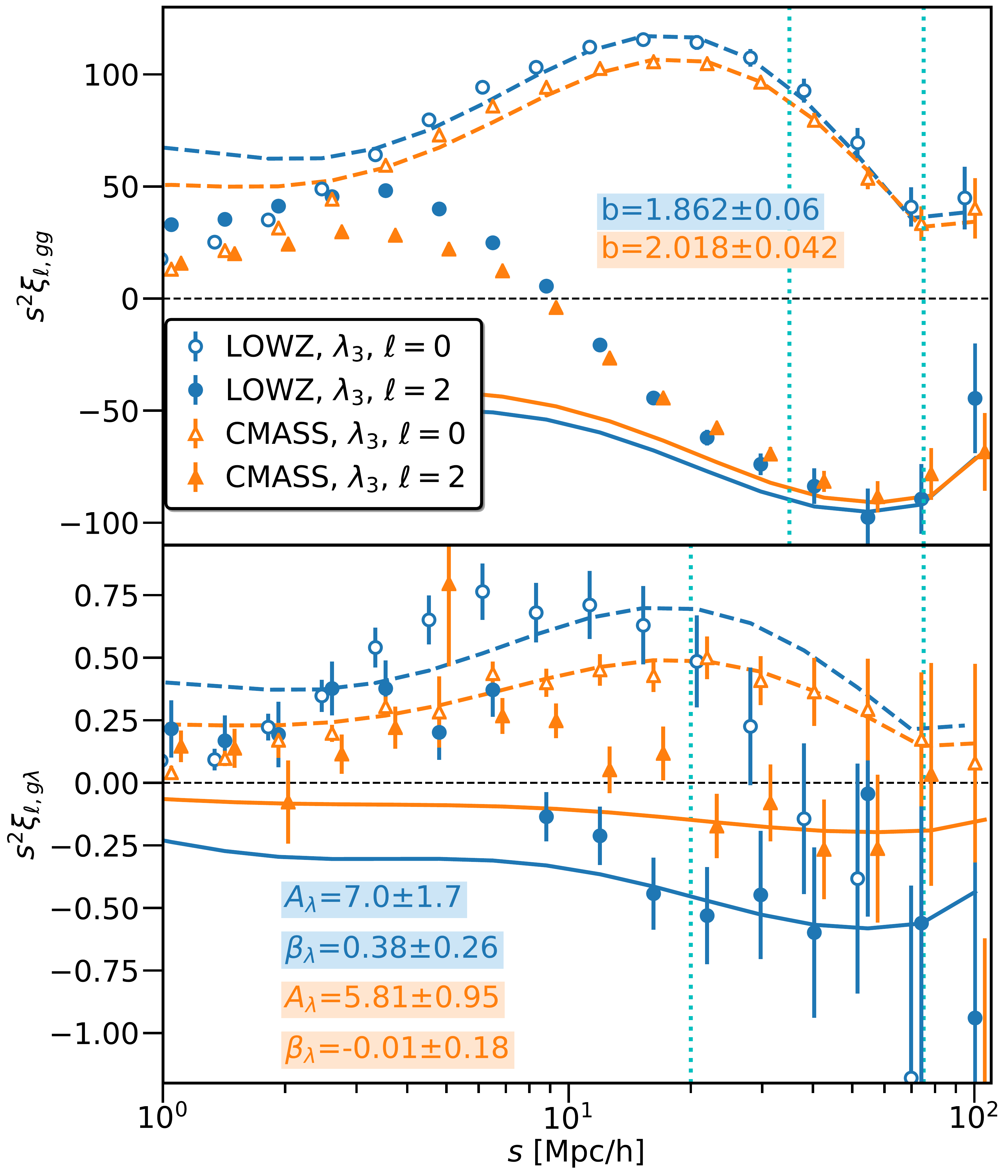}
				\caption{Measurements of the multipoles of galaxy clustering (upper panels) and galaxy-$\lambda$ (lower panels) 
					cross correlation functions for LOWZ (blue) and CMASS (orange) samples. Open points and dashed line shows the monopole and the best 
					fit model for the monopole ($\ell=0$) while closed points and lines show the same for quadrupole ($\ell=2$). Vertical cyan lines 
					mark the range over which the model was fit.
					}
				\label{fig:lowz_cmass_multipoles}
		   \end{figure}
		   
		   In figure~\ref{fig:lowz_cmass_multipoles}
		   we show the measurements of the multipoles of the galaxy clustering and galaxy-FP cross correlation measurements. The monopole and quadrupole of the
		   galaxy clustering are consistent with the expectations from the redshift space distortion measurements. Here we fit the
		   simple Kaiser model to fit both the monopole and quadrupole moments and we will present more detailed analysis of galaxy clustering multipoles in the
		   next section. The galaxy bias obtained from these fits is consistent with the values obtained from the projected correlation functions.
		   Since the model does not include the effects of non-linear corrections, it does not fit data well on small scales and hence we only use
		   $30<s<70\mpch$ to fit the galaxy-galaxy correlation functions and we further fix the growth rate to the value expected from our fiducial cosmology
		   $f=0.665 (0.77)$ for LOWZ (CMASS). 
		   Since our focus in this section is to study the anisotropy of the galaxy-FP 
		   correlation function, we prefer to use a simpler model with few parameters over a more detailed RSD model presented in the next section. 
		   
		   In the lower panel, we present the measurements and the fits for the galaxy-FP cross correlation function. We detect both the monopole and 
		   the quadrupole 
		   moments of the correlation functions, pointing to significant line of sight anisotropy in these measurements. Some level of anisotropy is 
		   expected as the 
		   the galaxy positions are measured in redshift space and the FP residuals are weighted with galaxy density field in the redshift space. 
		   To estimate the anisotropy  contributions from the FP residuals, we fit for the anisotropy factor, 
		   $\beta_\lambda$ (see eq.~\eqref{eq:lambda_beta}) (galaxy anisotropy $\beta_g$ is obtained from clustering). The best fit values of $\beta_\lambda$ we obtain 
		   are consistent with zero, contrary to the expectations from the model which predicts $\beta_\lambda=-3$ ($\beta_\lambda=-3$ predicts positive quadrupole 
		   moments at large scales ). These conclusions 
		   are not changed even if we fit with growth rate $f$ as a free parameter, if we vary the minimum scale used in the fits ($r_{p,\text{min}}=20$ or 
		   $40\mpch$) and even if we include the hexadecapole measurements.
		   
		   While our measurements appear to rule out the influence of IA on the FP residuals, we note here that this is only true within the model we assumed in this
		   work. Both IA and FP residuals are also weighted by the galaxy density field, which is measured in the redshift space and introduces higher order 
		   terms which can have significant contributions to the measured correlation functions. Furthermore, these higher order terms also contain the 
		   line of sight anisotropy terms which can in principle affect the $\beta_\lambda$ constraints. 
		   Modeling these higher order terms accurately is out of the scope of this work and can be attempted in a future work.


	\subsection{IA effects on RSD}\label{ssec:results_FP_RSD}
	
\begin{figure}
        \includegraphics[width=\columnwidth]{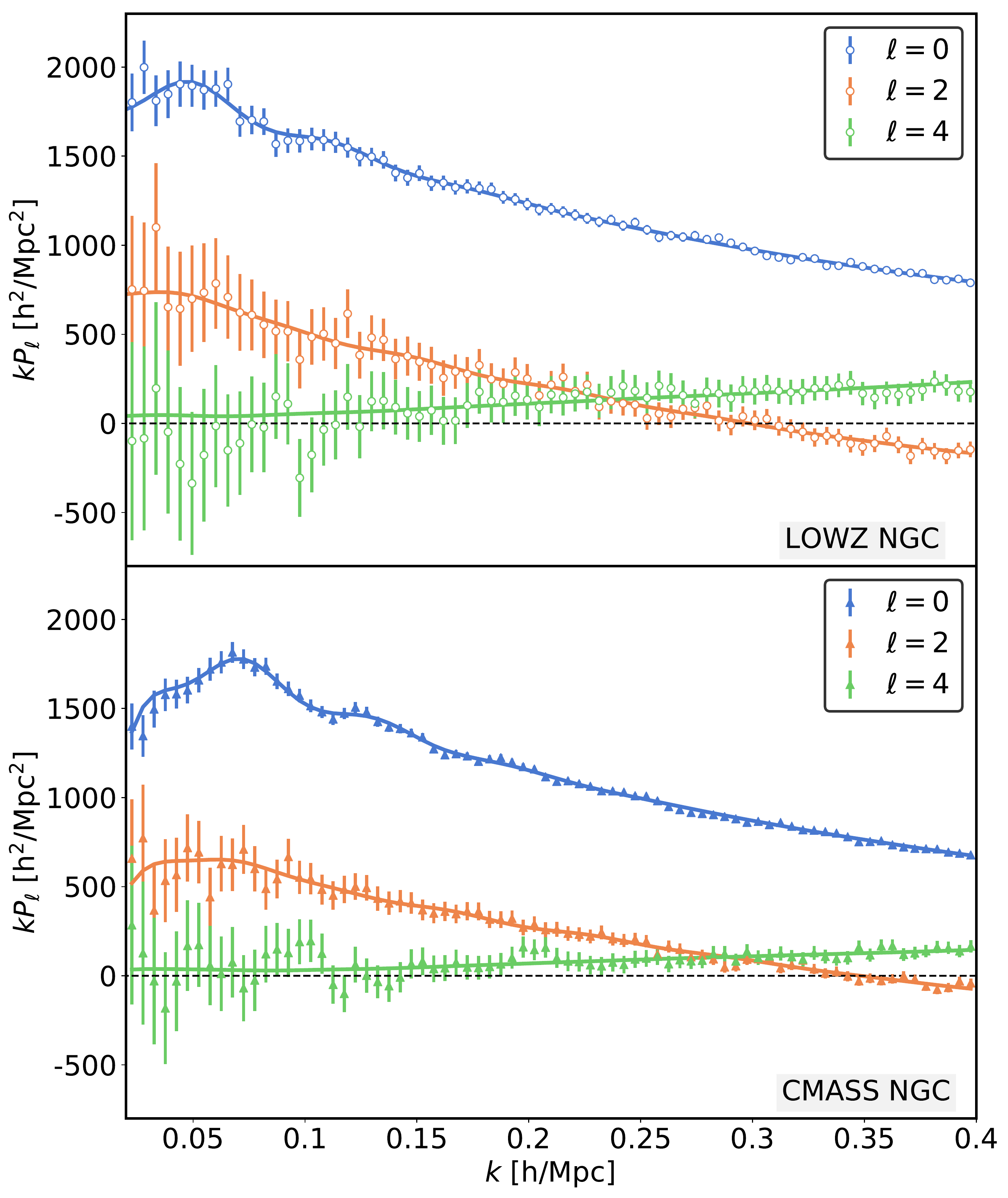}
			\caption{ The measured galaxy power spectrum multipoles in Fourier space (data points) and the best-fit theory curves (solid lines) for LOWZ NGC (upper panel) and CMASS NGC (lower panel) samples. We fit the model to the monopole (blue), quadrupole (orange), and hexadecapole (green), over the wavenumber range $k = 0.05 - 0.4$ and $0.02 - 0.4 h$Mpc$^{-1}$ for LOWZ and CMASS galaxies, respectively. Multipoles are accurately modelled, down to $k=0.4 h$Mpc$^{-1}$. Although not shown in the figure, we also find an excellent model fit to the SGC samples.
				}
			\label{fig:multipoles_LOWZ_CMASS}
	\end{figure}

	In this section we present the measurements of the galaxy power spectrum multipoles in Fourier space. In figure~\ref{fig:multipoles_LOWZ_CMASS}, we show the measured monopole $P_0(k)$, quadrupole $P_2(k)$, and hexadecapole $P_4(k)$ of the LOWZ NGC and CMASS NGC galaxies, using the FFT-based galaxy power spectrum estimator described in section~\ref{sssec:rsd}. We then fit the RSD model presented in section~\ref{sssec:rsd} to the measured multipoles and find that the power spectrum multipoles are accurately modeled, down to scales of $k=0.4 h$Mpc$^{-1}$, in agreement with~\cite{Hand:2017ilm}. The fits to the SGC galaxies are not shown in the figure, but we also find a good agreement between the model and the SGC samples. We include the hexadecapole $P_4(k)$ because this improves RSD constraints significantly, as reported earlier in \mbox{\cite{Beutler:2016arn}, \cite{Grieb2016},} and \cite{Hand:2017ilm}. In our fits, we set the minimum wavenumber $k_{\mathrm{max}}$ to 0.05 and 0.02$\ h$Mpc$^{-1}$, respectively for LOWZ and CMASS, in order to minimize any large-scale effects of the window function. As described in section~\ref{sssec:rsd}, we fix the AP distortion parameters to their fiducial values and constrain 11 free parameters, of which two are primarily of our interests: the growth rate $f$ and the amplitude of matter fluctuations $\sigma_8$. Fitting this RSD model to the BOSS DR12 multipole measurements, we obtain a tight constraint on the growth of structure, and more detail can be found in \cite{Yu2019}.

	\subsubsection{Fundamental plane cuts}\label{subsubsection:FPcut}
	
	In this section we fit FP to the LOWZ and CMASS samples, for NGC and SGC regions separately, and then split each galaxy sample into two subsamples according to the sign of the FP residuals, with the mean FP residual subtracted from each sample, following \cite{Martens2018}. Samples with positive (negative) FP residuals correspond to galaxies larger (smaller) than the FP-predicted size. We then fit the galaxy power spectrum model to each of the two subsamples, constraining 11 free parameters in the RSD model presented in section~\ref{sssec:rsd}. 
	
\begin{figure}
        \includegraphics[width=\columnwidth]{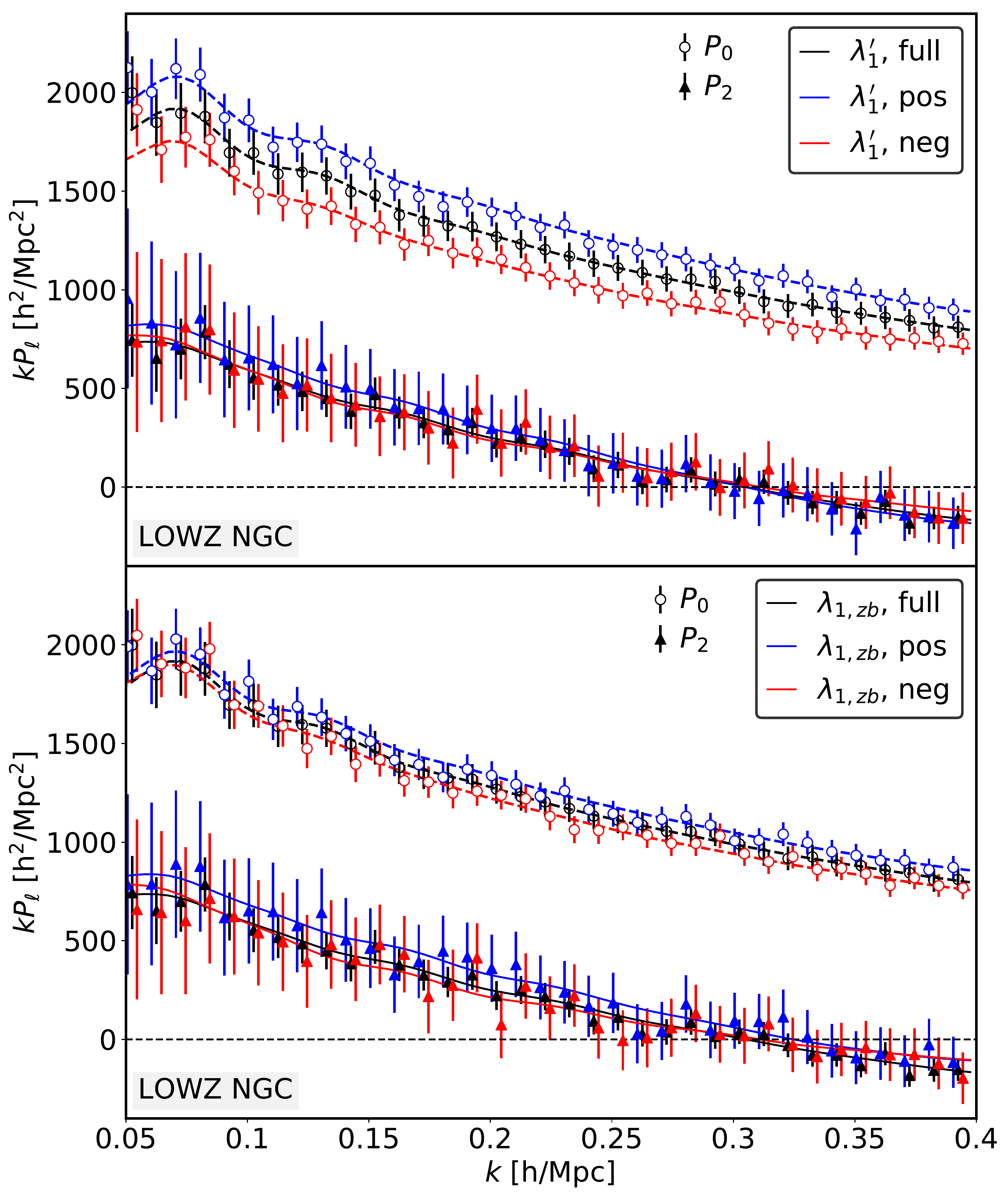}
			\caption{ Multipole measurements of the LOWZ NGC subsamples with positive (blue) and negative (red) FP residuals, with different types of FP fits. Multipoles of the full LOWZ NGC sample (black) are also provided as a reference, and open circular points and closed triangular points display the monopoles and quadrupoles, respectively. We show the measured multipoles (data points) and the best-fit theory curves (solid lines) for positive and negative subsamples with two different FP definitions: FP fit without velocity dispersion ($\lambda_{1}^I$; upper panel) and FP fit in narrow redshift bins ($\lambda_{1,zb}$; lower panel). The RSD model fits well to all subsamples, and other subsamples with different FP definitions similarly have good model fits, although not shown in the figure. The monopoles of positive and negative subsamples clearly have different amplitudes, suggesting the difference in their galaxy biases.
			 }
			\label{fig:FPfit1}
	    \end{figure}

In the analysis, we consider different types of FP residuals: FP fit with $N_z = 1$ ($\lambda_1$), FP fit with $N_z = 1$ in narrow redshift bins ($\lambda_{1,zb}$), FP fit only dependent on the surface brightness with the velocity dispersion measurements ignored ($\lambda_{1}^I$; still with with $N_z = 1$), FP fit equivalent to the one in \cite{Martens2018} ($\lambda_1^M$), FP fit with $N_z = 3$ ($\lambda_3$). Figure~\ref{fig:FPfit1} shows how the model fits to each of the two subsamples of the LOWZ NGC galaxies. The subsample with positive FP residuals (henceforth called the ``positive" subsample) has higher galaxy bias than the subsample with negative FP residuals, in 
agreement with the results in \cite{Martens2018}.  We also find that disregarding the velocity dispersion in the FP definition causes a larger deviation in the galaxy bias between 
subsamples. Some deviation in the galaxy bias is expected from the correlations between the FP and the galaxy properties as discussed
	in section~\ref{ssec:results_FP_fits}. Since the ``positive" sample preferentially selects brighter galaxies,
	 it is expected to have larger bias.

\begin{figure}
        \includegraphics[width=\columnwidth]{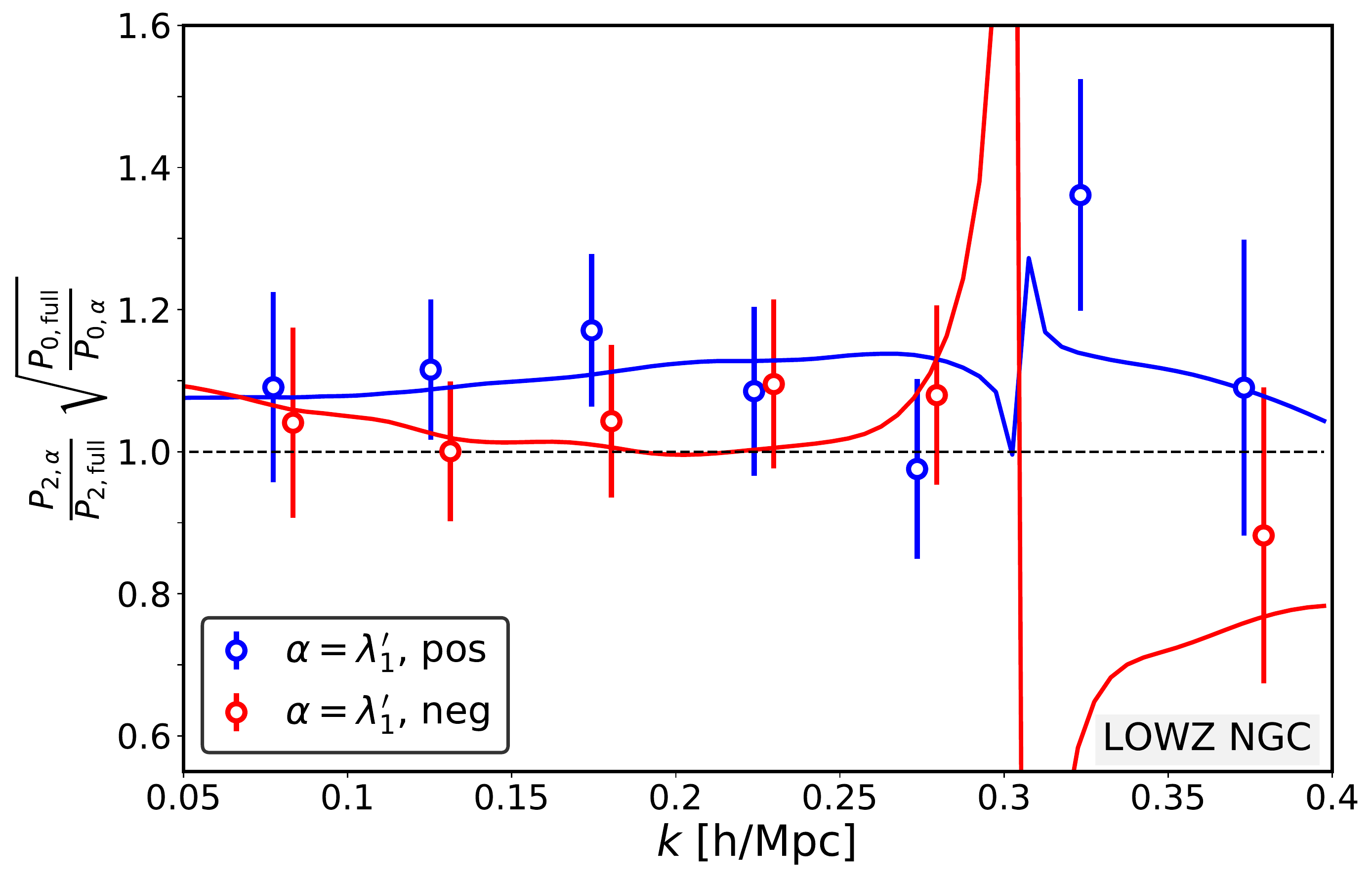}
			\caption{ A rough estimate of the ratio of $f\sigma_8$ values between the full LOWZ NGC sample and FP fit subsamples, as quantified by the quadrupole ratio $(P_{2,\alpha})/(P_{2,\mathrm{full}})$ divided by $\sqrt{(P_{0,\alpha})/(P_{0,\mathrm{full}})}$, where $\alpha$ denotes the FP fit subsample. Open circular points show the measured ratios, while the dotted lines show the ratios from the best-fit theory curves. The difference in the measured ratios of the two subsamples is not statistically significant, especially on the scales where non-linear, small effects are not important.
			 }
			\label{fig:FPfit3}
	    \end{figure}

	The monopole $P_0(k)$ and the quadrupole $P_2(k)$ scale as $(b_1\sigma_8)^2$ and $b_1 f\sigma_8^2$, respectively. Hence, we can roughly estimate the ratio of $f\sigma_8$ and $b_1 \sigma_8$ values between two subsamples of the FP fit from the ratios of the monopoles and quadrupoles. The square rooted ratio of the monopoles scales as $b_1\sigma_8$; for the LOWZ NGC sample, the subsample with positive $\lambda_1^I$ has this ratio $\approx$ 10\% larger than the subsample with negative $\lambda_1^I$. Adding the velocity dispersion term to the FP 
	definition reduces such deviation to $\approx$ 6\%. We can also take the quadrupole ratio between two samples and divide it by the square rooted ratio of their monopoles to remove the bias dependence. This quantity roughly determine the ratio of $f \sigma_8$ values between two samples. Figure~\ref{fig:FPfit3} plots the ratios 1) between the full LOWZ NGC sample and the subsample with positive FP residuals (blue) and 2) between the full LOWZ NGC sample and the subsample with negative FP residuals (red). The measured ratios are well within 1$\sigma$ from each other, particularly on the scales where non-linear, small effects are not important, suggesting that measurements of the difference in RSD constraints between the FP fit subsamples are not statistically significant.
	
\begin{figure}
        \includegraphics[width=\columnwidth]{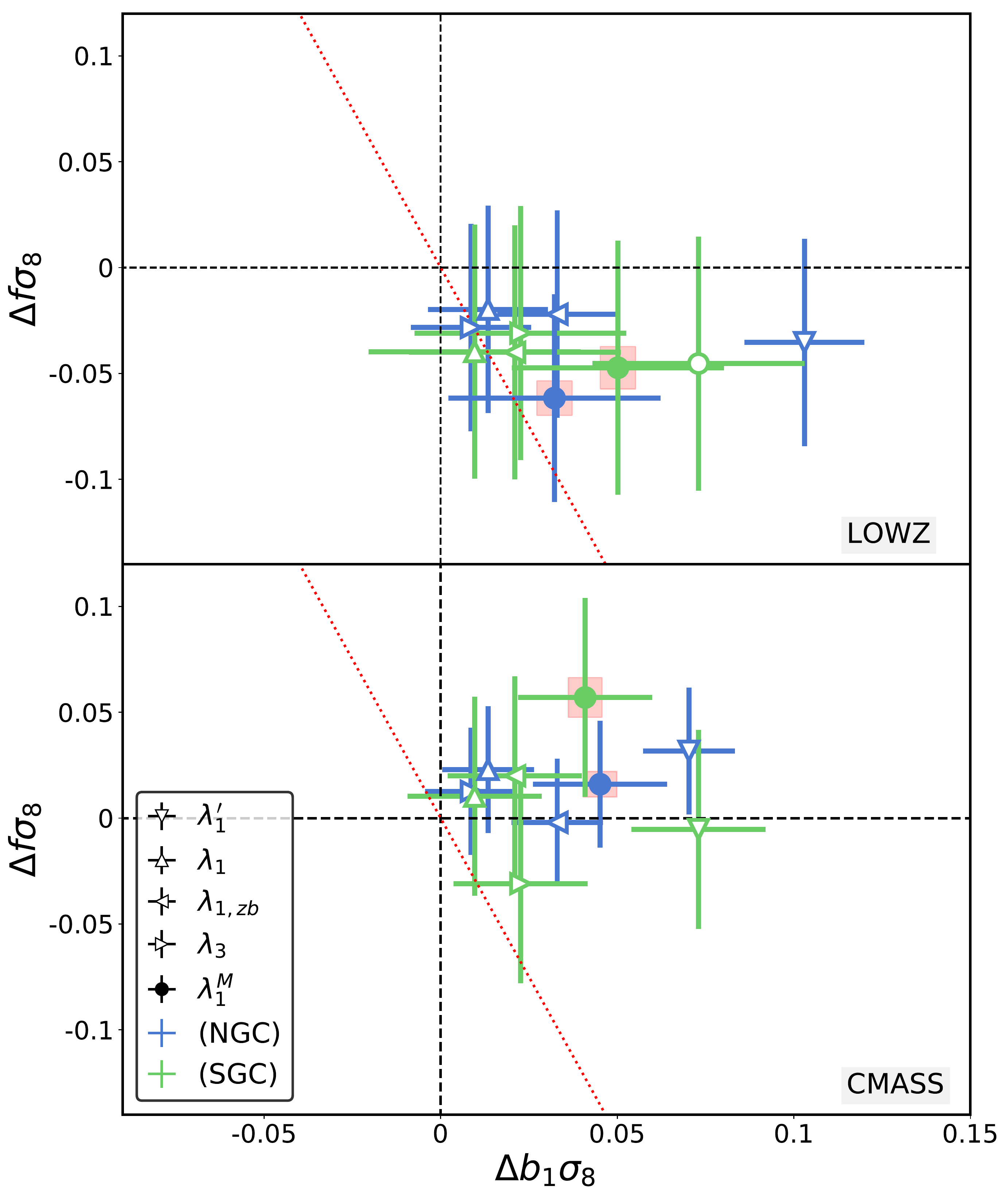}
			\caption{ The measured difference in $f \sigma_8$ and $b_1 \sigma_8$ between positive and negative subsamples, with different definitions of FP fits (indicated with different markers), for the LOWZ (upper panel) and CMASS (lower panel) galaxies in NGC (blue) and SGC (green) regions. Data points highlighted with red rectangles are using the FP measurements from Martens et al. (2018). Note that samples in the same color are all correlated with one another; we take the same galaxy sample and split them into the FP fit subsamples based on different FP definitions. The differences in $f \sigma_8$ values are all statistically consistent with the null results, while the signs of $\Delta b_1\sigma_8$ is consistently positive. For both LOWZ and CMASS, the theory prediction (red dotted lines) from section~\ref{sssec:IARSD}, $\Delta f \approx -3\Delta b_1$,  is not favored by the data points relative to the null model ($\Delta f\sim0$).
				}
			\label{fig:FPfit5}
	    \end{figure}

	RSD model fits in figure~\ref{fig:FPfit5} are in agreement with the above observation. For each galaxy sample, we fit the RSD model to the multipoles of positive and negative (FP fit) subsamples, for both NGC and SGC regions, and then measure the difference in $f \sigma_8$ and $b_1 \sigma_8$ constraints between two subsamples. With the convention that the value of the negative subsample is subtracted from the value of the positive subsample, we consider different types of the FP residuals ($\lambda_{1}^I$, $\lambda_{1}$, $\lambda_{1,zb}$, $\lambda_3$, and $\lambda_1^M$). In figure~\ref{fig:FPfit5}, FP fit subsamples selected in the same redshift range and sky region are marked in the same color, and such samples are are not independent, all correlated with one another. First, we find that all positive subsamples have larger galaxy biases than negative subsamples, as expected from figure~\ref{fig:FPfit1}, thereby resulting in the sign of $\Delta b_1\sigma_8$ positive in all FP fits. On the contrary, the signs of $\Delta f\sigma_8$ measurements are not consistent across all samples and all within 1$\sigma$ of the measurements, statistically consistent with the null result. This suggests that there is no evidence of significant bias in RSD measurements due to IA, in tension with the results from \cite{Martens2018}, which showed a consistent offset in $\Delta f$ between the FP fit subsamples. In section~\ref{sssec:IARSD}, the model expects that $\Delta b_g + \frac{1}{3}\Delta f \approx 0$, where $b_g$ and $f$ are galaxy bias and growth rate parameters, respectively. However, as shown in figure~\ref{fig:FPfit5}, the null model ($\Delta f\sim0$) is favored over such theory prediction by the data points. Similarly, figure 16 and 18 in \cite{Martens2018} also show a deviation between the model and measured data points.

	\cite{Martens2018} quantifies how well the measurements agree with the theoretical predictions in the following way, 
	
\begin{equation}
\frac{\mathrm{Observed}}{\mathrm{Theory}} = \frac{\Sigma_i \Big[ \frac{B_i^{\mathrm{Theory}}}{\sigma_i} \Big]^2 \frac{B_i^{\mathrm{Observed}}}{B_i^{\mathrm{Theory}}} }{\Sigma_i \Big[ \frac{B_i^{\mathrm{Theory}}}{\sigma_i} \Big]^2 } \pm \frac{1}{ \sqrt{\Sigma_i \Big[ \frac{B_i^{\mathrm{Theory}}}{\sigma_i} \Big]^2 } },
\end{equation}
where $B_i$ is the parameter measuring the amplitude of intrinsic alignments for each galaxy sample $i$, and $\sigma_i$ is the error on the measured $B$. 
\cite{Martens2018} measures Obs/Theory = $0.61 \pm 0.26$. Assuming the model in \cite{Martens2018}, we repeat the analysis and obtain Obs/Theory = $0.05 \pm 0.30$, consistent with zero. 

	In figure~\ref{fig:FPfit5}, we use the MultiDark-Patchy mock catalogues to estimate the size of the error bars of RSD constraints for the FP fit subsamples. In this work, we consider four different galaxy samples: LOWZ NGC, LOWZ SGC, CMASS NGC, and CMASS SGC. For each sample, we take 100 PATCHY mocks and separate each mock into two subsamples, depending on their stellar masses, as the stellar mass is correlated with the luminosity, which in turn is correlated with the FP residuals and also correlated with the galaxy alignment strength \citep{Joachimi2011,Singh2015}). Consequently, we have 100 subsamples with their stellar masses larger than the mean stellar mass of the sample and 100 subsamples with their stellar masses smaller than the mean. For each galaxy sample, we then perform fits to the measured multipoles of PATCHY mocks, and the best-fitting parameters for each of the 200 subsamples are obtained by maximum a posterior (MAP) estimation using the LBFGS algorithm. Subsequently, we measure the standard deviation of the best-fitting values of $f \sigma_8$ and $b_1 \sigma_8$ and obtain the propagated error for $\Delta f \sigma_8$ and $\Delta b_1 \sigma_8$. In appendix~\ref{appendix:nulltestRSD}, we also show the statistical error obtained by random splits; however, the sample variance and noise are generally treated better with the mock results, especially given the way samples are split. 
		   
		   
	\subsubsection{Luminosity/color cuts}
	
\begin{figure}
        \includegraphics[width=\columnwidth]{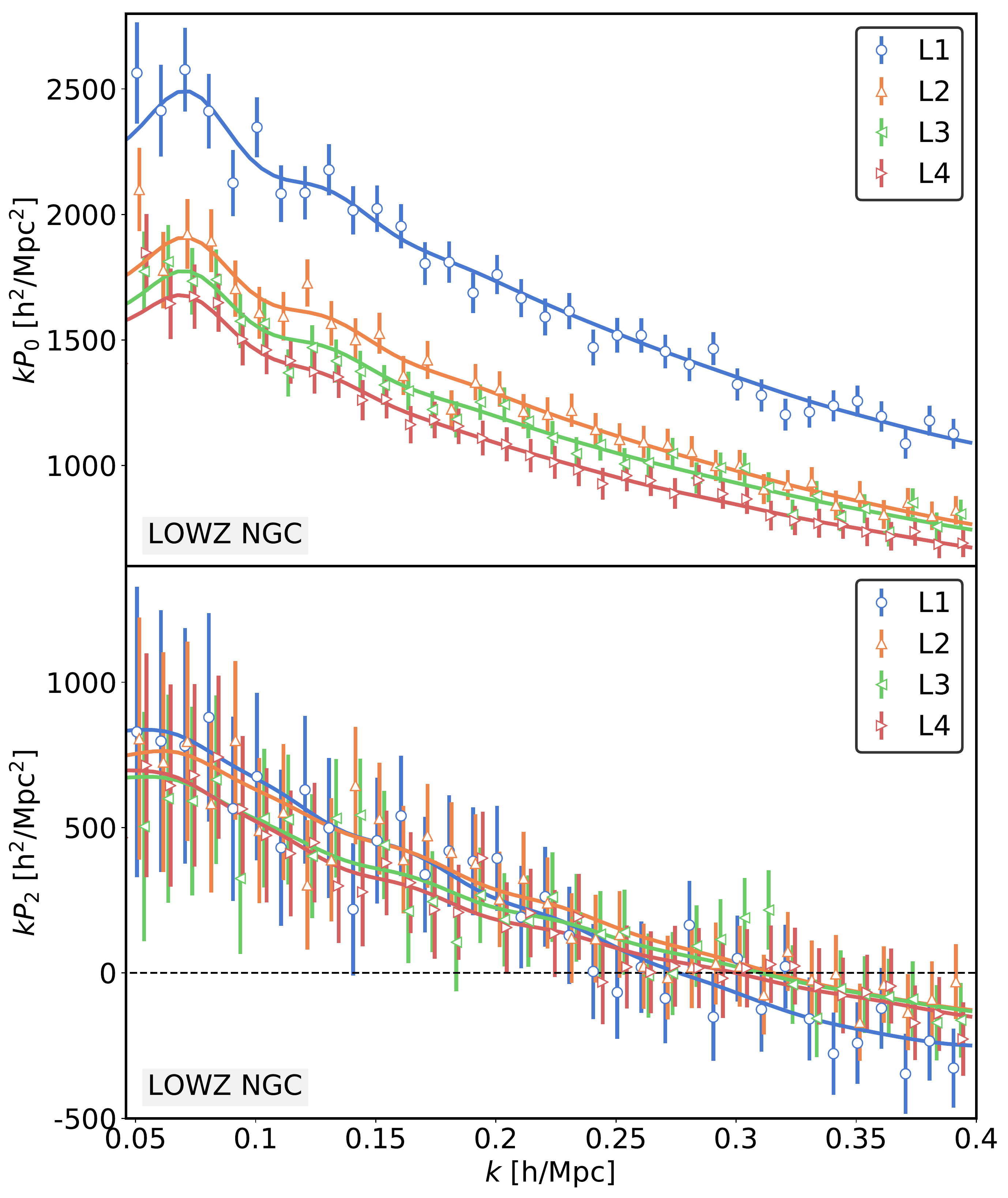}
			\caption{ The measured galaxy power spectrum multipoles (data points) of four luminosity subsamples (as in figure~\ref{fig:lowz_cmass_lum_w}), for the LOWZ NGC sample. Solid lines indicate the best-fit theory curves. The monopole measurements (upper panel) show a decreasing trend of bias with luminosity; this suggests that a brighter subsample has a higher galaxy bias. The quadrupole measurements (lower panel) of all luminosity subsamples are within 1 sigma of the quadrupole of the full sample.
				}
			\label{fig:LCcut1}
	    \end{figure}

	In fig~\ref{fig:LCcut1}, with luminosity cuts as described in section~\ref{sec:data}, we divide the LOWZ NGC sample into four subsamples, with $L_1$ being brightest and $L_4$ being faintest, and show how the RSD model fits the galaxy power spectrum multipoles of all subsamples. The measurements shown in the upper panel generally agrees with the results in appendix~\ref{appendix:lcdep}: as there is a decreasing trend of bias with luminosity. 
Similarly the lower panel plots the quadrupole measurements of all subsamples. The measured multipoles and the best-fit theory model of other samples, such as LOWZ SGC, CMASS NGC, and CMASS SGC, show a similar trend and therefore not shown in the figure. 
	
\begin{figure}
        \includegraphics[width=\columnwidth]{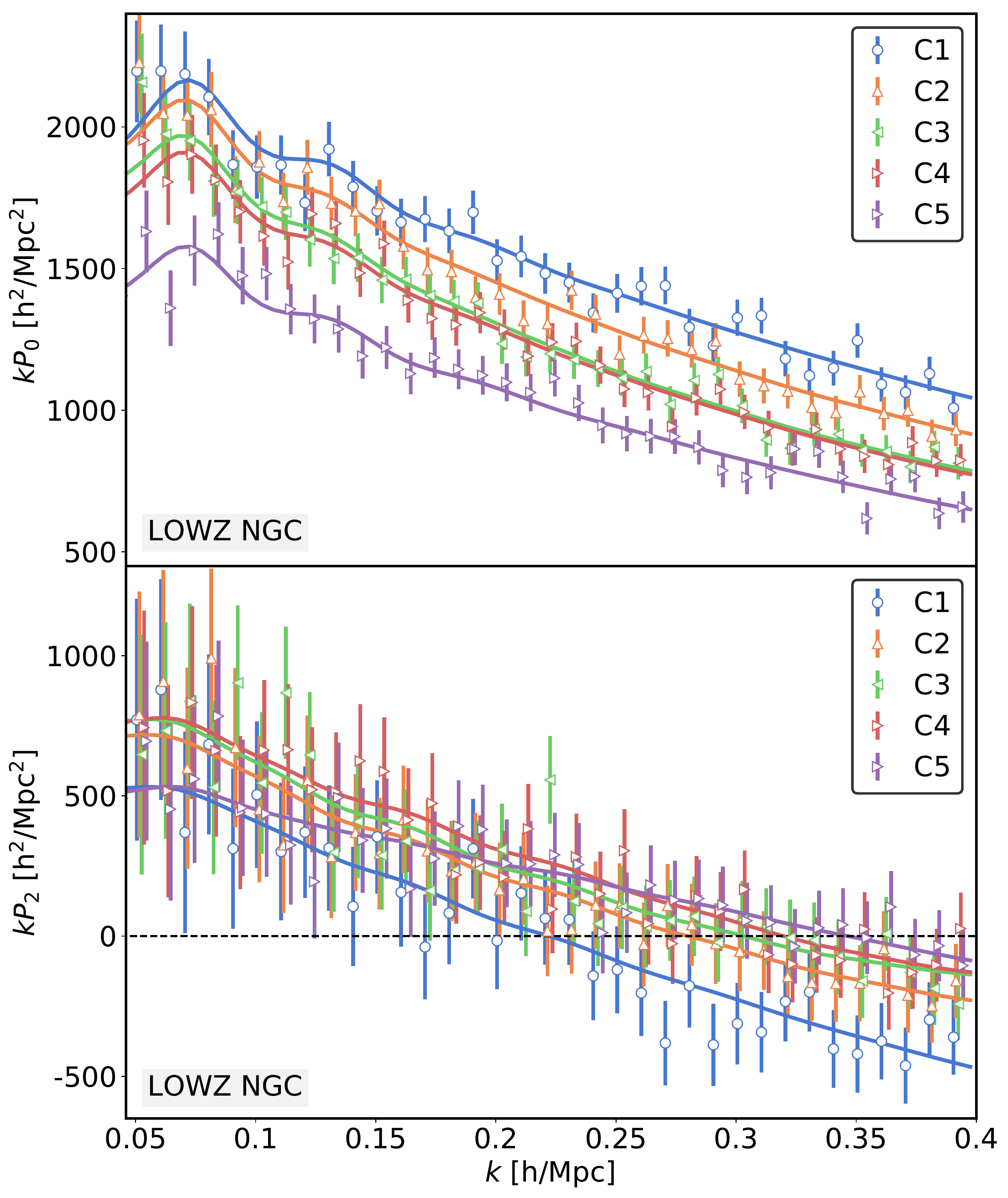}
			\caption{ Similar to figure~\ref{fig:LCcut1}. The multipoles of five color subsamples (as in figure~\ref{fig:lowz_cmass_color_w}) for LOWZ NGC and the best-fit theory model (solid lines). The monopole measurements (upper panel) show that a redder subsample has a higher bias.
				}
			\label{fig:LCcut2}
	    \end{figure}

	Fig~\ref{fig:LCcut2} presents the monopole and quadrupole measurements of the LOWZ NGC subsamples based on the color cut, with colors going redder from $C_5$ to $C_1$. As in figure~\ref{fig:lowz_cmass_color_w}, a redder subsample is shown to have a higher bias. However, the quadrupole measurement of $C_1$ clearly deviates from those of other subsamples, and fitting the RSD model to all color subsamples, we find that its $f \sigma_8$ constraint is significantly different from $f \sigma_8$ measurements of other subsamples. More detailed analysis can be found in appendix~\ref{appendix:FP_correlations}.
	
\begin{figure}
        		\centering
         		\includegraphics[width=\columnwidth]{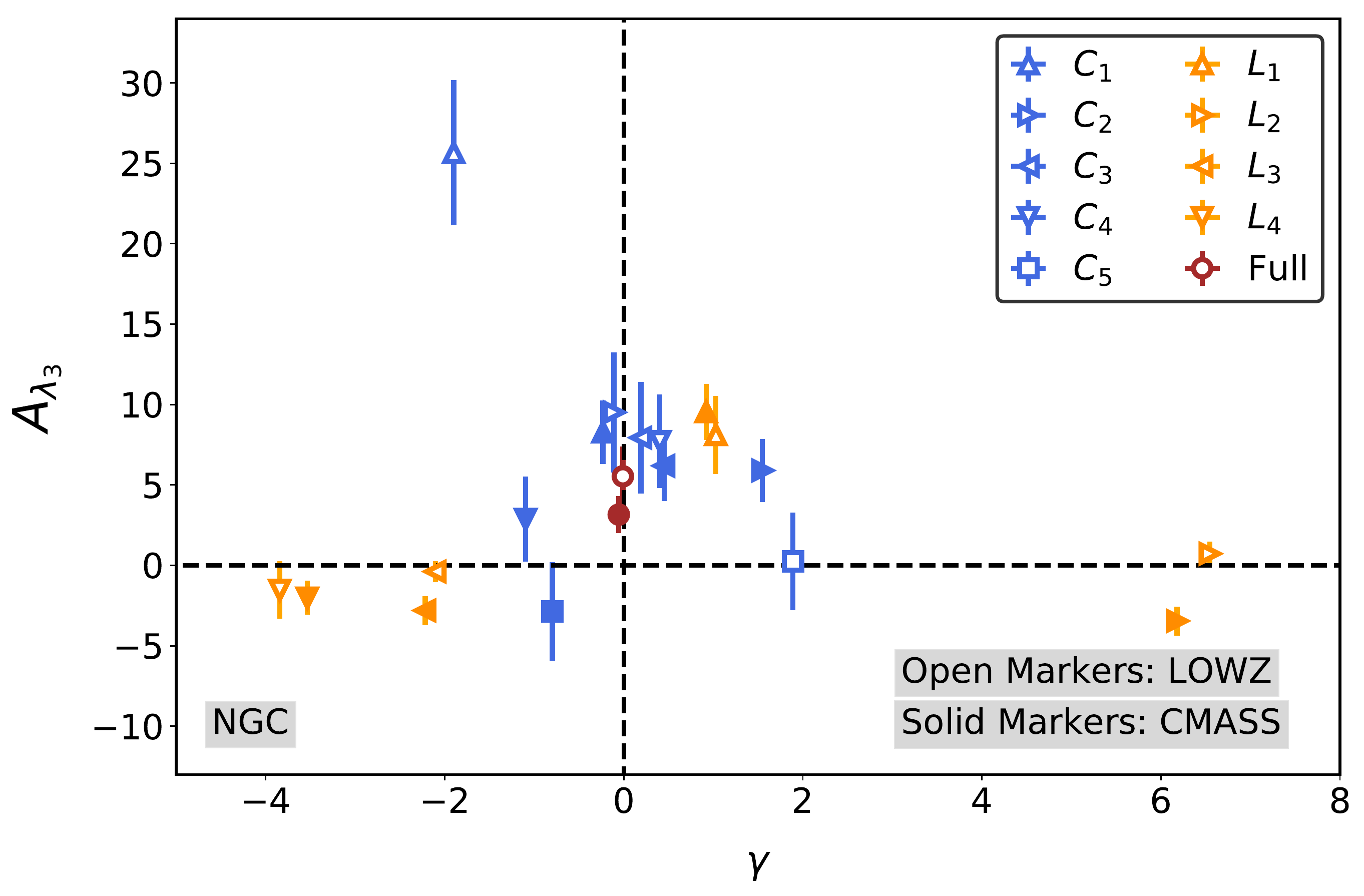}
				\caption{ Comparison of the FP residuals $A_{\lambda_3}$ and the selection dependence factor $\gamma$ for different subsamples of LOWZ NGC and CMASS 
				NGC. Different colors represent different splits, and different markers represent different subsamples. No correlation between $A_{\lambda_3}$ and $
				\gamma$ is evident in the figure.  
				}
	    	     \label{fig:Al_vs_gamma}
		     \end{figure}

	In figure~\ref{fig:LCcut3}, we show the correlations between the constraints on $f\sigma_8(z_{\mathrm{eff}})$, the product of $f$ and $\sigma_8$, each evaluated at the effective redshift of each sample, and the rescaled FP residual amplitude $A_{\lambda_3}$, computed at the fiducial $\sigma_8$. We divide each $f\sigma_8(z_{\mathrm{eff}})$ measurement with the predictions from the Planck 2015 data, to present $f\sigma_8$ measurements independent of the effective redshifts of our samples. To make direct comparisons with the model in section~\ref{sssec:IARSD}, we rescale $A_{\lambda_3}$ and then multiply it with a factor $\gamma$/$f_{\mathrm{fid}}$, where $\gamma$ is the response parameter as described in section~\ref{sssec:IARSD}. The measured $\gamma$ values (assuming $\lambda_{3}^{\mathrm{NS}}$, where `NS' denotes fitting NGC and SGC seprately.) are shown in figure~\ref{fig:Al_vs_gamma}.
	\footnote{Assuming the FP definition in \cite{Martens2018}, we get $\gamma$ = -0.21$\pm 0.03$, 0.11$\pm 0.04$, 0.18$\pm 0.02$, 0.22$\pm 0.04$ for LOWZ NGC, LOWZ SGC, CMASS NGC, and CMASS SGC respectively.}
	
	As shown in figure~\ref{fig:LCcut3}, we also find that there is only a weak evolution of the growth rate measurements with the FP residual amplitude. This agrees with the conclusion in section~\ref{subsubsection:FPcut} that no significant bias in RSD measurements due to IA is evident. We quantify this correlations by fitting a linear relation between the growth rate $f$ and the FP residual amplitude, and the following are the best-fit models: $(f\sigma_8/f\sigma_{8, \mathrm{fid}}) = (-0.05\pm0.02) - (-0.36\pm0.14) \cdot [-\gamma A_{\lambda_3}\zeta/f_{\mathrm{fid}}]$ for NGC and $(f\sigma_8/f\sigma_{8, \mathrm{fid}}) = (-0.07\pm0.03) + (-0.10\pm0.04) \cdot [-\gamma A_{\lambda_3}\zeta/f_{\mathrm{fid}}]$ for SGC, clearly in tension with the model in section~\ref{sssec:IARSD}, which predicted that the growth rate $f$ is larger for a larger FP amplitude; we find such correlations in the opposite direction for both NGC (left panel) and SGC (right panel) samples. Moreover, the slope of this fit is largely driven by the LOWZ $C_1$ outlier. Without the LOWZ $C_1$ sample, the slope would be closer to zero. Comparing the measurements to the null detection of IA effects on RSD measurements (shown by brown lines), we obtain $\chi^2$ values of 27.6 and 14.6 for 18 NGC and 18 SGC sub-samples, respectively. For the NGC samples, this corresponds to probability-to-exceed (PTE) of $\approx$ 7\%, which indicates that the measurements are consistent with the null result.

	   \begin{figure*}
        \includegraphics[scale=.315]{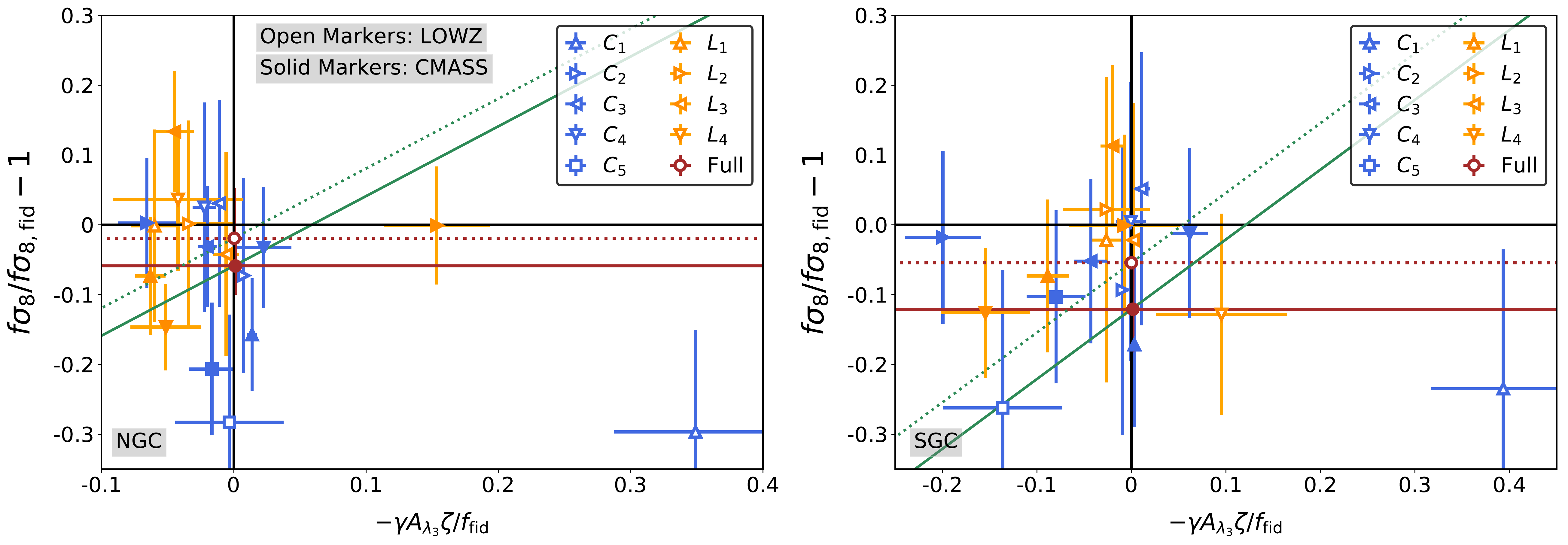}
			\caption{ Comparison of the RSD measurements of the growth of structure and the rescaled FP residual amplitude, for different subsamples of LOWZ (open markers) and CMASS (solid markers) galaxies in NGC (left panel) and SGC (right panel) regions. Each measurement of $f\sigma_8(z_{\mathrm{eff}})$ is divided by the predicted value assuming the fiducial Planck 2015 cosmology ($y$-axis), and the amplitude of galaxy size correlations measured using the FP residuals ($A_{\lambda_3}$) is rescaled and multiplied by $-\gamma\zeta$/$f_{\mathrm{fid}}(z_{\mathrm{eff}})$ of each sample ($x$-axis) so that the expected correlation coefficient between two variables is 1, according to $\Delta f = -\gamma A_{\lambda}\zeta$ (eq.~\ref{eq:IARSD_eq}). Green lines represent the expected relation between the FP residual and RSD measurements for the full LOWZ (dotted lines) and CMASS (solid lines) sample, and similarly brown lines show the growth of structure measurements for the full LOWZ and CMASS samples. We find that there is only a weak evolution of the growth of structure measurements with the FP residual amplitude, thereby suggesting that no significant bias in RSD measurements due to IA is evident. }
			\label{fig:LCcut3}
	    \end{figure*}

\section{Conclusions}
  
    In this work we have presented the estimations of the FP of the BOSS galaxies and the dependence of FP residuals on the galaxy redshift, environment and luminosity. We 
    show that the redshift evolution of the FP observed in earlier works \citep{Joachimi2015,Saulder2019} is primarily driven by the redshift evolution of the surface 
    brightness of the galaxies and correcting for this redshift implies that the FP is primarily a relation between the size, luminosity and velocity dispersion of the 
    galaxies. 
    The FP residuals are also strongly correlated with the luminosity of the galaxies and the luminosity evolution of the FP is primarily responsible for the
    apparent environment dependence of the FP residuals as was first detected by \cite{Joachimi2015}. 
    In appendix~\ref{appendix:FP_systematics} 
     we also show that the FP residuals are correlated with the 
     observational systematics, most notably the
    goodness of the galaxy profile fits, the PSF flux and in the case of CMASS sample the stellar density weights.
    
    In section~\ref{ssec:results_FP_correlations} we presented the measurements and analysis of the correlations between the FP residuals and the galaxy density field, 
    $w_{g\lambda}$. 
    FP residuals for BOSS Lowz and CMASS samples show similar correlations with the density field. We showed that these correlations are driven by the correlations between
    the galaxy properties and the galaxy density field. Galaxy luminosity, size and velocity dispersion are positively correlated with the density field while 
    the surface brightness is negatively correlated. The negative correlation of surface brightness is a non-trivial result and is potentially important for HOD modeling, 
    relating galaxies to the halos as well for the modeling of galaxy bias as function of redshift. 
    
    We also studied the dependence of the $w_{g\lambda}$ as function of galaxy environment, luminosity as well as color. Brighter galaxies show strong positive correlations
    between the FP residuals and the density field while the correlation amplitude is lower for fainter samples with the lowest luminosity sample showing negative 
    correlations. Similar trends are also observed for the color splits, with strong positive correlations for the red galaxies with lower correlations for bluer samples.
    Combining all the samples together, we show that there is strong correlation between the galaxy bias and the amplitude of $w_{g\lambda}$. This implies that the galaxies 
    in more over dense regions show stronger correlations between the FP/size residuals and the environment.
    
	We also compare the amplitude of $w_{g\lambda}$ measurements from model fits with amplitude of the intrinsic alignment (IA) measurements. Since our model assumes that
	the size correlations are sourced by the effects of IA in three dimensions projected onto the two dimensional plane of the sky, the amplitudes of IA and $w_{g\lambda}$ 
	measurements measurements are expected to be the same. Though there is considerable scatter, our results are in tension with this prediction. We further tested the 
	model by measuring the multipole moments of the correlation functions, finding again that the measured multipole moments are in tension with the model predictions. We 
	note that our modeling has a limitation as we do not include the density weighting effects. Thus while measurements are in tension with the incomplete model, it is 
	difficult to conclude that the IA does not have any impact on the size correlations as estimated using the FP.

    Furthermore, in section~\ref{ssec:results_FP_RSD} we presented the correlations between FP residuals and RSD constraints, in particular on the growth rate parameter. Splitting the BOSS LOWZ and CMASS galaxies into subsamples based on FP residuals, we fitted the RSD model to the measured multipoles of each subsample and showed that the differences in RSD constraints, across all subsamples, are statistically consistent with the null result. We hence conclude that there is no evidence of significant impact in RSD measurements due to IA, contrary to conclusions drawn by \cite{Martens2018}. Moreover, the RSD measurements of the BOSS samples split by luminosity and color further strengthens this argument; despite some scatter, we find only a weak evolution of the RSD constraints with the FP residual amplitude, and in comparison with the model from \cite{Martens2018}, we find this effect in the opposite direction. This suggest that there can be other overwhelming effects that impact the FP, and it can be difficult to simply disentangle the relation among FP, IA, and RSD measurements from other effects.

    Our work is a step towards improved understanding of FP to use it as a tool for studying galaxies as well as cosmology. The scatter in the FP is 
    similar to the scatter in the ellipticities of the galaxies, suggesting that it can be developed as a probe of gravitational lensing and beyond with similar potency 
    as the galaxy ellipticities (shear). There are notable impediments to such applications as the FP is dependent on the galaxy properties, environment, selection effects 
    and photometry errors. A deeper understanding of the impact of galaxy physics on the FP will require a similar study as ours using realistic cosmological simulations that we plan to pursue in near 
    future. A detailed understanding of observational effects will require image simulations as were performed for the context of galaxy 
    shape measurements \citep[e.g.][]{Mandelbaum2018}. A more detailed modeling for the FP correlations with density field also needs to be developed to capture the information from 
    small scales that we ignored in this work as well as to capture the impact of density weighting terms.
    The promise of galaxy sizes for cosmological applications and the upcoming large dataset from DESI, LSST, Euclid and WFIRST makes it an opportune 
    moment to study FP and  galaxy sizes in general in more detail.

\section*{Acknowledgments}
We thank Rachel Mandelbaum, Chris Hirata, Alexie Leauthaud and Song Huang for useful discussions related to this work.
We also thank Chris Hirata for providing access to data from \cite{Martens2018} and Rachel Mandelbaum for providing access to the galaxy shape measurements used for IA measurements.
This material is based upon work supported by the National Science Foundation under Grant Numbers 1814370 and NSF 1839217, and by NASA under Grant Number 80NSSC18K1274.

\bibliographystyle{mnras}
  \bibliography{sukhdeep_FP}

\appendix

\section{Comparison with Martens et al. (2018)}\label{appendix:Martens_comparison}
\subsection{Notation and parameters}
\cite{Martens2018} definition of size residuals, W, is related to our definition of $\lambda$ as 
\begin{align}
	W=-4\lambda
\end{align}
W is related to the tidal field as (following notation from \cite{Martens2018})
\begin{align}
	W=2B s_{33}&=2 \times1.74 b_\kappa s_{33},
\end{align}
$b_\kappa$ is written in terms of intrinsic alignments amplitude as 
\begin{equation}
W=-2\times1.74 A_I \zeta s_{33}
\end{equation}
From our model for $\lambda$, we can relate the model parameters as 
\begin{align}
	-4A_\lambda {\zeta}s_{33}&=-2\times1.74 A_I \zeta s_{33}\\
	A_\lambda&=\frac{1.74}{2}A_I
\end{align}

To relate the errors in the galaxy selection functions, we note that the error for a given galaxy should not depend on the definition of the FP residuals but do depend on the sign, following which we can write
\begin{align}
	\epsilon(\lambda)&\equiv-\epsilon(W)=>
	\gamma\lambda\equiv-(\eta\chi) W
\end{align}	
This allows us to relate $\gamma$ to $(\eta\chi)$ as
\begin{align}
\gamma=4(\eta\chi)\label{eq:gamma_eta_chi}
\end{align}

We can also derive the $\Delta(\eta\chi)$ parameter from \cite{Martens2018}, in terms of the derivative of the $S(\lambda)$ which can be written as  
\begin{align}
\frac{dS(\lambda)}{d\lambda}=&\frac{1}{(N_{\lambda+}+N_{\lambda-})^2}\left[2N_{\lambda-}\frac{dN_{\lambda+}}{d\lambda}-2N_{\lambda+}\frac{dN_{\lambda-}}{d\lambda}\right]\\
\approx&\frac{1}{2N_{\lambda}}\left[\frac{dN_{\lambda+}}{d\lambda}-\frac{dN_{\lambda-}}{d\lambda}\right]=2\Delta(\eta\chi)\label{eq:S_Deta_chi}
\end{align}

where we wrote $N_\lambda\approx N_{\lambda+}\approx N_{\lambda-}$. We note that eq.~\eqref{eq:gamma_eta_chi} and eq.~\eqref{eq:S_Deta_chi} are not necessarily 
consistent with each other and we only use them to compare the numerical quantities across the two studies.

\begin{figure}
        		\centering
         		\includegraphics[width=\columnwidth]{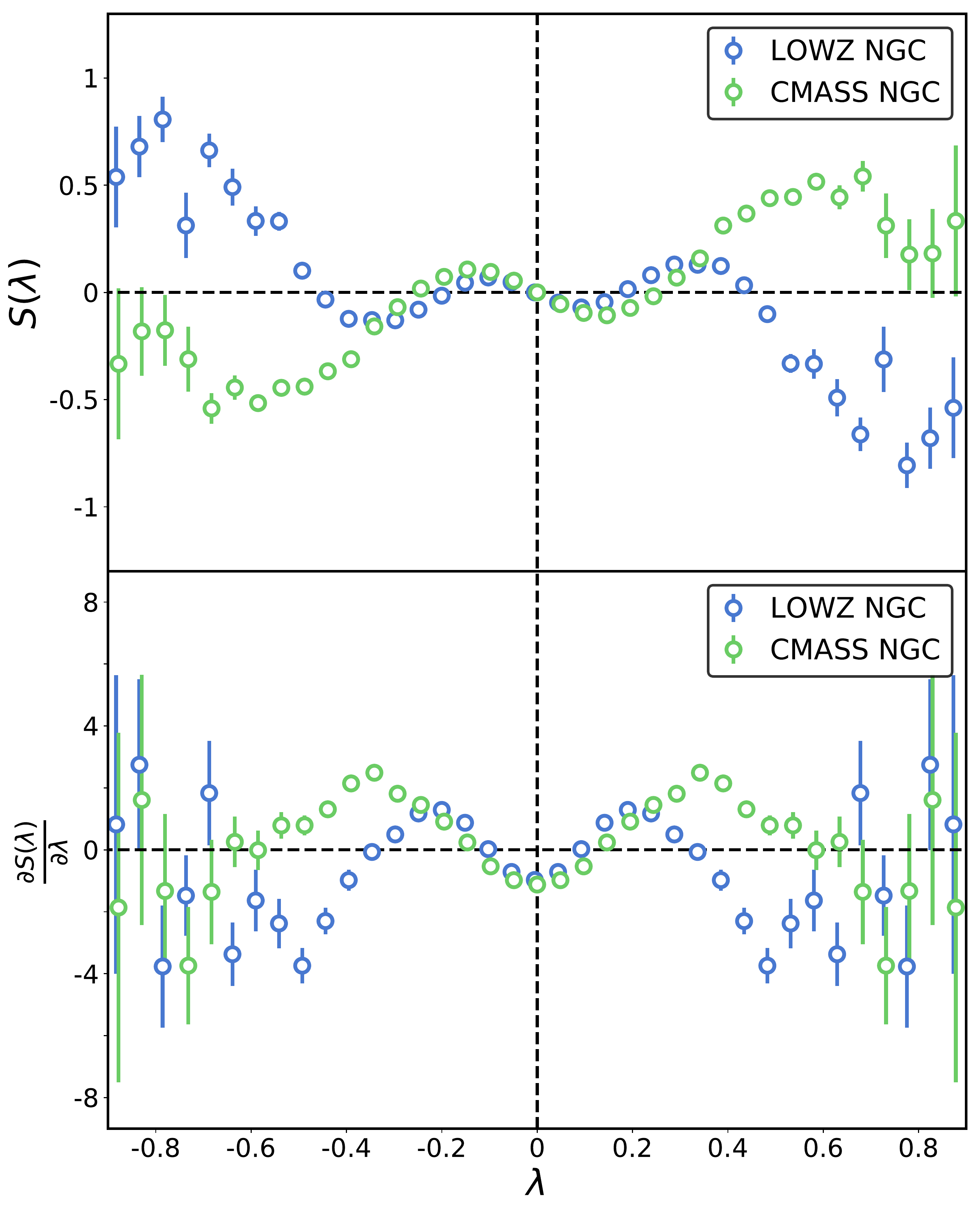}
				\caption{ Histogram showing the distribution of $S(\lambda)$ (upper pannel) and $\frac{\partial S(\lambda)}{\partial \lambda}$ (lower panel) values from LOWZ NGC and 
				CMASS NGC samples. We assume $\lambda=\lambda_3$.
				}
	    	     \label{fig:Slambda}
		     \end{figure}

In figure~\ref{fig:Slambda}, we show a histogram representing the distribution of $S(\lambda)$ and $\frac{\partial S(\lambda)}{\partial \lambda}$, produced by splitting galaxies into 41 $\lambda$ bins. Assuming the FP definition in \cite{Martens2018}, we obtain $\frac{dS(\lambda)}{d\lambda}\big|_{\lambda=0} = -1.69\pm0.09, -1.28\pm0.12, -1.22\pm0.06, -0.87\pm0.07$ for LOWZ NGC, LOWZ SGC, CMASS NGC, and CMASS SGC, respectively. Similarly with $\lambda = \lambda_3$, we measure $\frac{dS(\lambda)}{d\lambda}\big|_{\lambda=0} = -0.97\pm0.09, -0.68\pm0.13, -1.12\pm0.05, -0.87\pm0.08$ for LOWZ NGC, LOWZ SGC, CMASS NGC, and CMASS SGC, respectively.

%

	\subsection{Comparison of FP definitions}
\begin{figure}
        		\centering
         		\includegraphics[width=\columnwidth]{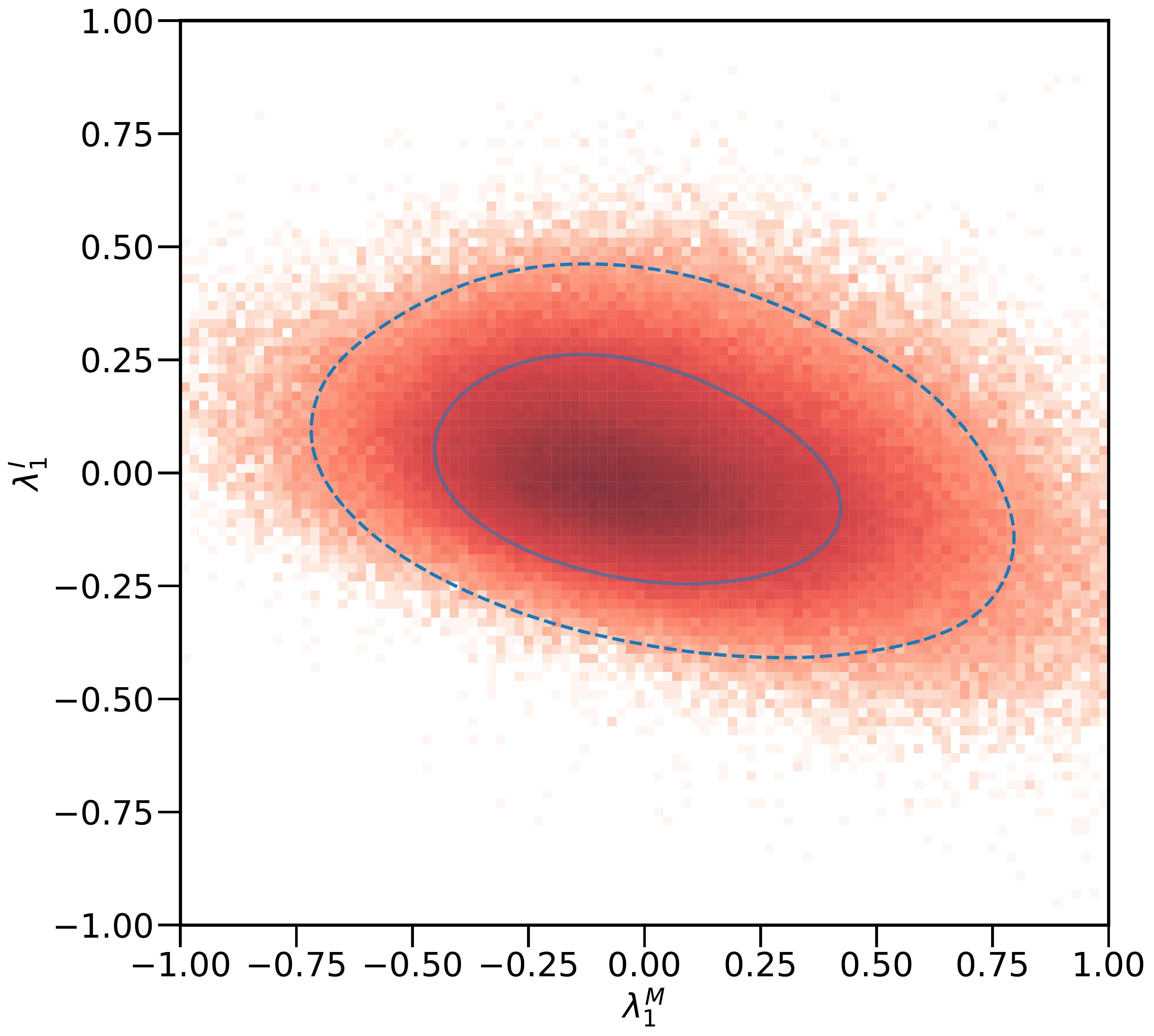}
				\caption{ Comparison of the FP residuals obtained using the surface brightness in the FP, $\lambda_1^I$ and the magnitudes in FP, $\lambda_1^M$. The solid and dashed blue lines show
				the 68\% and 95\% contours while the red color shows the underlying two dimensional histogram (color scaling is in log). The two FP definitions show only weak correlations and the 
				scatter in $\lambda_1^M$ is $\sim 2$ times larger.
				}
	    	     \label{fig:lambda_M_I_comp}
		     \end{figure}

		To rule out the differences due to the FP definitions, we also define the FP in terms of magnitude, $M$, instead of surface brightness as
		\begin{align}
            \log R_0&=b (M+10\log z)+c+\sum_{i=1}^{N_z} d_i z^i\\
            \lambda^M_{N_z}&=\ln \frac{R_0}{R^I_{\text{FP},N_z}}
        \end{align}

        where $\lambda^M_{1}$ is closest to the definition of FP used by \cite{Martens2018}. We note that the $\lambda^M_{N_z}$ and $\lambda^I_{N_z}$ obtained by the minimizing the scatter
        around the plane (minimize $\lambda_{rms}$) are not the same as the two planes get different weights and hence different contributions to scatter from the magnitude and size, $\log R$.
        The scatter plot between the two planes is shown in fig.~\ref{fig:lambda_M_I_comp}. 
        The $\lambda^M_{1}$ we obtain is consistent with the $-W_{33}/4$ from \cite{Martens2018}, with the very small scatter which 
        is sourced by different treatment of k-corrections and smaller differences in the cosmology across the two studies.
        We observe that the scatter in $\lambda^M_{1}$ is $\sim 2$ times larger than the scatter in $\lambda^I_{1}$. 
        We caution against interpreting $\lambda^I_{1}$ as better using scatter
        as metric since scatter of the $\lambda$ is simply determined by the weighted combination of the magnitude and size which are not perfectly correlated. 
        The correlation coefficient for $\log R$ and magnitude is $\sim0.5$ while for the $\log R$ and $\log I$ is $\sim -0.82$.

	\section{Analysis with \lowercase{$i$} band}\label{appendix:i_band}
			\begin{figure}
        		\centering
         		\includegraphics[width=\columnwidth]{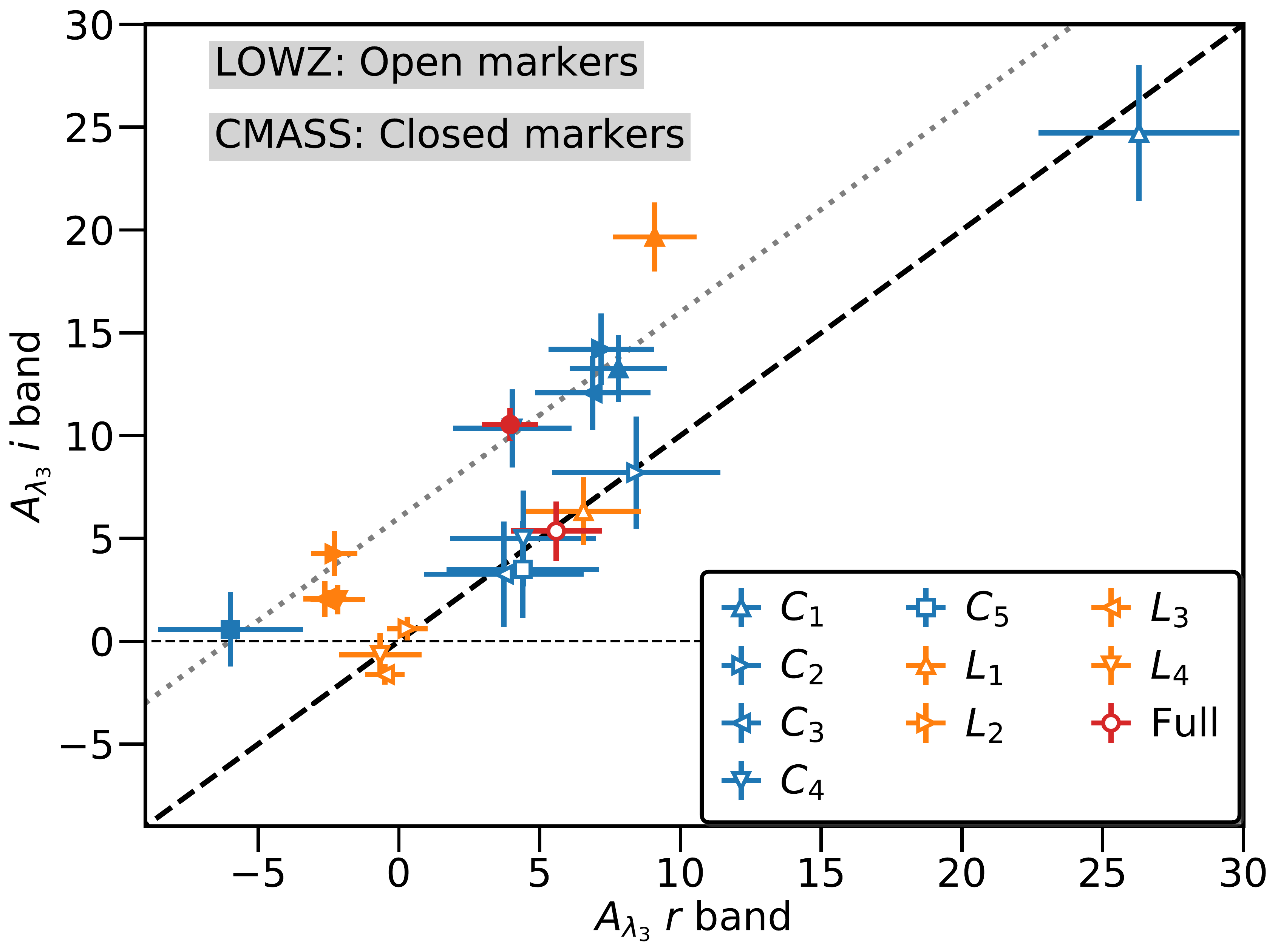}
				\caption{Comparison of $A_\lambda$ constraints obtained using $r$ band (x-axis) and $i$ band photometry. For the LOWZ sample, both $r$ band and 
					$i$ band give consistent results, while for the CMASS sample, there is an additive shift with $A_{\lambda,i}\approx A_{\lambda,r}+6$ 
					(shown by dotted gray line). The variation in CMASS sample is likely driven by the photometry differences due to redshift effects as
					4000 angstrom line break moves into $r$ band at $z\sim0.4$.
				}
	    	     \label{fig:A_lambda_r_i_comp}
		     \end{figure}
		     
		In fig.~\ref{fig:A_lambda_r_i_comp} we show the comparison of $A_{\lambda}$ values obtained from the fits in the $r$ band and the $i$ band. For the LOWZ sample, the results
		between the two bands are consistent. We also tested the comparison between the \dev\ and the model magnitudes in the $i$ band (they are same in the $r$ band except for a very small fraction 
		of galaxies that have exponential profiles). 
		The two magnitudes measure also give consistent results in the $i$ band for the LOWZ sample. For the CMASS sample on the other hand, we observe 
		differences in the $A_\lambda$ values between $r$ band and the $i$ band. These differences are consistent with an additive shift in $A_\lambda$. Such differences can be caused by the 
		variations in the photometry across two bands as the 4000 angstrom line break moves into $r$ band at $z\sim0.4$ and hence the measurements is $r$ band will be noisier and relatively biased
		with respect to $i$ band. We also observe similar differences when using 
		\dev\ and model magnitudes. For CMASS, a larger fraction of galaxies have exponential profiles and hence FP from \dev\ and model magnitudes can show larger differences leading to differences
		in $A_\lambda$.
		
		\section{Luminosity and Color Samples}\label{appendix:lcdep}
			\begin{figure*}
				\begin{subfigure}[t]{\columnwidth}
         			\includegraphics[width=\columnwidth]{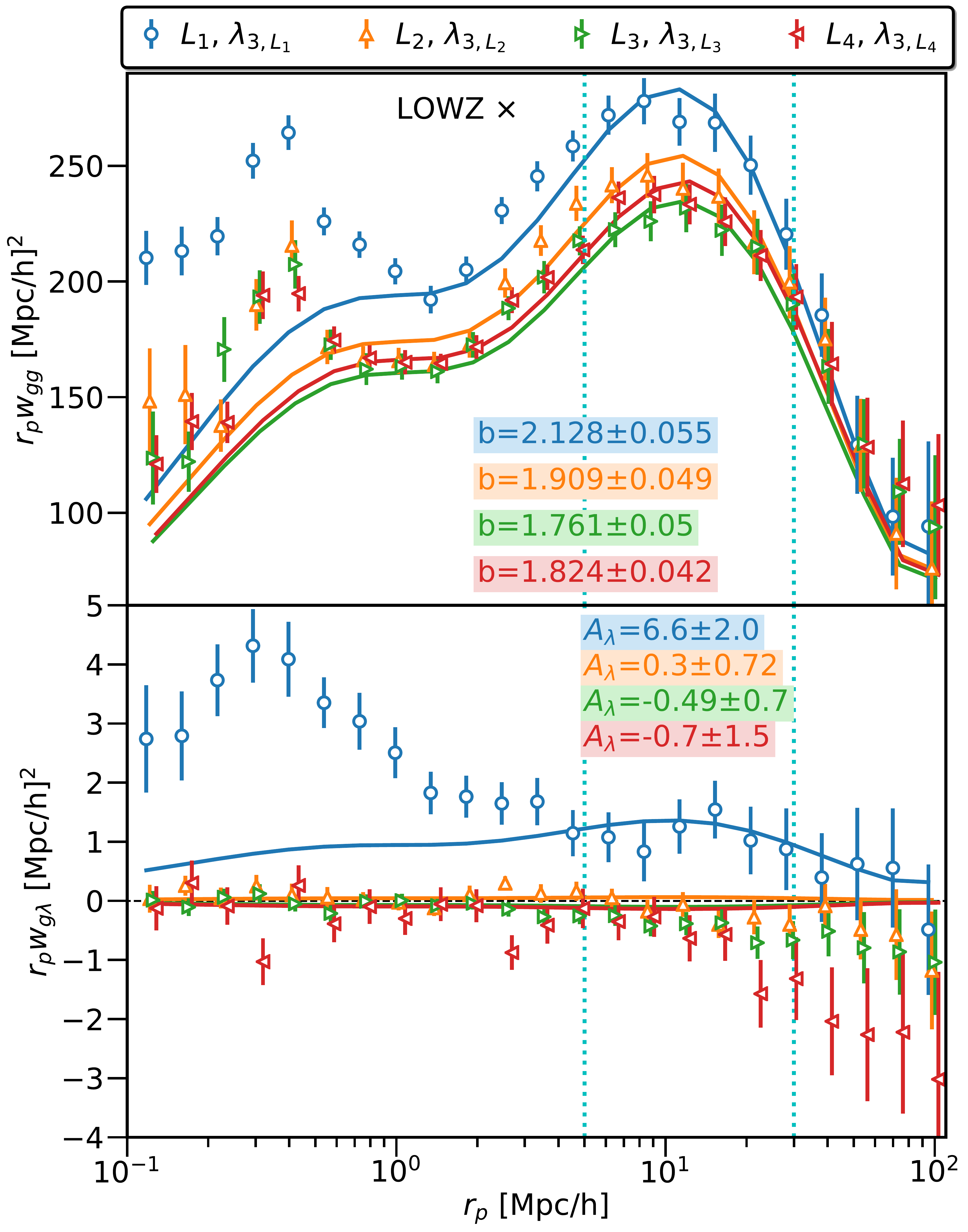}
		    	     \label{fig:}
			     	\caption{}
		    	 \end{subfigure}
			     \begin{subfigure}[t]{\columnwidth}
         			\includegraphics[width=\columnwidth]{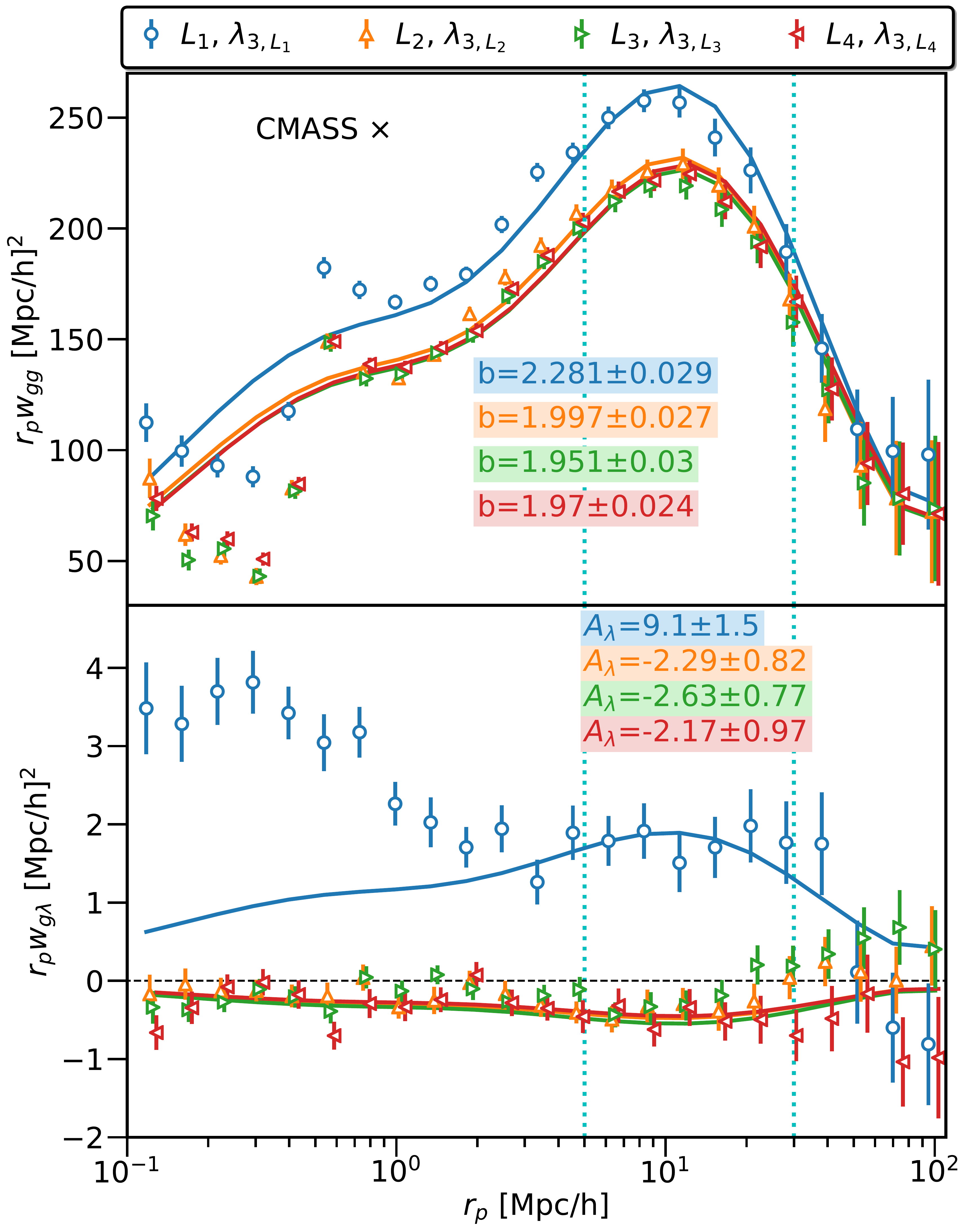}
	    	    	 \label{fig:}
				     \caption{}
				\end{subfigure}
				\caption{Galaxy clustering and galaxy-$\lambda$ cross correlation functions for different luminosity sub-samples for (a) Lowz  and 
				(b) CMASS samples. For both galaxy clustering and galaxy-$\lambda$ correlations, we use the full LOWZ or CMASS sample as the density 
				tracers.
				Here, $\lambda$ is obtained by fitting FP separately to each subsample. 
				Luminosity samples are arranged from brightest to faintest with $L_1$ being brightest and $L_4$ being faintest. $L_1$--
				$L_3$ each contain 20\% of the sample and $L_4$ contains 40\% of the sample and these samples are defined in narrow redshift bins such 
				that the overall redshift selection, $p(z)$ is same for all samples. FP residuals for $L_1$ sample show significant correlations with the
				galaxy density while $L_2$ and $L_3$ samples are consistent with no correlations. The faintest sample $L_4$ shows negative correlations.
				}
				\label{fig:lowz_cmass_lum_w}
		   \end{figure*}
		   In figure~\ref{fig:lowz_cmass_lum_w}, we show the galaxy-$\lambda$ cross correlations using subsamples based on luminosity cuts as described in 
		   section~\ref{sec:data}, with $L_1$ being brightest and $L_4$ being faintest. Note that we fit FP separately to each of these samples with the
		   residuals labeled as $\lambda_{3,L_i}$. 
		   
		   From galaxy clustering, we obtain largest galaxy bias for the brightest sample, $L_1$ with bias decreasing with luminosity for $L_2$ and $L_3$
		   though $L_4$ sample has higher bias. The decreasing trend of bias with Luminosity is consistent with the expectations that lower luminosity 
		   galaxies tend to be in lower mass halos which have lower bias. $L_4$ defies this trend, likely due to relatively higher fraction of satellite galaxies
		   within this sample, which tend to be in higher mass halos and hence have higher bias.
		   
		   The brightest sample show the largest correlations with the density, implying that within the brightest sample, 
		   larger galaxies reside in over dense regions. For the fainter samples, $L_2$ and $L_3$ samples give null signal, though we note that 
		   for both of these samples $rms$ of $\lambda_{3,L_i}$ is much smaller (by a factor of $\sim2$) than the $rms$ obtained for same galaxies when 
		   using the FP fit to
		   the full LOWZ or CMASS samples, unlike $L_1$ and $L_4$ samples for which $rms$ remains similar. Lower $rms$ of $\lambda$ also results in
		    lower errors in $A_\lambda$ for $L_2$ and $L_3$ samples. Furthermore, the amplitude of $A_\lambda$ for these samples also show low amplitude. We
		    confirmed that this is driven by the low amplitudes of the correlations of both size and surface brightness with the density field (plots not shown).
		   
		
			\begin{figure*}
				\begin{subfigure}[t]{\columnwidth}
         			\includegraphics[width=\columnwidth]{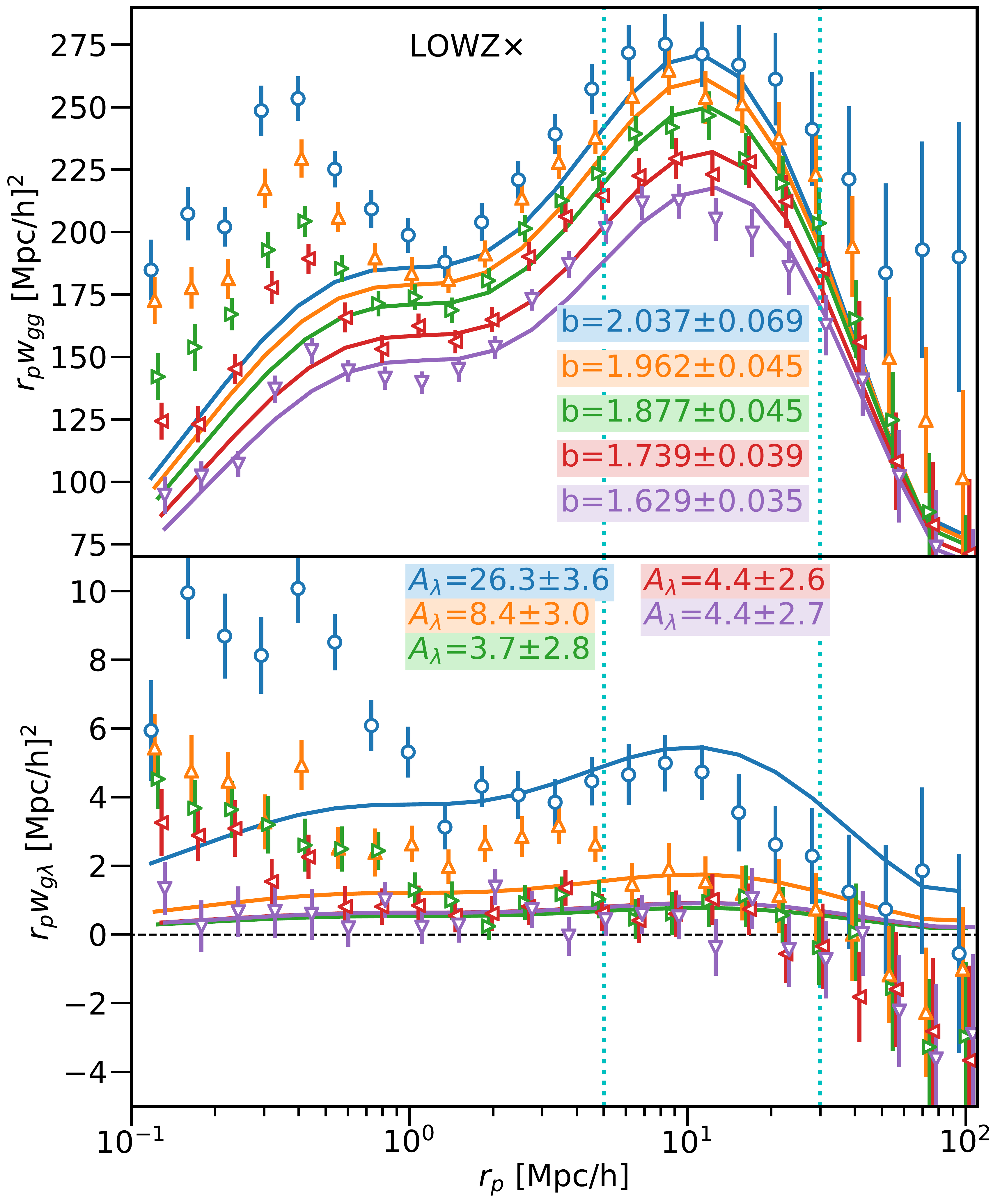}
		    	     \label{fig:}
			     	\caption{}
		    	 \end{subfigure}
			     \begin{subfigure}[t]{\columnwidth}
         			\includegraphics[width=\columnwidth]{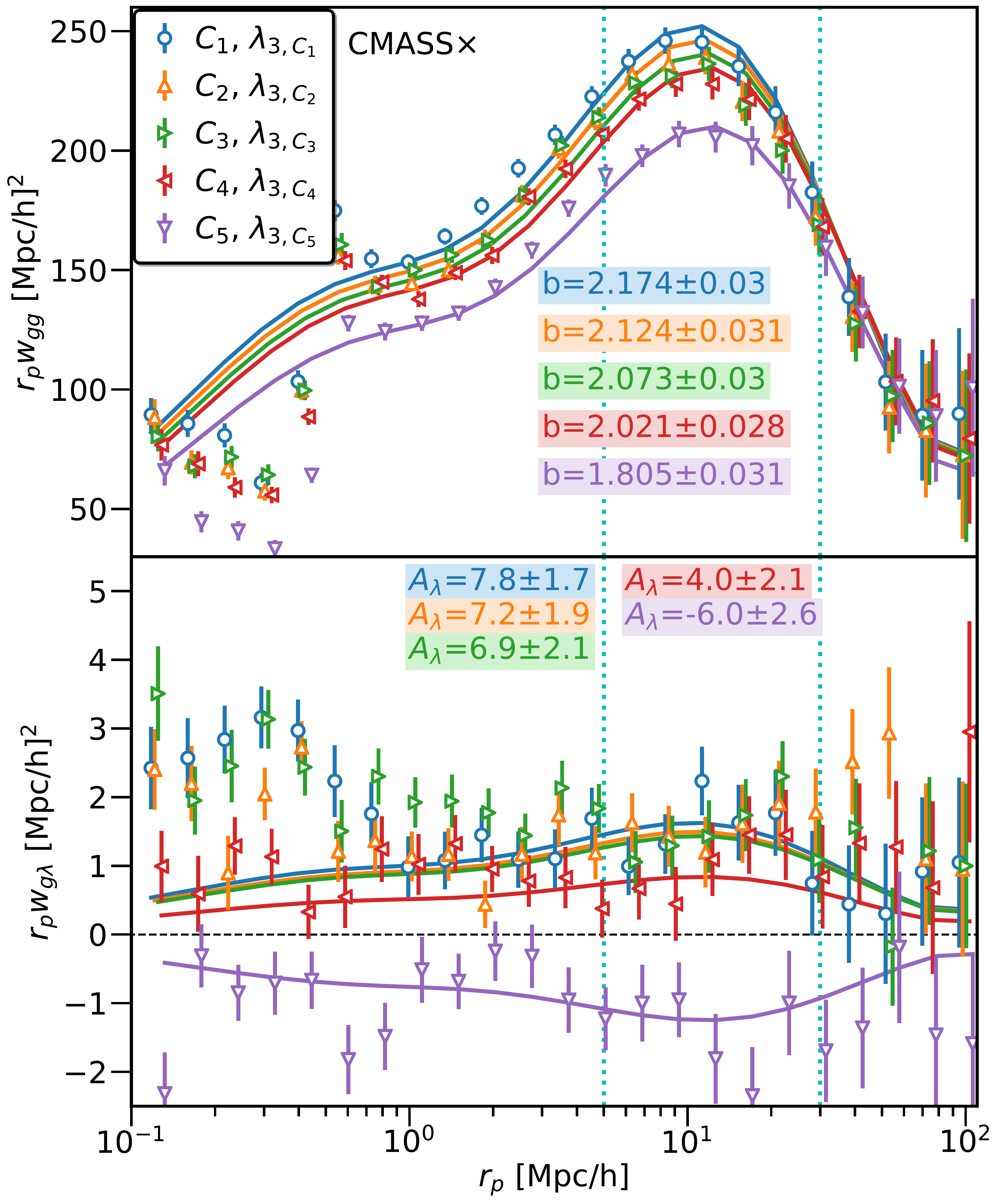}
	    	    	 \label{fig:}
				     \caption{}
				\end{subfigure}
				\caption{ Similar to figure~\ref{fig:lowz_cmass_lum_w}, the LOWZ and CMASS samples now split by color into five subsamples, with each
					sample containing 20\% of the sample. We observe variations in the galaxy-$\lambda$ correlations, with redder samples showing stronger 
					correlations.
					Note that the labelling of the sub-samples is same across the two panels.
				}
				\label{fig:lowz_cmass_color_w}
		   \end{figure*}
		   
		   In figure~\ref{fig:lowz_cmass_color_w}, we show the galaxy-$\lambda$ cross correlations using subsamples based on color cuts, 
		   with colors going from red to blue from $C_1$ to $C_5$ samples. We find that the redder samples show stronger 
		   correlations with the density though the trend is not very strong from $C_3-C_5$ samples and in the CMASS galaxies. 
		   The LOWZ $C_1$ sample is an outlier, as it has a very large relative to other subsamples of the LOWZ galaxies. 
		   We analyze this sample further in following section, appendix~\ref{appendix:lowz_C1}, and show that amplitude of $C_1$ sample is driven by
		   strong evolution of $A_\lambda$ in the redder part of the LOWZ sample. We also split $C_1$ sample by redshift but do not observe 
		   any significant redshift evolution.
		   
		   \subsection{Lowz $C_1$ sample}\label{appendix:lowz_C1}
		In this section we study the FP cross correlations of the LOWZ-$C_1$ subsample, which has high $A_\lambda$ when compared to the other samples 
		   and the expectations from bias evolution. 
		   
				\begin{figure}
         			\includegraphics[width=\columnwidth]{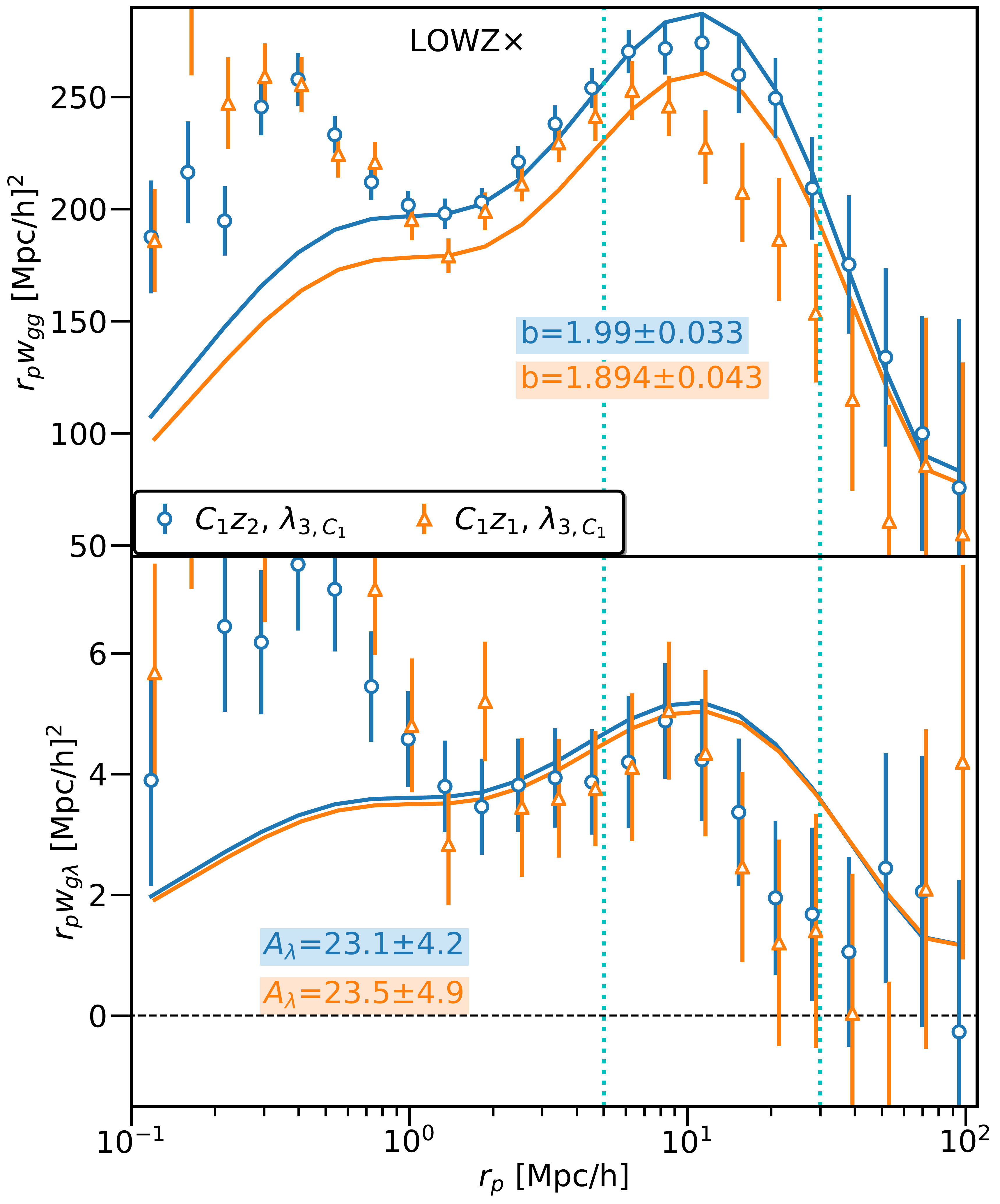}
			     	\caption{Splitting the Lowz $C_1$ sample into two subsamples based on redshift, with $z_1\in[0.16,0.3]$ and $z_2\in[0.3,.43]$. 
				Both samples give 
				consistent results, suggesting that there is no significant evolution with redshift.}
				\label{fig:C1_split_z}
		    	 \end{figure}
		   In figure~\ref{fig:C1_split_z}, we show the results from splitting the $C_1$ sample into two subsamples 
		   based on the redshift cut, with $z_1\in[0.15,0.3]$ and $z_2\in[0.3,0.43]$. We detect no redshift evolution of the results, suggesting that the 
		   $C_1$ constraints are robust against any redshift evolution and the high $A_\lambda$ is not dominated by any small redshift slice.
		   
			     \begin{figure}
         			\includegraphics[width=\columnwidth]{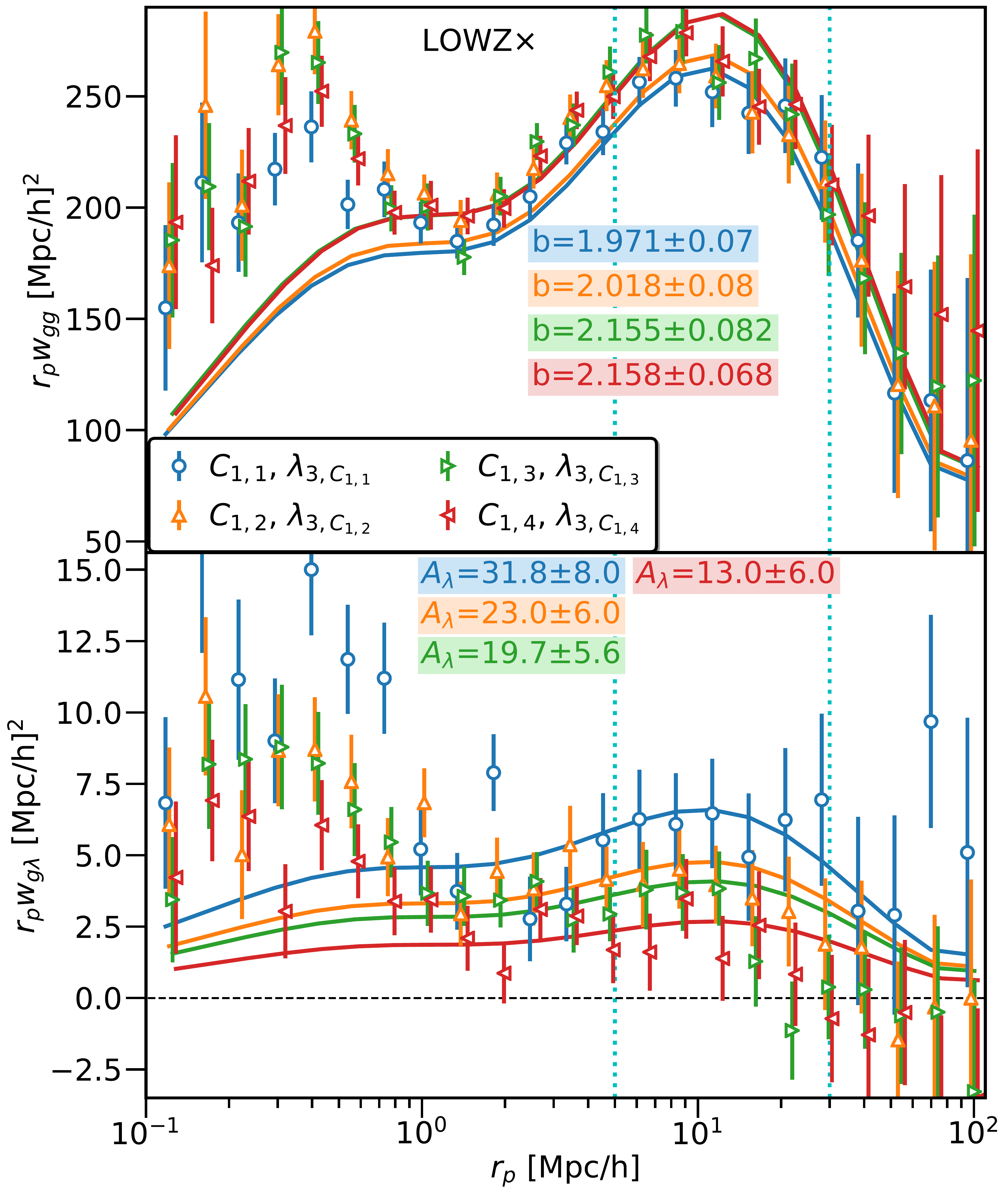}
				     \caption{Splitting the Lowz $C_1$ sample into four subsamples based on color, similar to the original color split. Similar to the original 
					color split, 
					we find a strong evolution with color even within the $C_1$ sample, with redder samples showing larger FP amplitude, $A_\lambda$.}
					\label{fig:C1_split_s4}
				\end{figure}
						   
		   Next we further split the $C_1$ sample into four smaller subsamples based on the color, using the same methodology as the original color splits. We 
		   denote these samples as $C_{1,i}$, where $i$ denotes the index of the sub-subsample. In figure~\ref{fig:C1_split_s4}, we observe a strong trend in 
		   $A_{\lambda}$ for the $C_1$ subsamples, with redder samples having larger amplitude and all of subsamples have larger amplitude than the $C_2$ sample.
		   These result hint that on the redder end of the LOWZ sample, there is a strong and smooth evolution in $A_\lambda$ with color and high amplitude 
		   observed for $C_1$ sample is a manifestation of this trend.

\section{Systematics tests}\label{appendix:FP_systematics}
			\begin{figure*}
				\begin{subfigure}[t]{.66\columnwidth}
         			\includegraphics[width=\columnwidth]{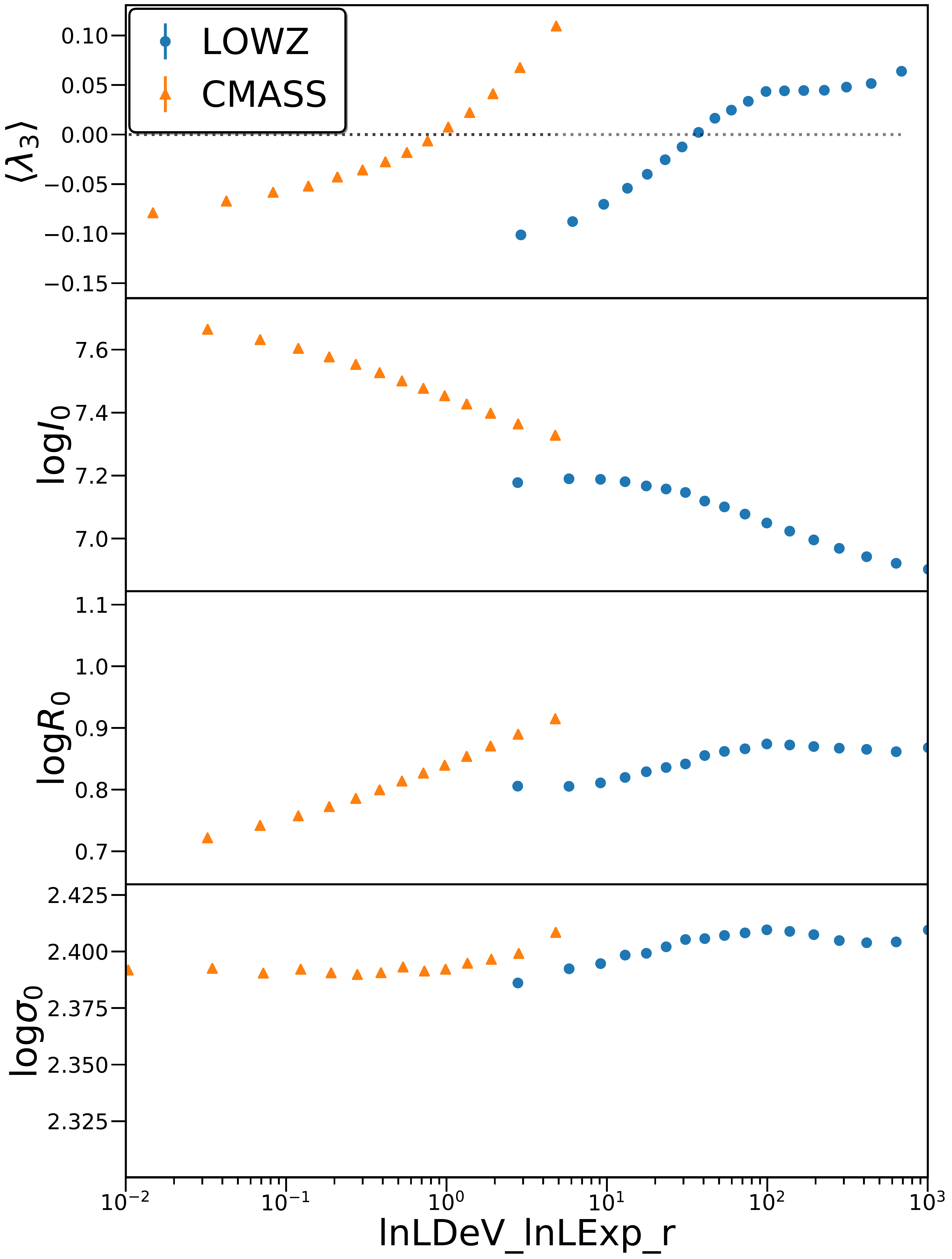}
	    	    	 \label{fig:lambda_mean_lnL}
				     \caption{}
				\end{subfigure}
			     \begin{subfigure}[t]{.636\columnwidth}
         			\includegraphics[width=\columnwidth]{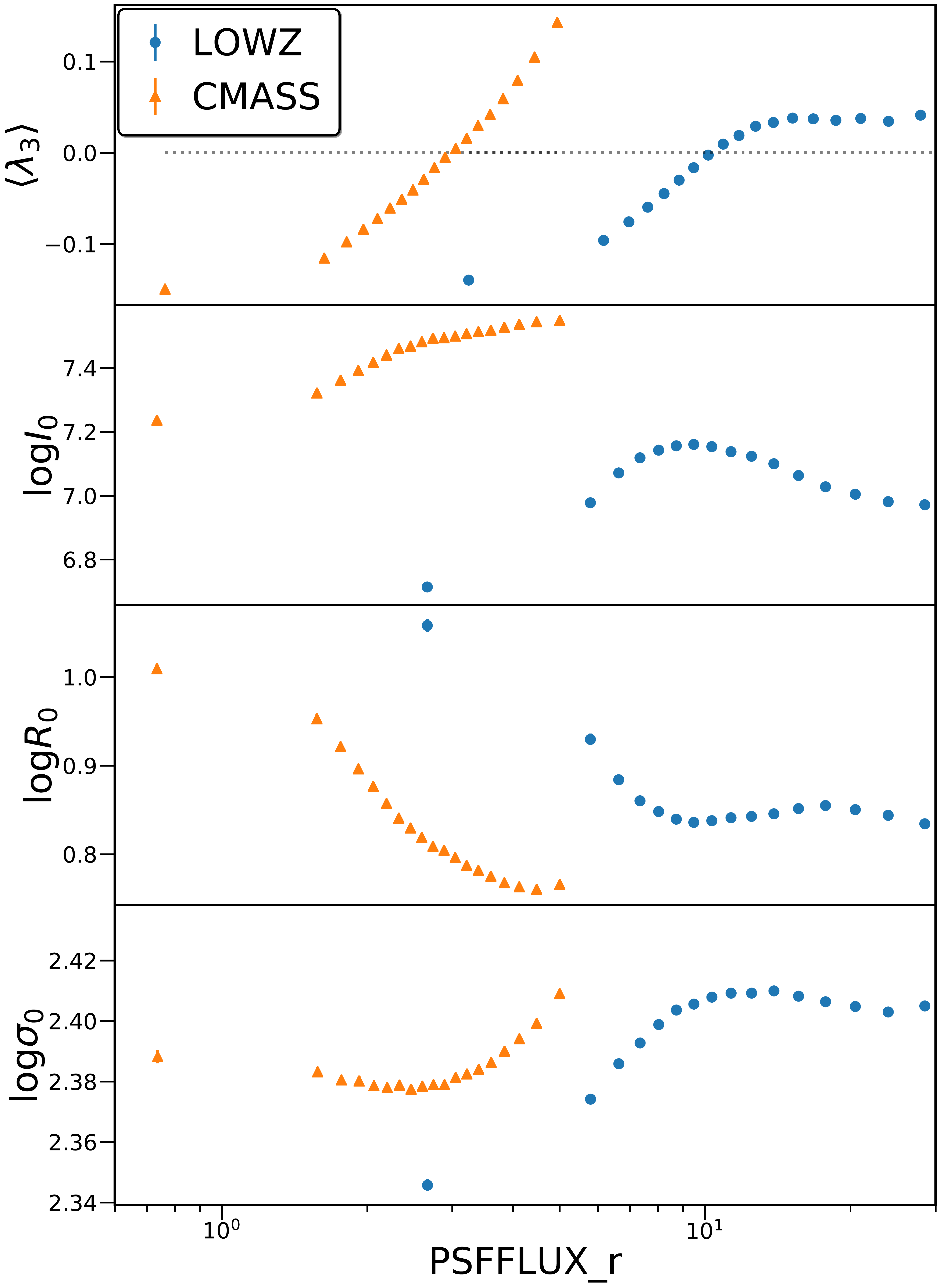}
	    	    	 \label{fig:lambda_mean_PSFFLUX}
				     \caption{}
				\end{subfigure}
				\begin{subfigure}[t]{.66\columnwidth}
         			\includegraphics[width=\columnwidth]{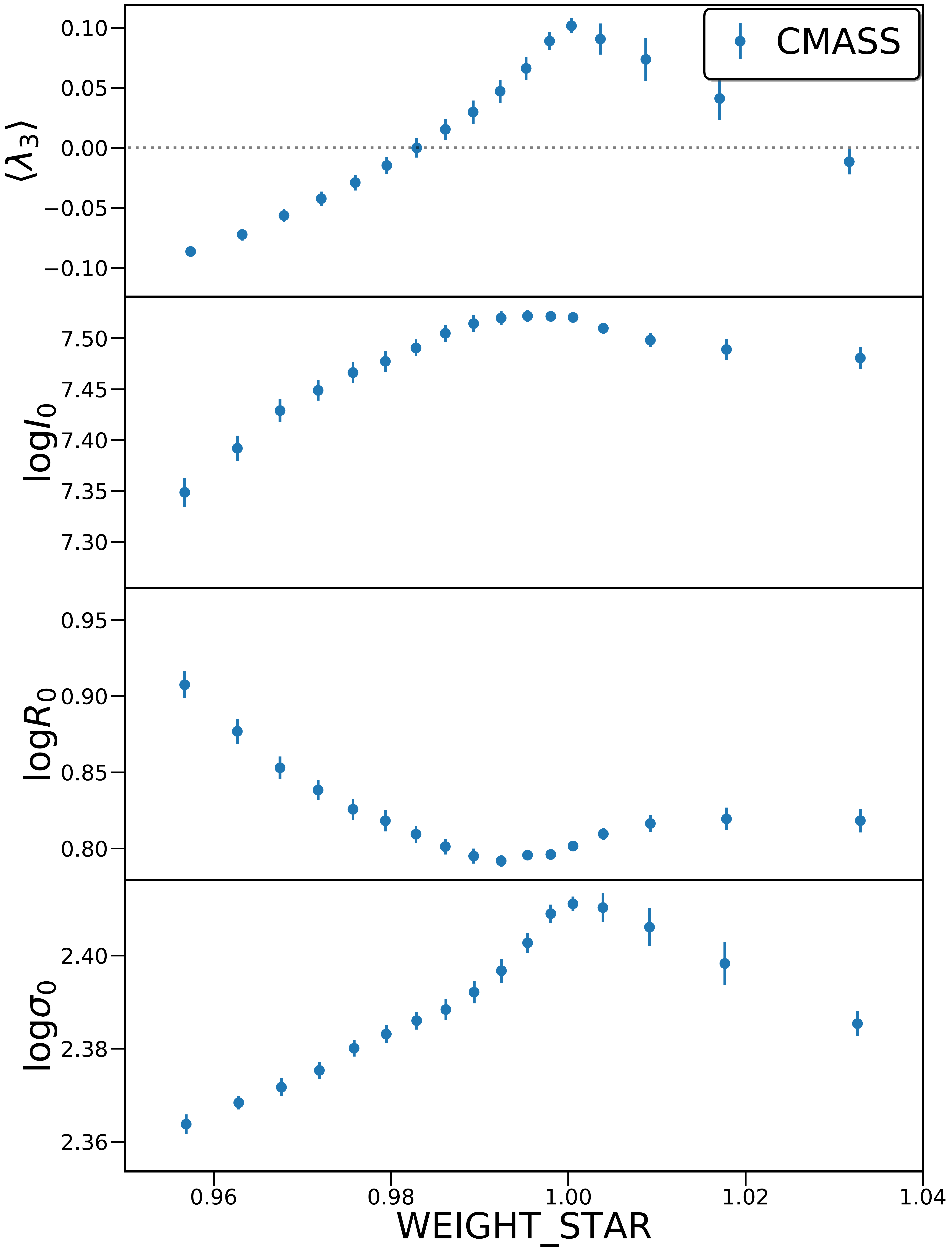}
		    	     \label{fig:lambda_mean_WEIGHT_STAR}
			     	\caption{}
		    	 \end{subfigure}
				\caption{Mean FP residuals (top panel) and galaxy properties (lower 3 panels) as function of various systematics, 
				a) the difference in log likelihood of \dev\ (DeV) and exponential (Exp)
				 profile fits to the galaxies, b) PSF flux in $r$ band, c) the star weights assigned to CMASS galaxies (LOWZ galaxies are not assigned 
				 these weights). There is considerable dependence of the FP residuals on these systematics, though this is primarily driven by the 
				 dependence of the estimated luminosity and size on these systematics.
				}
				\label{fig:lambda_mean_sys}
		    \end{figure*}
			In figure~\ref{fig:lambda_mean_sys},  we show the dependence of the FP residuals on some measures of the photometric quantities or systematics,
			namely the log likelihood difference between the \dev\ and exponential profile fits, PSF flux as well as stellar weights for CMASS galaxies, 
			which depend on the stellar density on the sky. While we tested for the impact of other observational systematics, we are only showing results
			for the ones where we detected significant variations in $\mean{\lambda_3}$, i.e. $\mean{\lambda_3}$ is same order as the $rms$ of $\lambda_3$.
			Correlations in figure~\ref{fig:lambda_mean_sys} can be understood in terms of the impact of these properties on the galaxy size and luminosity 
			estimates as shown in the same figure. 

\section{Systematics Tests-II: Correlation functions}\label{appendix:FP_correlations}
    \subsection{Randoms re-weighting}\label{appendix:randoms_reweight}
    	\begin{figure}
      		\centering
             \includegraphics[width=\columnwidth]{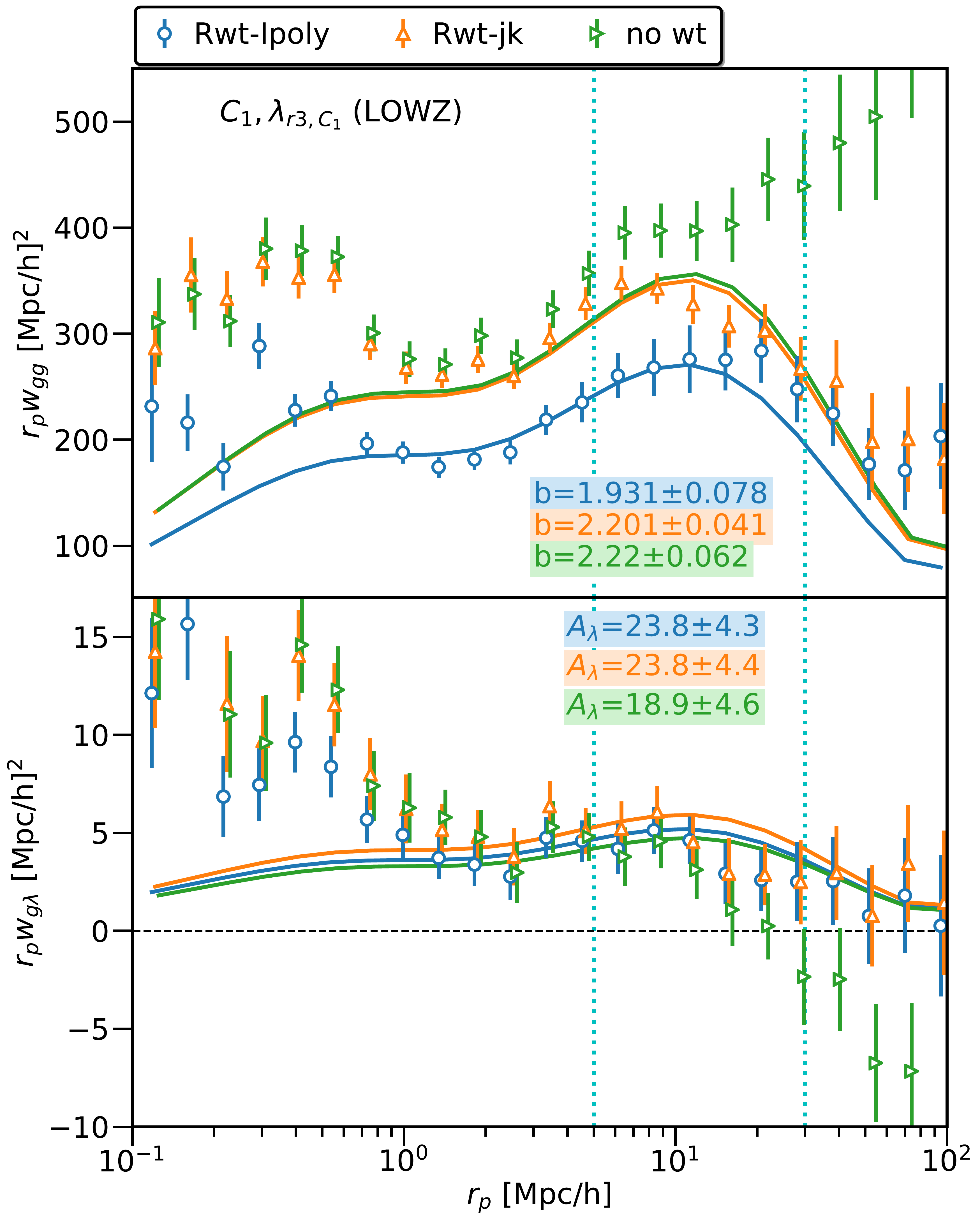}
    		\caption{Auto correlations of the LOWZ-$C_1$ sample and the impact of the re weighting the randoms. Splitting the LOWZ sample by
				color affects the selection function (in the plane of sky). This results in bias in the measurements at large scales, as shown by the 
				green (no wt) points. We attempt to correct this by re weighting the randoms by the fraction of galaxies that are in $C_1$ sample compared 
				to the full sample ($Rwt=N_{\text{gal},C_1}/N_{\text{gal},\text{LOWZ}}$). For blue points (Rwt-IPoly), we compute weights in the each 
				IPoly region while for orange point (Rwt-jk), we compute weights in each jackknife region ($\sim80$ deg$^2$). For the case of (Rwt-IPoly), 
				clustering is biased low which is because IPoly regions are too small and the re-weighted randoms are also clustered similar to 
				galaxies. Re weighting on size of jackknife regions on the other hand makes clustering nearly unbiased. Note that even though signal is 
				biased, the galaxy-bias ($b_g$) is fairly similar across all three cases. This is likely because the systematics from selection function 
				also affect the covariance, down weighting the bins that are biased and also increasing the error on $b_g$.
    				}
    	   	\label{fig:C1_rand_rewt}
    	\end{figure}
		Since the estimates of galaxy properties such as color and luminosity are affected by the observational systematics, which can vary over the sky, 
		splitting the samples by these properties can then lead to subsamples with significantly different selection functions. While our splitting method ensures
		that the selection function with redshift is not affected, we do not impose any such control in the sky coordinates (RA,DEC). To lowest order, the 
		incorrect selection function (randoms) changes the observed galaxy over-density field as
		\begin{equation}
			\widehat\delta_g=\delta_g+\delta_\text{sys}
		\end{equation}
		which in turn biases the correlation function and the covariance as (here we only show the equation in symbolic notation. See for example \cite{Singh2017cov} for a derivation of a similar effect)
		\begin{equation}
			\mean{\widehat\delta_g\widehat\delta_g}(\vec r)=\mean{\delta_g\delta_g}(\vec r)+\mean{\delta_\text{sys}\delta_\text{sys}}(\vec r)
		\end{equation}
		\begin{align}
			Cov{(\widehat\delta_g\widehat\delta_g,\widehat\delta_g\widehat\delta_g)}(\vec r_1,\vec r_2)=&
			Cov{(\delta_g\delta_g,\delta_g\delta_g)}(\vec r_1,\vec r_2)\nonumber\\
			+&Cov{({\delta_\text{sys}\delta_\text{sys},\delta_\text{sys}\delta_\text{sys}})}(\vec r_1,\vec r_2)
		\end{align}
		To reduce the impact of correlations from the systematics, for the subsamples we re-weight the randoms for subsamples by the ratio of 
		number of galaxies within the subsample to the number of galaxies in the parent sample, i.e.
		\begin{equation}
			Rwt(\vec x)=\frac{n_\text{subsample}}{n_\text{full}}(\vec{x},s)
		\end{equation}
		where $s$ denotes the pixelization scale of density field. For $s$, we try pixelization imposed by the Ipoly (pixelization scheme provided in BOSS 
		catalogs) or the $\sim80\deg^2$ regions we used for the jackknife covariances. 
		
		In figure~\ref{fig:C1_rand_rewt} we show an example demonstrating the effects of selection functions and the re-weighting of the randoms to correct for it, 
		for the Lowz $C_1$ sample. Without any corrections, the correlation function measurements show large bias on the large scales. However, measurements 
		on those scales are also very correlated (systematics impact on the covariance), and the best fit galaxy bias we obtain from this measurement is nearly unbiased 
		(agrees with 
		bias from cross correlations with full sample). To correct the bias on large scales, we then perform the measurements with the re-weighted randoms. 
		Re-weighting using Ipoly pixelization biases the measurement on small scales, likely because the pixelization from is too fine and randoms start 
		following the true underlying distribution of galaxies (if randoms perfectly followed galaxies, correlation function will be zero). Pixelization using 
		jackknife regions on the other hand is good enough to correct for the systematics while keeping the measurement un-biased. Measurement using this 
		pixelization thus leads to unbiased and more optimal measurement, with erorrs on the galaxy bias estimates lower by $\sim30\%$.
		
		Thus for subsamples, we will re-weight the  randoms using the weights from jackknife pixelization. Also, to avoid biases we will use the cross correlations
		between the sub-sample and the parent sample whereever possible as the $\delta_{sys}$ for full sample is very small. Using re-weighted randoms is 
		advantageous for cross correlations as well to obtain the optimal measurements.

	\subsection{Null test}\label{subsection:nulltest1}
		\begin{figure}
      		\centering
             \includegraphics[width=\columnwidth]{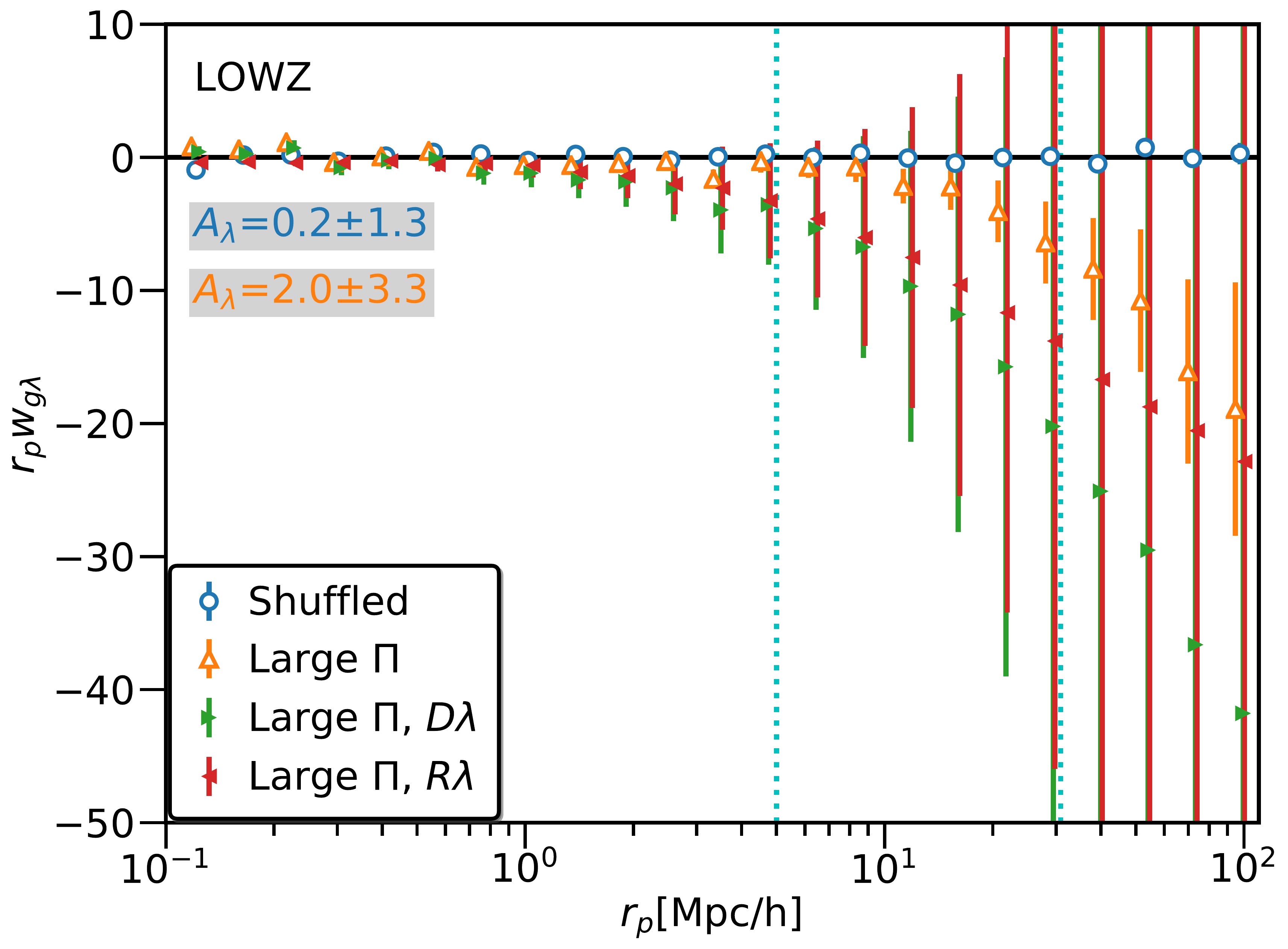}
    		\caption{ Null tests for the cross correlation measurement $w_{g\lambda}$. Blue points (Shuffled) show the results from shuffling the 
				$\lambda$ between different galaxies, which gives results consistent with zero as expected. Note that the shuffling in done in the 
				narrow redshift bins, so that we donot null the impact of variations in the $\mean{\lambda}$ as function of redshift. Our results 
				suggest that such effects are much smaller than the size of the noise in our measurements. Orange points (large $\Pi$) 
				show the measurements in the
				large line of sight bins, where we expect null signal in the absence of significant contributions from lensing. There is significant 
				non-zero signal at large scales, though the bins are very correlated and the overall significance of the signal is small. We fit 
				out model to the measurement (fixing galaxy bias $b_g=1.85$) and $A_\lambda$ is consistent with zero at $\lesssim1\sigma$. We also 
				show the measurements of $D\lambda/RR$ and $R\lambda/RR$ separately (solid red and green points) that contribute to the estimator of 
				$w_{g\lambda}=D\lambda/RR-R\lambda/RR$.
    				}
    	   	\label{fig:cross_corr_null}
    	\end{figure}
		In FP cross correlation measurements for the LOWZ sample, the signal is lower than the best fit model on large scales, which is likely due to 
		some large scale photometric systematics. To test source and potential impact of such a contamination, 
		we show the two null tests for the cross correlation measurements in 
		figure~\ref{fig:cross_corr_null}. 
		
		In the first test, we randomly shuffle the measurement of $\lambda_3$ between galaxies to break any possible correlations
		between $\lambda$ and the galaxies. However, this shuffling is done in narrow redshift bins, so that any possible contributions from the redshift 
		variations in  $\mean{\lambda_3}$ will still be present in the cross correlation measurement (see eq.~\eqref{eq:xigl_density_weight}). For the randomly
		shuffled  $\lambda_3$, we detect null correlation and fitting our model for the FP cross correlations, 
		the $A_\lambda=0.2\pm1.3$ value obtained is consistent with 
		zero as expected, ruling out any significant contributions from redshift dependent variations in $\mean{\lambda_3}$ to our main results.
		
		In the second test, we measure the cross correlations at large line of sight separations, $100<|\Pi|<500\mpch$. For such large line of sight separations, 
		we expect the cross correlation signal to be small (including contributions from weak gravitational lensing). As shown in figure~\ref{fig:cross_corr_null}, 
		we detect a negative signal on large scales, though the measurement is also very correlated across the bins. When fitting our fiducial model (which assumes 
		measurement is at small $\Pi$), we obtain $A_\lambda=2\pm3.3$ to be consistent with zero. This is due to the fact that the systematics causing the non-zero 
		measurement also contribute to the covariance and hence down weighting those scales in the model fits. Contributions from systematics to the 
		covariance are also the reason that the error on $A_\lambda$ for large $\Pi$ test is significantly larger than compared to the shuffled test. 
		
		These tests suggest that the impact of systematics is likely to be negligible on the $A_\lambda$ constraints we obtained in section~\ref{sec:results}.

	\subsection{$\upsilon$ Estimator}
		To remove the impact of any additive systematic with slowly varying correlation funcation, we define
		\begin{equation}
			\upsilon(r_p,r_0)=\frac{2}{r_p^2}\int_{r_0}^{r_p} dr_p' r_p' w_{g\lambda}(r_p') -w_{g\lambda}(r_p)+\frac{r_0^2}{r_p^2}w_{g\lambda}(r_0)
		\end{equation}
		This parameter is independent of the additive systematics with a scale independent correlation function and is in general less sensitive to the 
		systematics with weak scale dependence, as we observe in the measurements of $w_{g\lambda}$.
		Note that the $\upsilon$ parameter is analogous to the $\Upsilon$ parameter defined in the context for galaxy-galaxy lensing by \cite{Baldauf2010}. 
		Under the assumption of angular symmetry, $\upsilon$ with $r_0=0$ is analogous to $w_{g+}$ measured for the intrinsic alignments.
		In this section we will present results with $r_0=0.2\mpch$.

		\begin{figure}
      		\centering
             \includegraphics[width=\columnwidth]{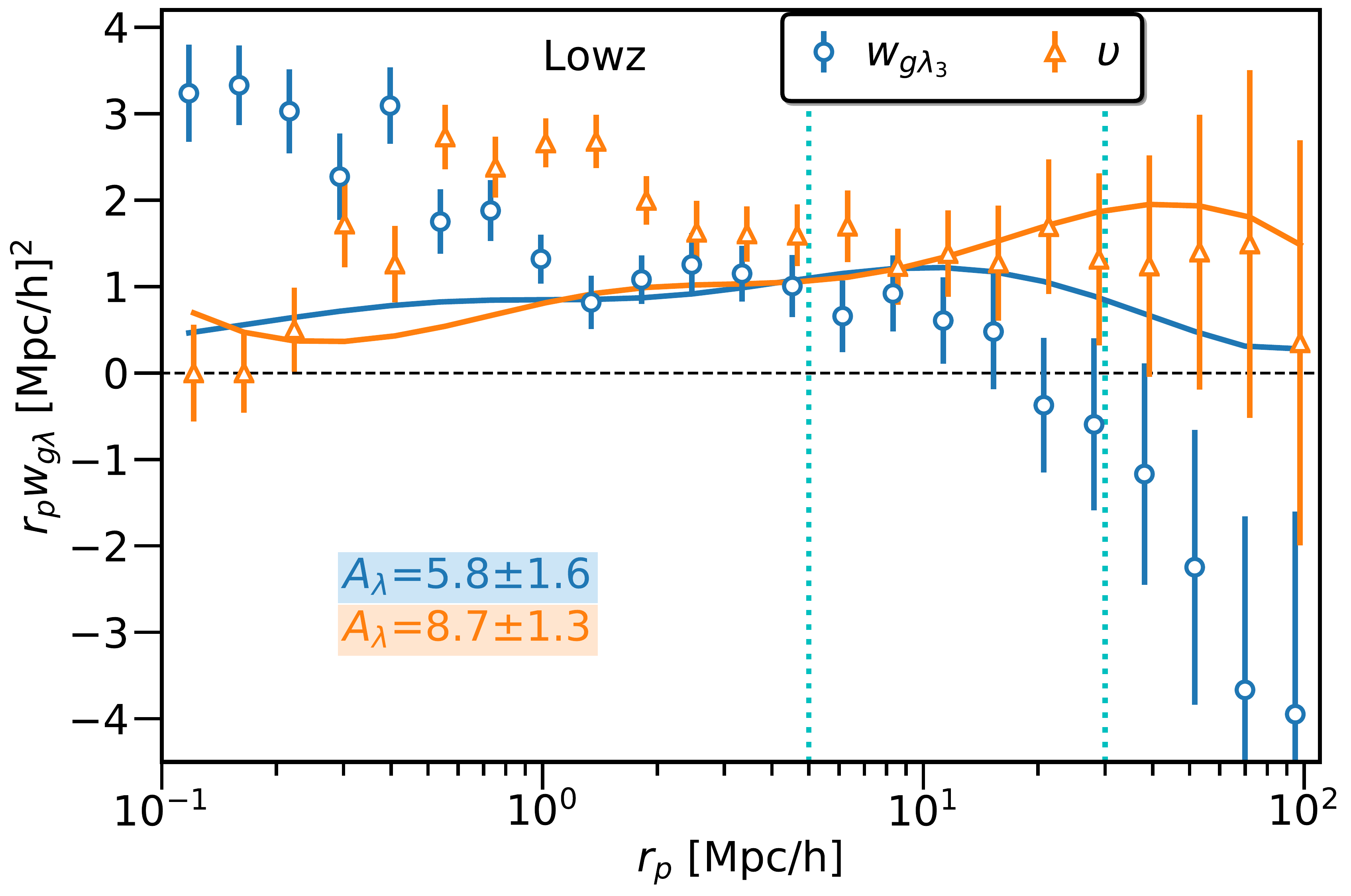}
    		\caption{Comparison of FP cross correlation measurement and model fits, using $w_{g\lambda}$ and $\upsilon$ estimators. Converting $w_{g\lambda}$ to $\upsilon$ reduces
				the impact of systematics with weakly varying correlation function and leads slightly higher $A_\lambda$.
    				}
    	   	\label{fig:lowz_fp_est}
    	\end{figure}
		
		\begin{figure}
      		\centering
             \includegraphics[width=\columnwidth]{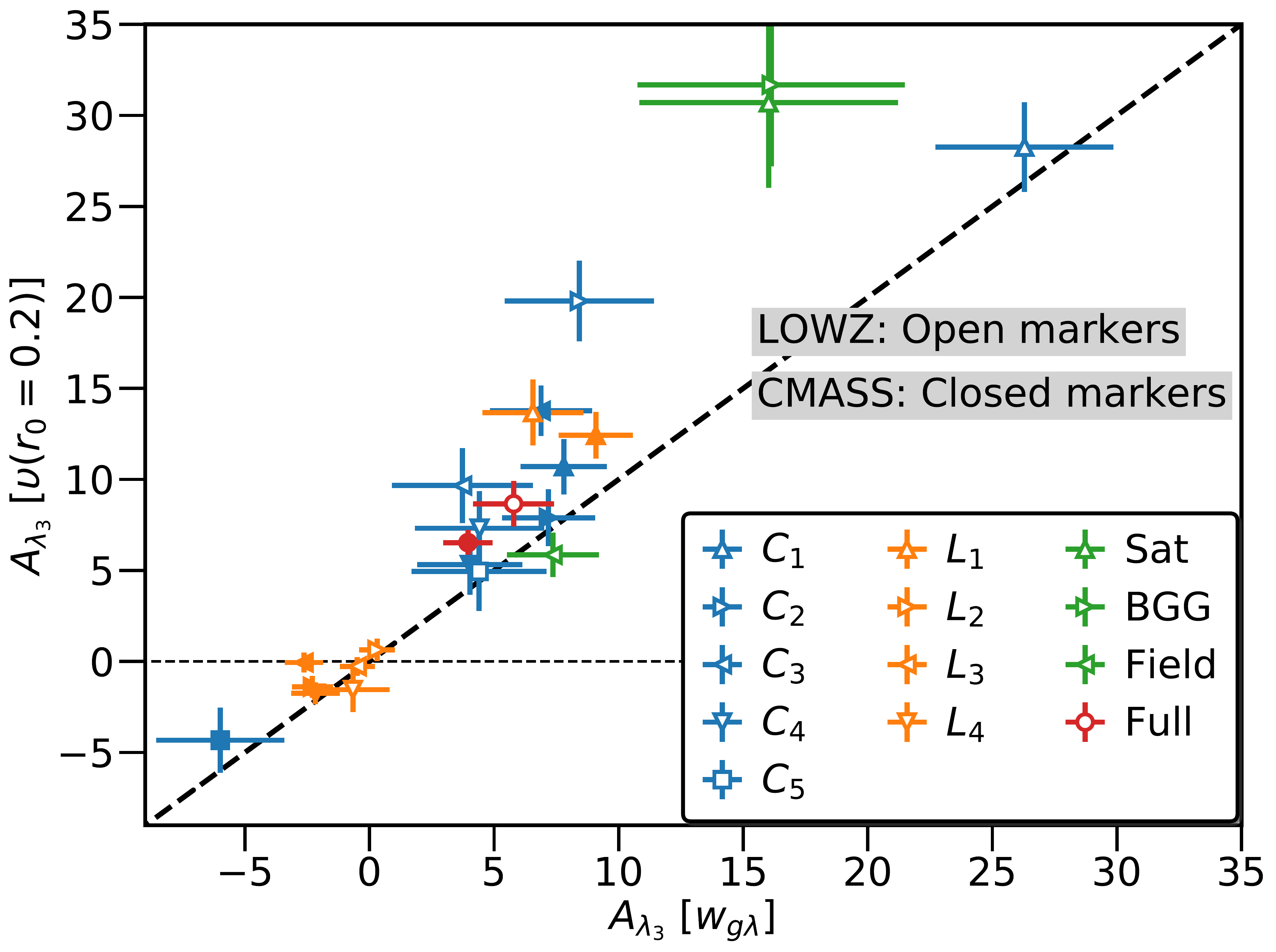}
    		\caption{ Comparison of $A_\lambda$ values obtained using $w_{g\lambda}$ compared to the values obtained using $\upsilon$. Using the $\upsilon$ parameter removes some effects 
		of the additive systematics and hence leads to relatively higher values of $A_\lambda$. However, these values donot change any of the conclusion in the main part of the paper 
		derived using $w_{g\lambda}$ as the qualitative trends do not change.
    				}
    	   	\label{fig:Alambda_est}
    	\end{figure}
		In figure~\ref{fig:lowz_fp_est} we show the comparison of $w_{g\lambda}$ and $\upsilon$ parameters. We fit the same model as $w_{g+}$ to the $\upsilon$ with the correct constant 
		factor as expected for the FP instead of intrinsic alignments. $\upsilon$ measurements are relatively more consistent with the expected 
		scaling from the model as it is expected to be less susceptible to the effects of the systematics. 
		In figure~\ref{fig:Alambda_est}, we show the comparison of the amplitude $A_\lambda$ obtained after fitting $w_{g\lambda}$ versus the $A_\lambda$ obtained after fitting 
		$\upsilon$. $\upsilon$ parameter leads to higher $A_\lambda$. This is likely because $\upsilon$ is less susceptible to the impact of additive systematics, which have negative 
		sign leading to lower $w_{g\lambda}$ (negative $w_{g\lambda}$ on larger scales).

		\begin{figure}
      		\centering
             \includegraphics[width=\columnwidth]{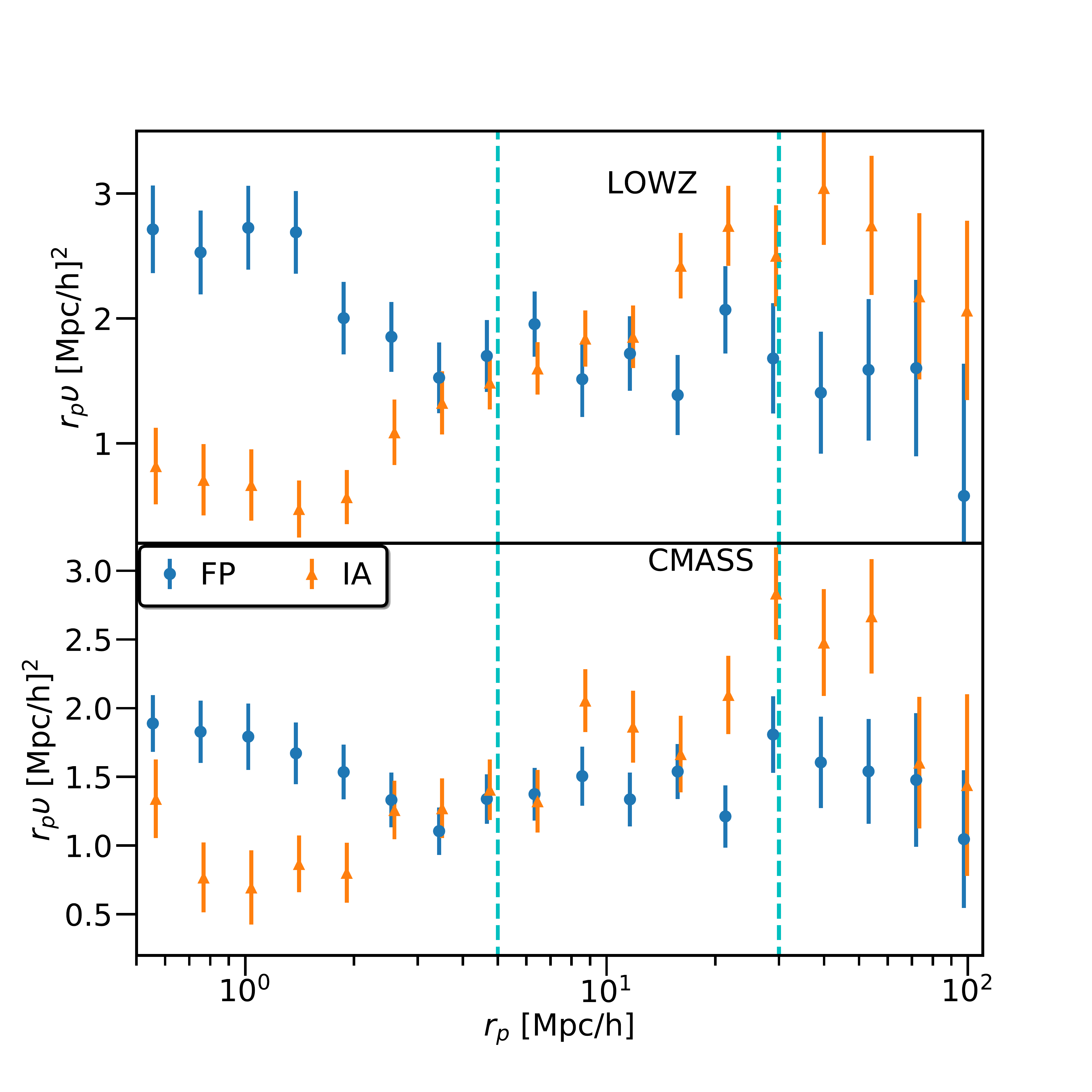}
    		\caption{ Comparison of $\upsilon$ parameter from the fundamental plane to that obtained with the intrinsic alignments of the galaxies (we convert $w_{g+}$ to also 
					$\upsilon$). Under the assumptions of the model, i.e. intrinsic alignments sourcing the FP residuals, the two measurements are expected to have the same scaling
					but different amplitude by factor of 6. On small scales, we expect some deviations as the satellites show very small IA on small scales.
    				}
    	   	\label{fig:IA_fp_est_comp}
    	\end{figure}
		In figure~\ref{fig:IA_fp_est_comp}, we show a comparison of the $\upsilon$ obtained from FP cross correlations and also from the intrinsic alignment of the galaxies. 
		The two measurements are expected to have same scaling, with FP correlations having lower amplitude by factor of 3 (the constants in the model for IA and FP differ by factor of 
		3). On small scales, IA is expected to be lower as the satellite galaxies show lower alignments because the orbital motion can lower the alignments but does not 
		significantly 
		affect the size correlations (assuming angular symmetry of mass distribution). On large scales, we observe FP and IA to have similar amplitude, which is consistent with the 
		observation that $A_\lambda$ is larger than $A_I$ by a factor of $\sim 3$ (note that our model fits are most sensitive to scales $5\lesssim r_p\lesssim 10$ \mpch). Finally we 
		note that the comparison is not strictly fair as the effects of higher order terms from density weighting can still be different in $\upsilon$ derived from FP and IA.

%


\section{RSD analysis}
	\subsection{Posterior distribution for the model parameters}
	
	In this work, we use three different methods to obtain parameter posteriors: 1) Maximum a posteriori (MAP) estimation and Laplace approximation, using the hessian of the log posterior to obtain the inverse covariance matrix, 2) MCMC sampling of the likelihood, assuming Gaussian covariance between multipoles, and 3) optimization-based posterior inference method called EL$_2$O from \cite{Seljak:2019rta}. We obtain the full posterior distributions for each of our 11 model parameters using three aforementioned methods. We find that the posteriors obtained from EL$_2$O agree well with the MCMC posteriors, while MAP can get the peak of the posterior incorrect. This is more evident in the case of the posteriors for parameters which are most non-Gaussian. More detailed analysis about the comparison can be found in \cite{Seljak:2019rta}. In this work we use the EL$_2$O method to obtain the model parameter posteriors.
	

	\subsection{Null test}\label{appendix:nulltestRSD}
	
\begin{figure*}
        \includegraphics[scale = 0.315]{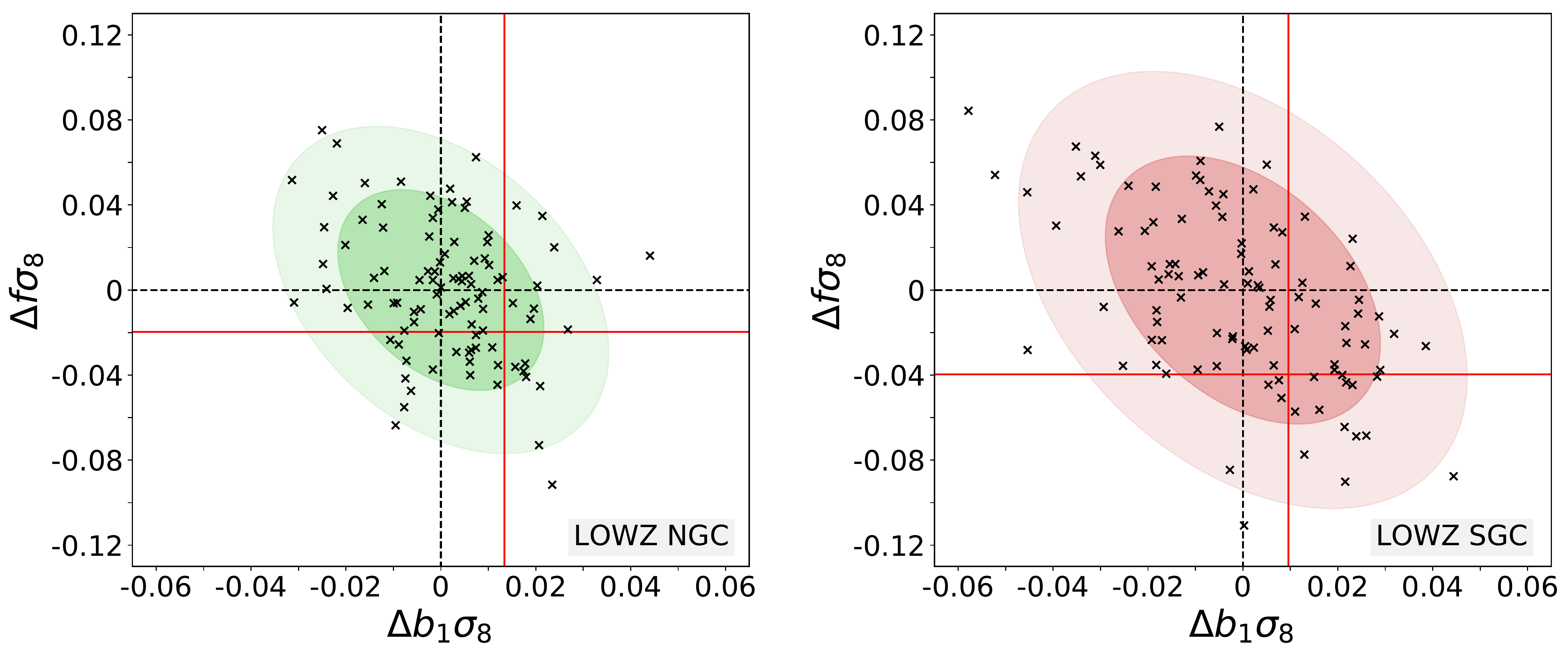}
         \includegraphics[scale = 0.315]{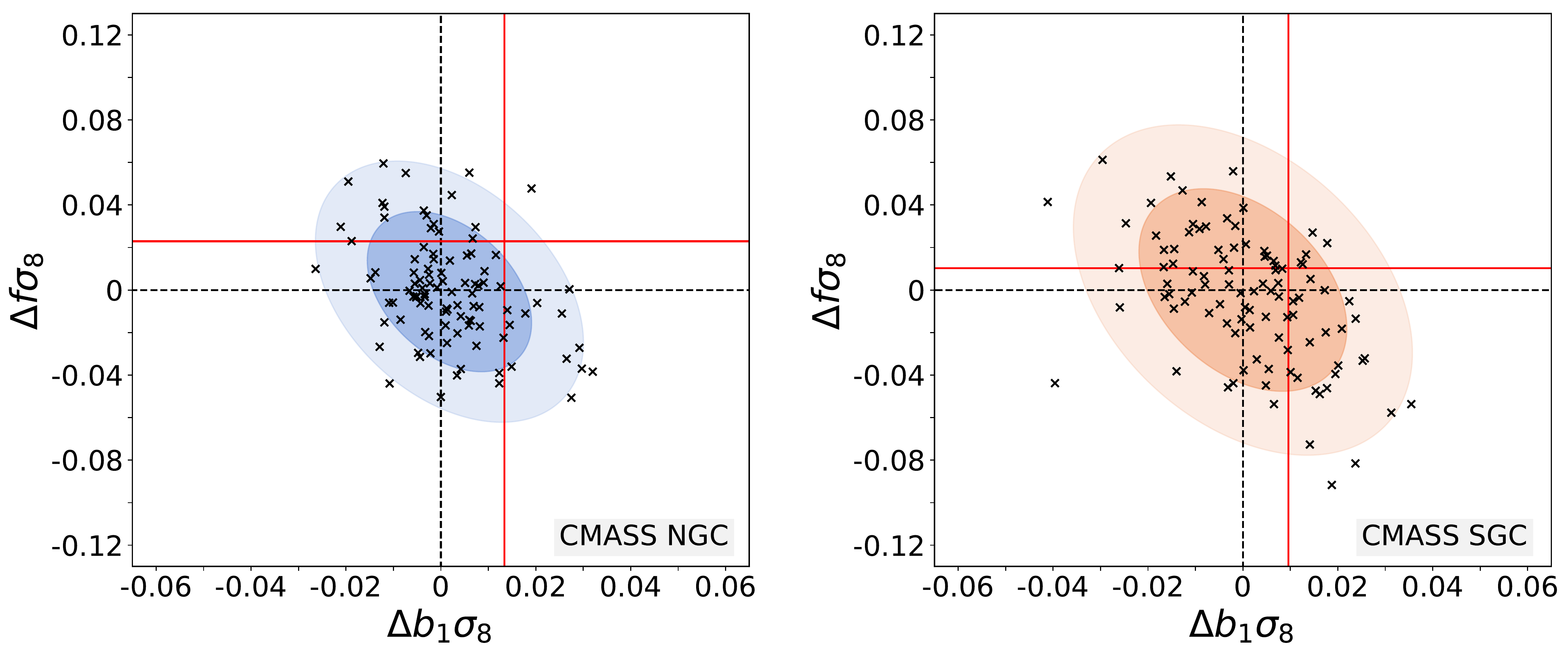}
			\caption{Null test for $\Delta f \sigma_8$ and $\Delta b_1 \sigma_8$ measurements. We take 100 random splits for each galaxy sample and measure the difference in $f \sigma_8$ and $b_1 \sigma_8$ for each pair of random split subsamples. Black dotted points present the measurements for each pair, showing that both measurements are consistent with the null detection, as expected. Colored confidence ellipses correspond to 68\% and 95\% probability contours. Red solid lines indicate the measurement from the $\lambda_1$ split in figure~\ref{fig:FPfit5}. All $\Delta f\sigma_8$ measurements in figure~\ref{fig:FPfit5} are consistent with being drawn from the null distributions.
			}
			\label{fig:ap2}
	    \end{figure*}

	Following appendix~\ref{subsection:nulltest1}, in narrow redshift bins we randomly shuffle the measurements of FP residuals between galaxies. We perform 100 random splits of each galaxy sample (LOWZ and CMASS for NGC and SGC regions), creating a pair of subsamples per each split. Finally, we measure the difference in the parameters $f \sigma_8$ and $b_1 \sigma_8$ for each pair. As shown in figure~\ref{fig:ap2}, for all galaxy samples the difference in both parameters is consistent with the null detection, as expected: $\Delta f\sigma_8$: -0.009 $\pm$ 0.031 and -0.007 $\pm$ 0.041 for LOWZ NGC and SGC respectively and 0.006 $\pm$ 0.024 and -0.008 $\pm$ 0.031 for CMASS NGC and SGC respectively. In figure~\ref{fig:FPfit5}, we find that $\Delta f\sigma_8$ measurements are statistically consistent with the null result, and consequently they are equivalent to being drawn from the distributions in figure~\ref{fig:ap2}.
	
\begin{figure*}
        \includegraphics[scale = 0.26]{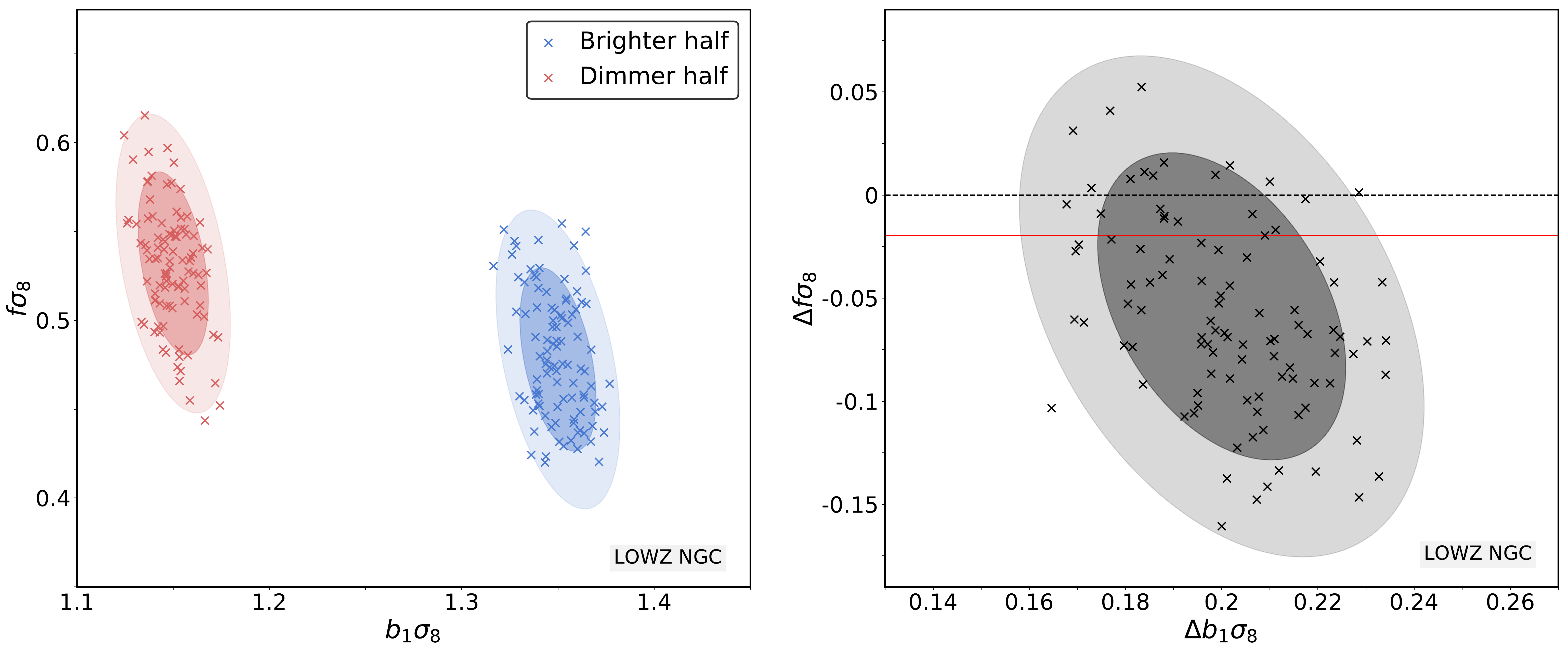}
			\caption{Measuring errors on $\Delta f \sigma_8$ and $\Delta b_1 \sigma_8$. \textit{Left}: We mimic the FP cut on the MultiDark-Patchy mock catalogues by separating each mock into two subsamples according to their stellar masses: ``brighter half," (blue) the subsample with its stellar mass greater than the sample mean, and ``dimmer half," (red) the remaining subsample with smaller stellar mass. We note the FP cut is a `noisy' version of this cut as $\lambda$ is correlated with luminosity with considerable scatter. We show the measurements of $f \sigma_8$ and $b_1 \sigma_8$ parameters from both subsamples. In agreement with appendix~\ref{appendix:lcdep}, we find that a more luminous galaxy has a higher bias. $f \sigma_8$ measurements, on the other hand, are consistent across two subsamples. \textit{Right}: $\Delta f \sigma_8$ and $\Delta b_1 \sigma_8$ measurements, with the values of the dimmer half subtracted from the brighter half. Similar to figure~\ref{fig:ap2}, red solid line indicates the measurement from the $\lambda_1$ split in figure~\ref{fig:FPfit5}.
			}
			\label{fig:ap5}
	    \end{figure*}

	In addition to the random splits of the galaxy sample, we take 100 MultiDark-Patchy mock catalogues to estimate the size of the error bars of RSD constraints. We separate each PATCHY mock into two subsamples based on their stellar masses. We call the subsample with its stellar mass greater than the sample mean the ``brighter half", and the remaining subsample with smaller stellar mass is called the ``dimmer half". This mimics the FP cut in section~\ref{subsubsection:FPcut} because stellar masses are strongly correlated with the galaxy luminosity and consequently FP residuals as observed in section~\ref{ssec:results_FP_lum}. We note that there is considerable scatter between FP residuals and galaxy luminosity and hence the FP cuts are a `noisy' version of the cuts based on stellar mass.
	Combined with random cuts presented earlier, these cuts provide the two limit for the cuts based on FP residuals.
In the left panel of figure~\ref{fig:ap5}, we show the best-fit $f \sigma_8$ and $b_1 \sigma_8$ parameters we obtain by fitting the RSD model to the measured multipoles from subsamples of 100 LOWZ NGC mocks. In agreement with appendix~\ref{appendix:lcdep}, the subsample with a higher luminosity, the brighter half, has a higher galaxy bias. $f \sigma_8$ measurements, on the other hand, are consistent across two subsamples. In the right panel of figure~\ref{fig:ap5}, we measure $\Delta f\sigma_8$ and $\Delta b_1\sigma_8$ between subsamples by subtracting the values of the dimmer half from that of the brighter half. For LOWZ NGC, we obtain $\Delta f\sigma_8 = -0.054 \pm 0.049$. Similarly, we measure the propagated errors of $\Delta f\sigma_8$ for all galaxy samples: $\sigma(\Delta f\sigma_8) =$ 0.049, 0.060, 0.030, and 0.047, for LOWZ NGC, LOWZ SGC, CMASS NGC, and CMASS SGC, respectively. We also create 4 luminosity subsamples, $L_1 - L_4$ with $L_1$ being brightest and $L_4$ being faintest, from the PATHCY mocks, according to their stellar masses. We follow the convention that $L_1 - L_3$ contain 20\% of galaxies each, while $L_4$ contains 40\%. As shown in figure~\ref{fig:ap6}, we find the decreasing trend of galaxy bias with luminosity, confirming our observation in appendix~\ref{appendix:lcdep}. Moreover, figure~\ref{fig:ap6} shows that $f \sigma_8$ measurements are consistent across all luminosity subsamples.

    	   \begin{figure}
        \includegraphics[width=\columnwidth]{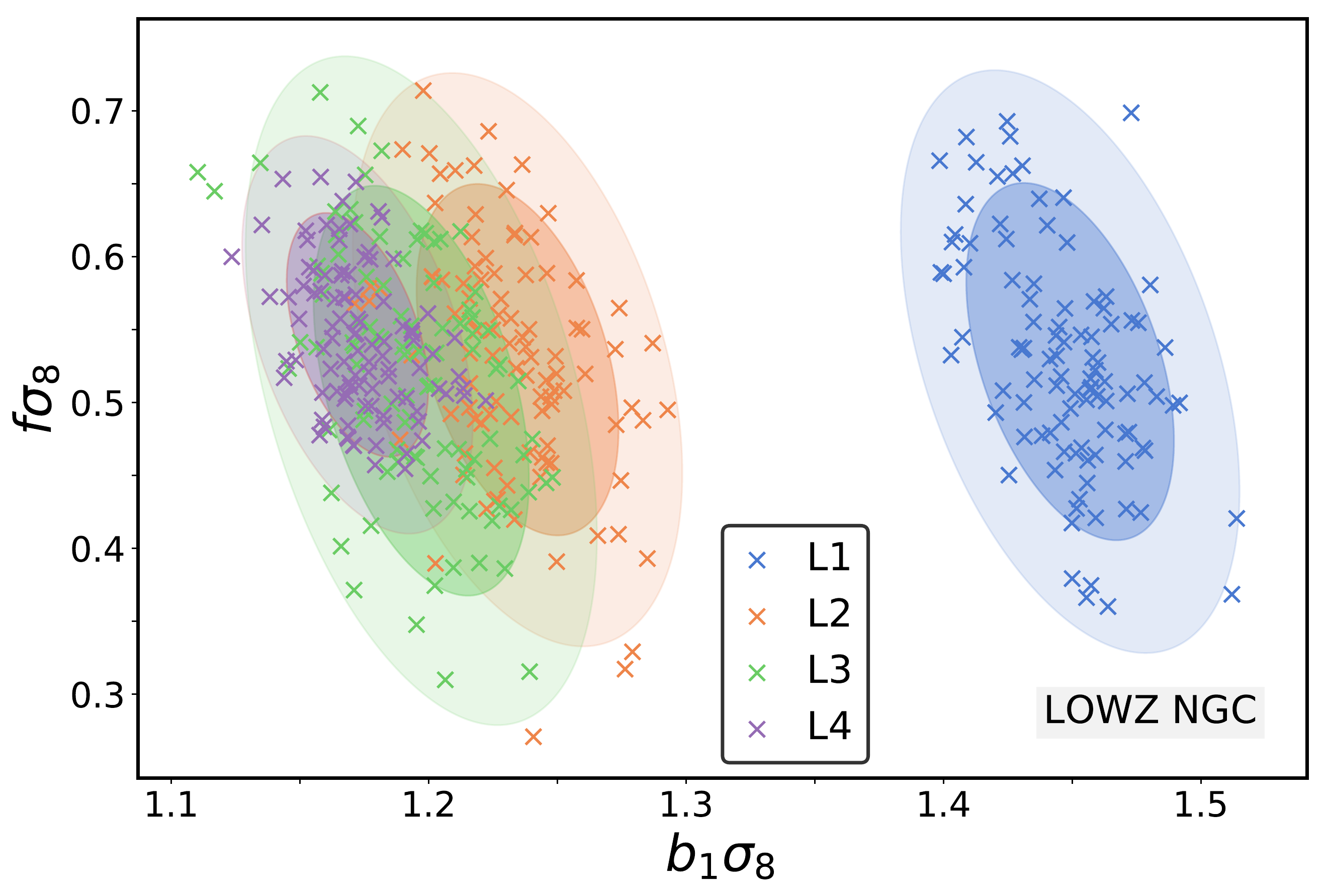}
			\caption{Similar to figure~\ref{fig:ap5}. Measurements of $f \sigma_8$ and $b_1 \sigma_8$ from 4 luminosity subsamples, $L_1 - L_4$ with $L_1$ being brightest and $L_4$ being faintest, of the MultiDark-Patchy mock catalogues. We find the decreasing trend of galaxy bias with luminosity, while $f \sigma_8$ measurements are consistent across all subsamples.
			}
			\label{fig:ap6}
	    \end{figure}

\end{document}